\documentclass[12pt, preprint]{aastex}

\newcommand{\ha}{H$\alpha$}
\newcommand{\hb}{H$\beta$}
\newcommand{\kms}{ \ifmmode{\rm km\thinspace s^{-1}}\else km\thinspace s$^{-1}$\fi}
\newcommand{\dg}{\ifmmode{^{\circ}}\else $^{\circ}$\fi}

\newcommand{\hi}{\ion{H}{1}}
\newcommand{\hii}{\ion{H}{2}}

\newcommand{\lb}{\ifmmode{(\ell,b)} \else $(\ell,b)$\fi}
\newcommand{\nii}{[\ion{N}{2}]}
\newcommand{\niiblue}{[\ion{N}{2}]$~\lambda5755$}
\newcommand{\sii}{[\ion{S}{2}]}
\newcommand{\oiii}{[\ion{O}{3}]}
\newcommand{\hei}{\ion{He}{1}}

\newcommand{\vlsr}{\ifmmode{v_{\rm{LSR}}}\else $v_{\rm{LSR}}$\fi}
\newcommand{\av}{\ifmmode{A(V)}\else $A(V)$\fi}
\newcommand{\ebv}{\ifmmode{E(B-V)}\else $E(B-V)$\fi}
\newcommand{\iha}{\ifmmode{I_{\rm{H}\alpha}} \else $I_{\rm H \alpha}$\fi}
\newcommand{\ihb}{\ifmmode{I_{\rm{H}\beta}} \else $I_{\rm H \beta}$\fi}

\slugcomment{ApJ, accepted}

\begin{document}

\title{A Multiwavelength Optical Emission Line Survey of Warm Ionized 
Gas in the Galaxy}

\shorttitle{Survey of Warm Ionized Gas}

\author{G. J. Madsen\altaffilmark{1}}
\affil{Anglo-Australian Observatory, P.O.~Box 296, Epping, NSW 1710, Australia}
\altaffiltext{1}{NSF MPS Distinguished International Postdoctoral Research Fellow}
\email{madsen@aao.gov.au} 
\author{R. J. Reynolds and L. M. Haffner}
\affil{Department of Astronomy, University of Wisconsin--Madison,
475 N. Charter Street, Madison, WI 53706}

\begin{abstract}

We report on observations of several optical emission lines toward a 
variety of newly revealed faint, large-scale \ha-emitting regions in the 
Galaxy. The lines include \nii$~\lambda6583$, \nii$~\lambda5755$, 
\sii$~\lambda6716$, \oiii$~\lambda5007$, and \hei$~\lambda5876$\ obtained 
with the Wisconsin H-Alpha Mapper (WHAM) toward sightlines that probe 
superbubbles, high latitude filamentary features, and the more diffuse 
warm ionized medium (WIM).  Our observations include maps covering 
thousands of square degrees toward the well-known Orion-Eridanus
bubble, a recently discovered $60\dg \times 20\dg$ bipolar superbubble
centered in Perseus, plus several classical \hii\ regions surrounding
OB stars and hot evolved stellar cores.   
We use the emission line data to explore the temperature 
and ionization conditions within the emitting gas and their variations 
between the different emission regions. We find that in the diffuse WIM 
and in the faint high latitude filamentary structures the line ratios of 
\nii/\ha\ and \sii/\ha\ are generally high, while \oiii/\ha\ and \hei/\ha\ 
are generally low compared to the bright classical \hii\ regions. This 
suggests that the gas producing this faint wide-spread emission is warmer, 
in a lower ionization state, and ionized by a softer spectrum than gas in 
classical \hii\ regions surrounding O stars, the presumed ionization 
source for the WIM. In addition, we find differences in physical 
conditions between the large bubble structures and the more diffuse WIM, 
suggesting that the ionization of superbubble walls by radiation from 
interior O associations does not account entirely for the range of 
conditions found within the WIM, particularly the highest values of 
\nii/\ha\ and \sii/\ha.

\end{abstract}


\section{INTRODUCTION}
\label{sec:intro}

The interstellar medium (ISM) plays a vital role in the ongoing cycle of
stellar birth and death and galactic evolution.  However, the role of
interstellar matter, from how its properties are influenced by stars
to how in turn its properties influence star formation, is poorly
understood and is arguably the least understood portion of the cycle.
Warm diffuse ionized hydrogen has become recognized as a major phase
of the ISM of our Galaxy; see, for example, reviews by \citet{KH87,
  Cox89, Reynolds91b, Mathis00}.   
This phase consists of regions of warm (10$^{4}$ K),
low-density (10$^{-1}$ cm$^{-3}$), nearly fully ionized hydrogen that
occupy approximately 20\% of the volume within a 2 kpc thick layer
about the Galactic midplane \citep[e.g.,][]{Reynolds91a, NCT92, TC93, HRT99}.
Near the midplane, the rms density of \hii\ is less than 5\% that of
\hi. However, because of its greater scale height, the total column
density 
of interstellar \hii\ along high Galactic latitude sight lines is
relatively large, with $N_{\rm{H\sc{II}}} \sim 1/3~N_{\rm{H\sc{I}}}$. 
One kiloparsec above the midplane, warm \hii\ may be the dominant
state of the interstellar medium in the Milky Way.  Widespread,
diffuse ionized gas is now firmly established as an important
constituent of the ISM in external galaxies as well
\citep[e.g.,][]{RKH90, HG90, Dettmar92,  WB94, Ferguson+96, RD00,
  CR01, MV03}.   

Despite its significance, the origin and physical conditions within
the warm ionized medium (WIM) remain poorly understood. In particular,
the ubiquitous nature of this gas is difficult to explain. Of the
known sources of ionization within the Galaxy, only O stars generate
enough power to sustain the WIM \citep{Reynolds92}. 
Therefore, it is generally believed that the O stars, confined
primarily to widely separated stellar associations near the Galactic
midplane, are somehow able to photoionize a significant fraction of
the ISM not only in the disk but also within the halo, 1-2 kpc above
the midplane. 
However, the need to have a large fraction of the Lyman continuum photons from
O stars travel hundreds of parsecs through the disk seems to conflict
with the traditional picture of \hi\ permeating much of the
interstellar volume near the Galactic plane.  
It has been suggested that extensive cavities in the neutral gas,
created either by ``superbubbles'' of hot gas from supernovae
\citep{Norman91}, or  carved out by O star photons in low-density
regions \citep{MC93}, may extend far above the midplane \citep{DS94,
  DSF00}. 
Although the existence of \hi\ superbubbles has long been established
\citep{Heiles84}, direct observational evidence that cavities are
actually responsible for the transport of hot gas and ionizing
radiation up into the Galactic halo is very limited.  One piece of
evidence for 
such transport has been provided recently by the WHAM \ha\ sky survey 
\citep[][also see \S6 below]{RSH01}.

The WHAM \ha\ sky survey is a velocity-resolved map of diffuse 
interstellar \ha\ emission at $1\dg$ angular resolution over the entire 
northern sky ($\delta > -30\dg$) within approximately $\pm100$ \kms\ of 
the local standard of rest (LSR) \citep{Haffner+03}. The survey maps show 
\ha\ emission covering the sky, with ionized gas associated with large 
scale loops, filaments, and bubbles superposed on a fainter, diffuse 
background, as well as the bright classical \hii\ regions near the 
Galactic plane.  Several of the high latitude structures appear to be 
associated with hot stars and OB associations; however, the diffuse
background and many features superposed upon it have no clear
association with known ionizing sources. This survey provides the
basis for studies of the physical conditions within these newly
revealed emission regions and the source of their ionization. 

Even though the primary source of ionization is believed to be O stars, 
the temperature and ionization conditions within the diffuse ionized gas 
differ significantly from conditions within classical O star \hii\ 
regions.  These conditions have been inferred by using optical line ratios 
diagnostic techniques. For example, anomalously strong 
\sii$~\lambda6716$/\ha\ and \nii$~\lambda6583$/\ha, and weak 
\oiii$~\lambda5007$/\ha\ emission line ratios (compared to the bright, 
classical \hii\ regions) indicate a low state of excitation, with few ions 
present that require ionization energies greater than 23 eV 
\citep{Reynolds85, HRT99, Rand97}.  This is consistent with the small 
value of
\hei~$\lambda5876$/\ha\ near the midplane, indicating that the ionization 
fraction of helium is low and suggesting that the spectrum of the diffuse 
interstellar radiation field that ionizes the hydrogen is significantly 
softer than that from the average Galactic O star population \citep{RT95, 
TuftePhD}. In addition, the elevated \nii/\ha\ and \sii/\ha\ ratios in the 
WIM suggest that this low density diffuse gas is significantly warmer 
than traditional \hii\ 
regions and may require spectral processing of the stellar radiation
\citep[e.g.][]{WM04} and/or an additional heating  
source beyond photoionization \citep{RHT99}. Recent observations of 
\niiblue/\nii$~\lambda6583$ have indeed confirmed that the WIM is about
2000 K warmer than \hii\ regions \citep{Reynolds+01}.

Below we present new WHAM observations of \nii$~\lambda6583$, \sii$~\lambda6716$, 
\oiii$~\lambda5007$, \hei$~\lambda5876$ and \niiblue\ toward large-scale 
emission structures as well as individual lines of sight, representing a 
substantial increase in the number of 
observations of these diagnostic emission lines in the Galaxy. This is 
primarily an empirical study. The emission regions examined span a wide 
range in location, environment, morphology, and scale, and we have 
compared the line intensity ratios in these different environments in 
order to explore the variations in physical conditions between them.  A 
more detailed analysis and deeper understanding of all these different 
regions, the relationships between them, and the reasons for the
observed differences will require combining these observations with
photoionization  
models, and is beyond the scope of this work. 

We begin with an overview  of the relationship between the emission
line ratios and the temperature  and ionization state of the gas in
\S\ref{sec:physconds}.  
Our observational techniques and data reduction procedure are discussed in 
\S\ref{sec:obs}. In \S\ref{sec:hii}, we present our results for several 
classical O-star \hii\ regions that form the basis for comparison to the 
fainter \ha\ emission structures. To illustrate the general spectral 
difference between the \hii\ regions and the WIM, we present in 
\S\ref{sec:wim} spectra toward one of the \hii\ regions, where the diffuse 
gas and the classical \hii\ region are along the same line of sight, but 
at separate radial velocities. Two large bubble-shaped features that each 
span more than 40\dg, the Orion-Eridanus bubble and the Perseus 
superbubble, are discussed in \S\ref{sec:bubbles} along with
comparisons to the \hii\ regions and the WIM. Observations of high
latitude filamentary structures are presented in \S\ref{sec:hlfil}.  A
direct measure of the temperature of ionized gas,  
through observations of \niiblue\ and \nii$~\lambda6583$, is discussed in 
\S\ref{sec:niiblue}, followed by a summary and conclusions in \S\ref{sec:summary}.

\section{EMISSION LINE RATIOS AND PHYSICAL CONDITIONS}
\label{sec:physconds}

Observations of optical emission lines and their relative strengths is a 
common diagnostic tool used to assess the physical conditions of ionized 
gas.  The WHAM sky survey has measured the \ha\ surface brightness, 
\iha, which is directly proportional to the emission measure. In the 
absence of extinction, this relationship is \begin{equation} EM \equiv 
\int n_e^2 dl = 2.75~T_4^{0.9}~\iha~\rm{(cm}^{-6}~\rm{pc)}, \end{equation} 
where $T_4$ is the temperature of the gas in units of 10$^4$~K, and \iha\ 
is measured in Rayleighs\footnotemark \footnotetext{1 R = $10^6$/4$\pi$ 
photons s$^{-1}$ cm$^{-2}$ sr$^{-1}$}
\citep{Haffner+03}.  

In the WIM, the collisionally excited lines of \nii$~\lambda6583$ and 
\sii$~\lambda6716$ are the next brightest optical lines that can be 
observed with WHAM. \citet{HRT99} presented the first velocity resolved 
maps of these lines in the Galaxy, toward a 40\dg$\times$30\dg\ region in 
Perseus. 
Radial velocity interval maps showed a strong trend in \nii/\ha\ and
\sii/\ha, in which these ratios were higher toward regions of low \ha\
emission, while \sii/\nii\ remained relatively  
constant. These line ratio variations can be interpreted as variations
in the temperature and ionization state of the gas as follows. Using
the standard formulation for the strengths of collisionally excited
lines, the \nii/\ha\ intensity ratio can be parameterized as 

\begin{equation} \frac{[\rm{N~\textsc{II}}]}{\rm{H}\alpha} = 
1.62\times10^5~T_4^{0.4}~e^{-2.18/T_4} 
\left(\frac{\rm{N}^+}{\rm{N}}\right) \left(\frac{\rm{N}}{\rm{H}}\right) 
\left(\frac{\rm{H}^+}{\rm{H}}\right)^{-1}, \label{eq:niieq} 
\end{equation} 

\noindent where the lines strengths are measured in energy units, $T_4$ is 
the temperature in units of 10$^4$~K, N/H is the gas phase abundance by 
number, and N$^+$/N and H$^+$/H are the ionization fractions of N and H, 
respectively \citep{HRT99, Osterbrock89}. The similar first ionization 
potentials of N and H (14.5 and 13.6 eV, respectively), along with N-H 
charge-exchange, mean that in the WIM the ionization fraction of N$^+$/N 
is expected to be similar to H$^+$/H. The fraction H$^+$/H is observed to 
be near unity in the WIM \citep{Reynolds+98, Hausen+02-aj}, and the high 
second ionization potential of N (29.6 eV) means that little N is likely 
to be in the form of N$^{++}$. 
This is supported by the weak \oiii/\ha\ ratios in the WIM (see below)
and by photoionization modeling \citep[e.g.,][]{Sembach+00}, which
have shown that N$^+$/N $\approx 0.8$ over a wide range of input
spectra and ionization parameters.  
As a result, N$^+$/H$^+$\ is likely to depend almost 
entirely on the gas phase abundance of N/H, which we have assumed to be 
the same for all the emission regions.
 Using this argument, \citet{HRT99} attributed the higher \nii/\ha\
 ratios in the WIM to higher temperatures of the gas, which has since
 been confirmed (see \S\ref{sec:niiblue}). They found in the Perseus
 spiral arm, for example, an increase in temperature from $T  
\approx 7000$ K close to the Galactic plane up to $T \gtrsim 9000$ K at 
$\sim$ 1 kpc from the midplane. 

The observed variations in \sii/\ha\ can 
also be interpreted as a change in temperature. However, the second 
ionization potential of S (23.4 eV) is just below the neutral He edge at 
24.6 eV.  Therefore, a significant fraction of S can be S$^{++}$, and the 
ratio of \sii/\ha\ is a combination of both temperature and ionization 
effects. The ratio of \sii/\nii, however, is insensitive to temperature 
because of the nearly identical energies required to excite the lines. 
This ratio can be parameterized as

\begin{equation}
\frac{[\rm{S~\textsc{II}}]}{[\rm{N~\textsc{II}}]} = 4.62~e^{0.04/T_4}~
        \left(\frac{\rm{S}^+}{\rm{S}}\right)
        \left(\frac{\rm{S}}{\rm{H}}\right)
    \left[\left(\frac{\rm{N}^+}{\rm{N}}\right)  
    \left(\frac{\rm{N}}{\rm{H}}\right)\right]^{-1},     
\label{eq:siieq}
\end{equation}

\noindent with the same conventions as equation \ref{eq:niieq} 
\citep{HRT99, Osterbrock89}.  By assuming that N$^+$/N does not change in 
the WIM, and adopting a value for the abundances of N and S, equation 
\ref{eq:siieq} can be used to estimate S$^+$/S.

A more direct method of measuring the temperature of ionized gas is 
through observations of multiple emission lines from the same ion. The 
extremely faint ``auroral" line of \nii$~\lambda5755$, along with 
\nii$~\lambda6584$, are two such diagnostic lines that are within the 
observational capabilities of WHAM. The ratio of the emissivity of these 
lines is given simply by

\begin{equation}
\frac{[\rm{N~\textsc{II}}]~\lambda5755}{[\rm{N~\textsc{II}}]~\lambda6584} = 0.192~e^{-2.5/T_4},
\label{eq:niiblueeq}
\end{equation}

\noindent \citep{Osterbrock89}. \citet{Reynolds+01} were the first to 
detect this auroral line in the warm ionized medium. Along a single 
sightline toward the Perseus spiral arm, \lb\ = (130\dg, -7.5\dg), this 
ratio was found to be twice as high as observed in traditional O-star 
\hii\ regions. They concluded that the WIM in this direction is $\approx$\ 
2000 K warmer than in \hii\ regions, and that the elevated ratios of 
\nii/\ha\ and \sii/\ha\ in the WIM are in fact due, at least in part, to 
higher temperatures.

Because the second ionization potential of oxygen is 35 eV, observations 
of \oiii$~\lambda5007$ can provide information about the higher ions.  In 
particular, in regions where \oiii/\ha\ is large, the assumption above 
that N$^{++}$ is small may not be valid. The ratio of strengths of \oiii\ 
and \ha\ can be parameterized as

\begin{equation}
\frac{[\rm{O~\textsc{III}}]}{\rm{H}\alpha} = 1.74\times10^5~T_4^{0.4}~e^{-2.88/T_4}
\left(\frac{\rm{O}^{++}}{\rm{O}}\right)
\left(\frac{\rm{O}}{\rm{H}}\right)
\left(\frac{\rm{H}^+}{\rm{H}}\right)^{-1}
\label{eq:oiiieq}
\end{equation}
\citep{Osterbrock89, OGR02}.
\citet{Reynolds85} searched for \oiii\ emission in the 
diffuse WIM along two lines of sight in the Galactic plane, and found that 
\oiii/\ha\ is very low, $\approx$ 0.06. We confirm this result.

A constraint on the hardness of the radiation field
is provided by observations of \hei$~\lambda5876$. This recombination line 
is the helium equivalent of Balmer-$\alpha$, and thus is related to the 
number of He-ionizing photons with $h\nu > 24.6$ eV. The strength of 
\hei\ relative to \ha\ is given by

\begin{equation}
\frac{\rm{He~\textsc{I}}}{\rm{H}\alpha} \simeq
0.47~T_4^{-0.14}~\left(\frac{\rm{He}^+}{\rm{He}}\right) 
\left(\frac{\rm{He}}{\rm{H}}\right)
\left(\frac{\rm{H}^+}{\rm{H}}\right)^{-1}
\label{eq:heieq}                                    
\end{equation}
\citep{RT95} and is therefore a measure of the relative flux of helium
ionizing photons to hydrogen ionizing photons.  
In directions at low Galactic latitude (where \iha $\approx$ 10 
R), \citet{RT95} and Tufte (1997) found \hei/\ha\ significantly 
below that measured for O star \hii\ regions. These observations implied 
that the spectrum of the diffuse radiation field in the WIM, at least 
along those low latitude lines of sight, is significantly softer than that 
from the average Galactic O star population.

We have extended these emission line analyses to many other directions in 
order to explore the variations in conditions within the different regions 
of interstellar ionized hydrogen mapped by the WHAM survey.  When 
estimating the physical conditions, we assume that H$^+$/H = 1,
and N$^+$/N = 0.8.  When estimating electron densities from the emission measure, we assume the gas is at $T=8000$K and completely fills a spherical or cylindrical volume defined by its appearance on the sky.
Except where noted, we also assume interstellar gas phase abundances of N/H = $7.5\times10^{-5}$ \citep{MCS97}, S/H = $1.86\times10^{-5}$ \citep{AG89}, O/H = 
3.19$\times10^{-4}$ \citep{MJC98}, and He/H = 0.1.

\section{OBSERVATIONS}
\label{sec:obs}

All of our observations were obtained with the Wisconsin H-Alpha Mapper 
(WHAM) spectrometer. WHAM was specifically designed to detect very faint 
optical emission lines from the diffuse interstellar medium, and consists 
of a 0.6~m siderostat coupled to a 15 cm dual-etalon Fabry-Perot system 
\citep{TuftePhD, Haffner+03}. It produces a spectrum at a resolution of 12 
\kms\ within a 200 \kms\ wide spectral window, integrated over circular, 
$1\dg$ diameter field of view. The spectrometer can be centered on any 
wavelength between 4800 and 7400 \AA. WHAM is located at the Kitt Peak 
National Observatory in Arizona and is completely remotely operated from 
Madison, Wisconsin.

The data presented here can be separated into two categories: survey mode 
observations and pointed mode observations.  The large-scale maps of \nii\ 
and \sii\ toward the Orion-Eridanus bubble ($\sim$ 400 deg$^2$) and the 
Perseus bipolar superbubble ($\sim$ 2400 deg$^2$) were taken in survey 
mode, similar to the manner in which the WHAM-NSS data were obtained 
\citep[see][]{Haffner+03}. For this mode the observations were divided 
into contiguous `blocks', with each block consisting of up to 49 
observations that cover an approximately $6\dg \times 7\dg$ area of the 
sky.  Each direction within a block was observed once for 60 s in each 
line, with several blocks observed per night. In addition, spectra of 
\nii$~\lambda6583$, \niiblue, \sii, \hei, \oiii, and were obtained toward 
a number of individual sightlines in pointed mode to probe selected parts 
of emission features at a high signal-to-noise ratio. Hereafter, \nii\
will refer to the $\lambda6583$  
line, unless otherwise indicated. Pointed observations were made by 
alternating between a given sightline toward a selected feature, an {\sc{ON}} 
direction, and an accompanying, usually nearby, {\sc{OFF}} direction.  Each pair 
of observations was observed for 120 s at a time (to get well above the effective 
$\pm$\ 4 $e^-$\ readnoise of WHAM's CCD) with a total {\sc{ON}} integration time 
of a few hours for the faintest lines (e.g., \nii$~\lambda5755$).  The {\sc{OFF}} 
directions were chosen to be as spatially close to the {\sc{ON}} direction as 
possible and to provide the off-source spectrum containing the atmospheric 
foreground and Galactic background emissions.  All of the observations 
were carried out during clear, dark of the moon nights to avoid the 
contribution of scattered solar and terrestrial lines in the spectra.

\subsection{Removal of Atmospheric Emission Lines}

The geocoronal \ha\ line, with \iha\ $\approx$ 5-10 R, is the strongest 
terrestrial emission line contaminating the Galactic spectra.  However, in 
addition to \ha, all of the spectra are contaminated by much weaker 
atmospheric lines, typically 5-7 with $I\approx 0.05 - 0.5$ R and FWHM 
$\lesssim 10$ \kms\ within the 200 \kms\ spectral window.  The positions 
of these lines, which are largely unidentified, are fixed with respect to 
a geocentric reference frame.  Their strength is observed to vary with 
both position in the sky and with time during the night, sometimes by up 
to a factor of two.  However, their relative strengths do not appear to 
change by more than $\approx$ 10\% \citep{Haffner+03, Hausen+02-apj}.

To remove this faint foreground emission in the survey mode observations, 
an atmospheric line template was fitted to and then subtracted from each 
spectrum, in the manner described by \citet{Haffner+03}. These
templates were constructed by observing the faintest  
direction in the \ha\ sky, in multiple emission lines, for an entire 
night.  This direction is near the Lockman Hole \citep{LJM86}, and has a 
total \ha\ intensity of \iha\ $\lesssim$\ 0.1 R \citep{Hausen+02-apj}.  
The average of the all the spectra had a sufficient signal-to-noise ratio 
to reveal the location, width, and relative intensities of these atmospheric lines.  
An examination of the changes among individual spectra taken through the 
night confirmed the terrestrial origin of the lines and provided a measure 
of their overall strength.  Since the relative strengths of the lines 
appear to be constant, the template was multiplied by a single number, a 
scaling factor, for each block of survey mode observations. The value of 
this scaling factor was determined by matching the strengths of 
atmospheric lines in parts of block-averaged spectra that did not show any 
Galactic emission. The scaled template was then subtracted from each 
spectrum in the block.

In pointed mode spectra, the atmospheric lines were removed by subtracting 
an appropriate {\sc{OFF}} from the {\sc{ON}} to produce a flat continuum. Past 
exprience with this technique has shown that the degree to which the 
atmospheric emission in the {\sc{OFF}} spectrum represents the emission in the {\sc{ON}} 
spectrum is usually the dominant source of uncertainty in the resulting 
Galactic spectrum \citep[e.g.,][]{Madsen+01, Gallagher+03}.  The large 
number of short exposure time observations employed in this study, 
combined with 
the alternating {\sc{ON}}/{\sc{OFF}} observing technique yielded a very good subtraction 
of the atmospheric lines.  Some spectra were sensitive to lines as faint 
as $I\approx 0.005$ R.  This lower limit is set by random errors in the 
baseline as well as a slightly incomplete subtraction of the atmospheric 
lines.

\subsection{Intensity Calibration}
\label{sec:intcalib}

Intensity calibration involved several steps.  The \ha\ spectra were 
calibrated using synoptic observations of a portion of the North America 
Nebula (NAN), which has an \ha\ surface brightness of \iha = 800 R with an 
uncertainty of $\approx 10\%$ \citep{Scherb81, Haffner+03}. All of the 
other emission line data were calibrated initially by determining the throughput of 
the spectrometer at the different emission line wavelengths relative to 
\ha.  These included the quantum efficiency of the CCD, the transmission 
of the narrowband (FWHM $\approx$ 20\AA) interference filters, the 
transmission of the atmosphere, and a correction for the properties of the WHAM 
optical train (e.g. coatings, reflections, and transmission of the mirrors and lenses). The corrections for the CCD quantum efficiency and the 
interference filters were based on data provided by the manufacturers of 
those systems.  The transmission of the atmosphere could not be well 
determined on a night-by-night basis, due to a lack of observations of a 
large number of calibration targets taken at a variety of airmasses each 
night.  A few nights, however, were spent observing enough targets to 
determine the average relative transmission of the atmosphere.  We 
determined an average zenith transmission that ranges from 94\% at \sii\ 
to 85\% at \hb. These data agree well with the standard atmospheric 
transmission curve at Kitt Peak that is included in the popular IRAF data 
reduction package.  We assumed that the relative transmission of the 
atmosphere was the same for each night of observing and that the 
atmosphere is plane-parallel.

The correction for the transmission of the WHAM optics was 
determined empirically by using a combination of \ha\ and \hb\ 
observations toward a part of the large \hii\ region surrounding Spica 
($\alpha$ Vir), a nearby B1~III star. In the absence of extinction, the 
photon number ratio of \iha\ to \ihb\ of warm, low density photoionized 
gas is 3.94, set by the `Case B' recombination cascade of hydrogen 
\citep{Osterbrock89, HS87}. We assume that the emission from the Spica 
\hii\ region suffers no extinction because of its proximity 
\citep[$d \approx 80$ pc;][]{Hipparcos}, high Galactic latitude ($b 
\approx +50\dg$), and the low interstellar hydrogen column density to the 
exciting star \citep[$1.0 \times 10^{19} $ cm$^{-2}$;][]{YR76}.  After 
applying the CCD, interference filter, and atmospheric corrections mentioned above, we found that the \hb\ spectra 
needed to by multiplied by an additional factor of 1.36 for the ratio of 
\iha/\ihb\ to be equal to the expected value of 3.94.  We assume that this 
decrease in transmission from the red to the blue is linear with 
wavelength, and interpolate this correction for the other, redder, 
emission lines. As a consistency check, fully corrected \ha\ and \hb\ 
spectra toward NAN yielded an \iha/\ihb\ ratio of 5.1, consistent with 
observations of extinction toward stars in the nebula \citep{Cambresy+02}. 
The results presented in \citet{MR05} also confirm the validity of this 
calibration technique.  Note that nearly all the results presented here 
involve comparing line ratios in one emission region with the 
corresponding ratio in another region.  As a result, much of the analysis 
is insensitive to calibration errors between different wavelengths.

\subsection{Measurement Uncertainties}
\label{sec:errors}

The velocity calibrations of these spectra were derived from observations
of bright, narrow emission line HII regions, and are based on the
assumption that the emission from all of the lines from an individual   
\hii\ region are at the same velocity with respect to the local standard
of rest (LSR).  For \ha, \hb, and \sii, relatively bright terrestrial     
emission lines within the spectra were used to confirm the calibration.
The calibrations were also checked against an empirical prediction based on
the tunes of the Fabry-Perot etalons \citep[see ][Chapter 5]{MadsenPhD}.  The
resulting systematic uncertainty is estimated to be typically 2-3
kilometers per second. The uncertainty in the velocity calibration due to
random noise in the data is only a few tenths of a kilometer per second.

One of the contributions to the uncertainty in emission line strengths 
comes from the random errors in measuring the level of the continuum.  A 
least-squares linear fit was used to estimate the continuum level and 
remove it from each spectrum. However, there is scatter in the residual 
due to the Poisson statistics of the detected photons as well as from the 
incomplete subtraction of the atmospheric lines. This scatter introduces 
an uncertainty when integrating the area under the emission line. In the 
pointed mode observations, this dominant source of uncertainty was 
generally $I \approx 0.01$ R, and in some cases half of that value.  
The 1$\sigma$ values are listed in Tables~\ref{tab:hiibasic} through 
\ref{tab:niiblue}.  For some observations, the measured line strength is negative, indicating a non-detection.  In this case an upper limit to the line strength, equivalent to the 1$\sigma$ uncertainty, is given in the Tables.

In the survey mode spectra, on the other hand, the large area of the sky 
observed in this mode prohibited an alternating {\sc{ON}}/{\sc{OFF}} observing 
technique. Also, these spectra were obtained with considerably shorter 
total exposure times and often have Galactic emission present across much 
of the 200 \kms\ spectral window.  As a result, the dominant source of 
uncertainty in this mode is the removal of the atmospheric lines, which in 
turn depends upon the accuracy of the atmospheric line template and the 
uncerainty in its scaling factor. We conservatively estimate the 
uncertainty in the scaling factor to be 30\%, based on the visual 
appearance of the corrected spectra, as well as the values of the 
different scaling factors used in all of the spectra. The uncertainty
varies with position within each  
spectrum, because the atmospheric lines only appear in certain places in 
each spectrum. The error bars that appear in the figures below for survey 
mode data represent this 30\% uncertainty in the atmospheric scaling 
factor.

\section{OBSERVATIONS OF CLASSICAL \hii\ REGIONS}
\label{sec:hii}

The warm ionized medium is thought to be ionized primarily by Lyman 
continuum photons from hot stars, although the mechanism by which this 
happens is largely unknown (see \S\ref{sec:intro}). Therefore, to investigate 
the nature of the WIM, it is useful to compare the observed emission line 
ratios in the faint diffuse emission regions with the 
corresponding 
ratios observed in the bright classical \hii\ regions immediately 
surrounding hot stars.  In this section we discuss the results of emission 
line strengths of \hb, \nii, \sii, \oiii, and \hei\ relative to \ha\ for a 
collection of 13 O-star \hii\ regions, plus two regions of ionized gas 
surrounding hot evolved stellar cores.  Some of these \hii\ 
regions (immediately surrounding O stars in Orion OB1 and Cas OB6) are 
associated with much larger extended regions of filaments and loops, 
allowing us to explore the question of whether the diffuse WIM is the 
superposition of such extended structures surrounding some O stars and O 
associations (\S\S\ref{subsec:ori} and \ref{subsec:per}).

Tables~\ref{tab:hiibasic} and \ref{tab:hiimulti} summarize 
the \hii\ region observations, which were taken in pointed mode. The {\sc{OFF}}s 
were selected based on the \ha\ maps from the WHAM-NSS. They were chosen 
to be as close to the \hii\ region as possible, but in regions where the 
emission is diffuse and could be considered to be part of the WIM. This 
selection criteria was somewhat subjective, and many {\sc{OFF}}s were tens of 
degrees away from the \hii\ regions. However, the resulting line 
intensities are mostly insensitive to the selection of {\sc{OFF}}s, because the 
\hii\ region emission lines are much stronger than those of the background 
WIM and the atmosphere.

The first column in the top part of Table~\ref{tab:hiibasic} gives the 
names of the O-star \hii\ regions from the catalogs of 
\citet{Westerhout58}, \citet{Sharpless59}, and \citet{Sivan74}.  The names 
and spectral types of the stars or OB associations thought to be creating 
the \hii\ regions are listed in the second and third columns, 
respectively. The identification of the ionizing sources come from the 
angular proximity of the \hii\ regions to stars found in the databases of 
SIMBAD and the O-star catalog of \citet{Maiz+04}, and should not be 
necessarily considered secure. For the OB associations, the listed 
spectral type is for the hottest known member of the association. The 
exciting O stars have been sorted in order of increasing stellar 
temperature.  The bottom two rows of the table provide information about 
the two \hii\ regions near evolved stellar cores and are also identified 
by the spectral type of the likely ionizing source.  The Galactic 
coordinates of each observation direction appear in columns 4 and 5. The 
centroid LSR velocity of the \ha\ emission from each \hii\ region is 
listed in column 6.

Many of the \hii\ regions were observed in \hb\ as well as \ha\ in order 
to quantify the 
extinction to these sources, which lie primarily near the Galactic plane. 
In photoionized gas at a temperature of $T_e = 8000$ K, the ratio of the 
number of \ha\ to \hb\ photons emitted is 3.94 \citep{HS87,
  Osterbrock89}, which is only very weakly dependent on temperature
($\propto  T^{0.07}$) and density. 
Our observed values of \iha\ and \ihb\ can then be used to estimate the 
extinction to the nebulae. Assuming the dust has a total-to-selective 
extinction ratio $R_V = \av/E(B-V) = 3.1$, and the wavelength dependence 
of extinction characterized by \citet{CCM89}, we obtain

\begin{equation}
\av = 3.12\,\mbox{ln}(\frac{\iha/\ihb}{3.94})
\label{eq:aveq}
\end{equation}
\citep{MR05}.  These values of \av\ are listed in column 7 of
Table~\ref{tab:hiibasic}.  The uncertainties in \av\ are a
reflection of the uncertainties in the strength of the \ha\ and \hb\
lines as discussed in \S\ref{sec:errors}, with the errors propagated
according to equation \ref{eq:aveq}. 

For directions in which \av\ was determined, the intensity of the observed 
lines (and its uncertainty) has been adjusted to its extinction-corrected 
value, using the extinction at other wavelengths determined by 
\citet{CCM89}. A few \hii\ region observations were not observed in \hb, 
and hence no value of \av\ was derived.  Because of the uncertainties in 
the measurements of \av, and their relatively low values, we have not 
corrected the emission from the \hii\ regions that were not observed in 
\hb.  However, the corrections to the line {\it{ratios}} 
(Table~\ref{tab:hiimulti}) are not 
very sensitive to moderate values of \av, especially for \nii/\ha\ and 
\sii/\ha.  For a given value of \av, the correction factors are 
$e^{0.28\av}$, $e^{0.10\av}$, $e^{-.003\av}$, and $e^{-0.02\av}$ for 
\oiii/\ha, \hei/\ha, \nii/\ha, and \sii/\ha, respectively, using the ratio 
of optical depths at different wavelengths from \citet{CCM89}. Only one of 
the non-corrected O-star \hii\ regions, S292, which is ionized by the CMa 
OB1 association, lies near the Galactic plane where the extinction may be 
significant. \citet{Claria74} conducted a photometric study of this star 
cluster, and found considerable scatter in the visual absorption for 
member stars of the OB association, with a mean value of $\av = 0.81\pm 
0.30$ mag.  If the emission from S292 suffers this same average extinction 
within the WHAM beam, than the extinction corrected value for \iha\ is 514 
R. The corrections for the line ratios are all smaller than the 
uncertainties, except for \oiii/\ha\ which increases from $\approx$\ 0.10 to 0.13. 
The other two non-corrected O-star \hii\ regions, S264 and S276, are both 
part of the nearby Orion OB1 association, which has a lower mean 
interstellar extinction of $\av \approx 0.15$ mag \citep{WL78}. The two 
non-O-star \hii\ regions at the bottom of Table~\ref{tab:hiibasic} were 
also not corrected, but they are at very high latitudes, $b \gtrsim 
50\dg$, where the exinction is also likely to be low.

The data in the eighth column of Table~\ref{tab:hiibasic} lists the 
intensities of the \ha\ line within WHAM's $1\dg$ beam. 
We note that  this measurement is a lower limit for several \hii\
regions that do not fill the beam, as indicated in the Table. 
We see that these extinction corrected \ha\ intensities vary by three 
orders of magnitude, from $\approx$ 2 R to $\approx$ 3000 R, with the
brightest toward the OB association Cas OB6, and the faintest toward
the two faint evolved stellar cores.   

\begin{figure}[p]
\center{
\includegraphics[scale=0.8]{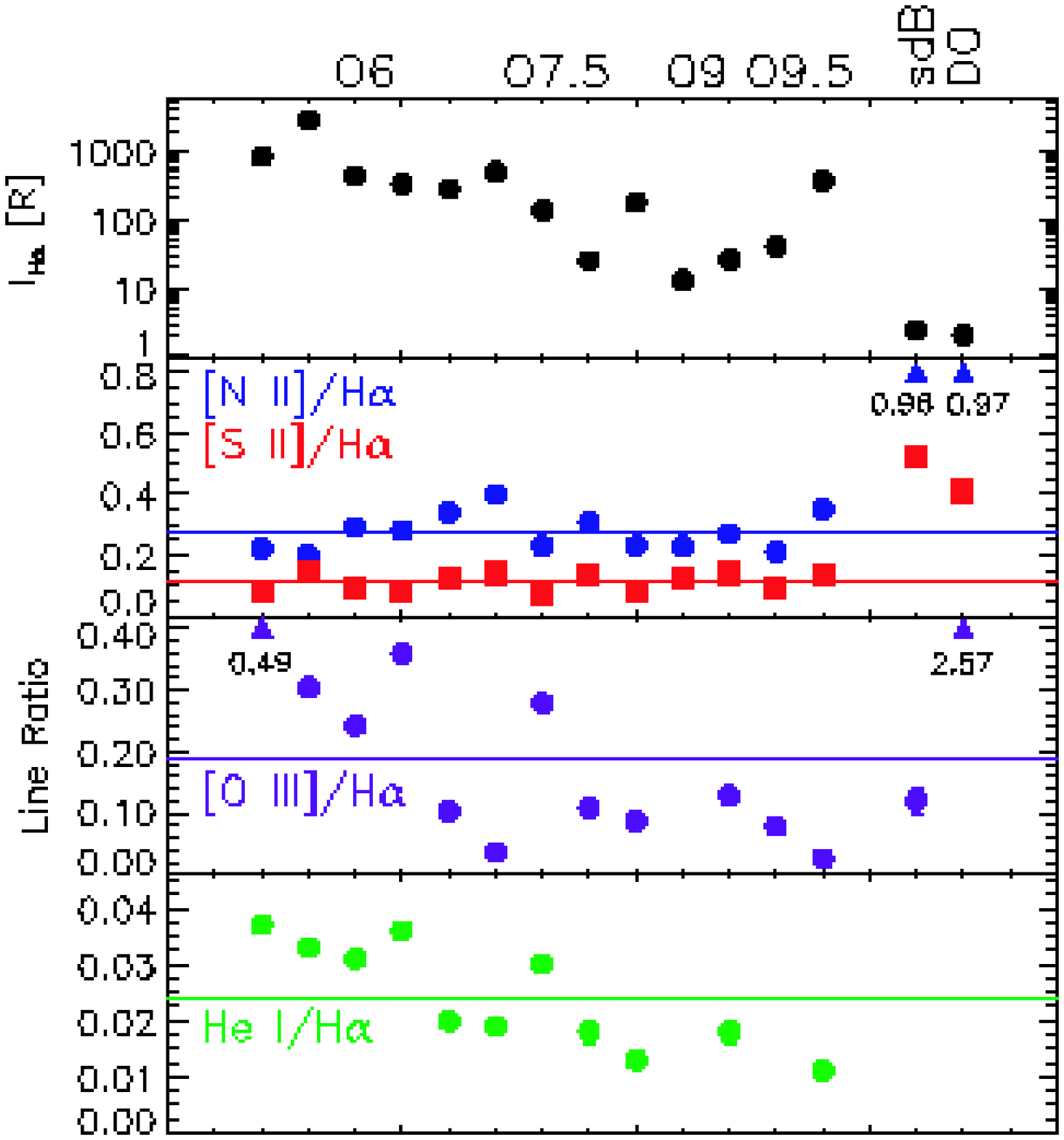}
}
\caption{Emission line strengths and line ratios toward \hii\ regions,
  from Table~\ref{tab:hiimulti}. The data have been sorted by spectral
  type of the ionizing star(s), as indicated at the top of the
  diagram. Data for two hot, evolved stellar cores are on the far
  right. Most of the error bars are smaller than the symbol sizes. 
The solid horizontal lines represent the average line ratio for the
O-star \hii\ regions.  Data that are off-scale are denoted by
triangles. The scatter in \nii/\ha\ and \sii/\ha\ is small and shows
no apparent trend with spectral type of the ionizing sources, in
contrast with the \oiii/\ha\ and \hei/\ha\ data. 
\label{fig:hiisummary}}
\end{figure}

The line strengths of \nii, \sii, \oiii, and \hei\ emission relative to 
\ha, (in energy units) are presented in Table~\ref{tab:hiimulti}.  For 
\nii/\ha, we find that the average value for the O-star \hii\ regions is 
0.27.  As will be seen in the following sections, this value is 
significantly lower than what is generally observed in the WIM, where a 
typical value of \nii/\ha\ is $\approx$ 0.5, but in some cases exceeds 
1.0.  In addition, there is no strong trend in \nii/\ha\ with spectral 
type.  If \nii/\ha\ is tracing the electron temperature of the gas, we 
might expect to see a slight decrease in \nii/\ha\ with decreasing stellar 
temperature. However, the ionizing radiation from the hottest stars may 
have a significant flux above 29.6 eV (as suggested by their higher 
\oiii/\ha), which can ionize N$^+$ and thus 
complicate the relationship between electron temperature and \nii/\ha\ 
(eq. \ref{eq:niieq}).  For the three observations taken near the very 
large \hii\ region ionized by the O7.5III star $\xi$ Per (S220 and Sivan 
4), \nii/\ha\ varies by almost a factor of two, from 0.23 to 0.40, with 
the brightest portion of the region, often referred to as the California 
nebula, having the highest ratio. The highest \nii/\ha\ ratios are
near unity and are associated with the \hii\ regions  
surrounding the hot stellar cores. These results are presented graphically 
in Figure~\ref{fig:hiisummary}.  The horizontal line in each panel 
denotes the average value for that particular line ratio.

For \sii/\ha, we find an average value of 0.11 for the O-star \hii\ 
regions. This value is also significantly lower than what is generally 
found in the WIM, consistent with previous studies. 
Because of the low ionization potential of S$^+$, \sii/\ha\ is more
sensitive to ionization effects that \nii/\ha. However we do not see
any strong trends in this ratio with spectral type of the ionizing
sources. Again, the highest \sii/\ha\ ratios are associated with the
regions that surround the hot stellar cores. 
On the other hand, line ratios that include ions with higher ionization 
potentials, namely, \oiii/\ha\ and \hei/\ha\, are observed to increase 
with increasing photospheric temperature of the O star. This is consistent 
with the gas near the hotter stars being subject to a harder incident 
spectrum.  We find a large scatter in \oiii/\ha\ with an average value of 
\oiii/\ha\ of 0.18.  Values for \hei/\ha\ show a strong trend with 
spectral type, ranging from about 0.011 for S276 (O9.5 V) to 0.037 for S132 
(WN6 + O6 I).

The last two rows of the Table~\ref{tab:hiimulti} summarize the 
observations toward two hot, evolved stellar cores. These 
lines of sight are centered on a sub-dwarf B star and a helium-rich DO hot 
white dwarf.  While they are not traditional \hii\ regions, they were 
included in this study to compare the emission-line characteristics of the 
faint ionized gas around these very hot but low luminosity stars with the 
WIM emission. These two \hii\ regions were first noted by 
\citet{Haffner01}, in a preliminary search of the WHAM-NSS for faint \ha\ 
emission near hot white dwarf and sub-dwarf stars.  A more thorough search 
of the \ha\ survey has revealed numerous small ($\lesssim\ 1\dg$) 
scale \ha\ 
enhancements, many of which are not associated with any known ionizing 
source \citep{Reynolds+05}. The detection of enhanced \ha\ emission around 
the $m_V = 13.5$ star PG 1047+003 is the first detection of ionized gas 
surrounding this sub-dwarf B star. The emission associated with DO star PG 
1034+01 near \lb = (248\dg,+48\dg) has been recently explored by 
\citet{Hewett+03} and \citet{RKP04}, who conclude that this region is a 
high-excitation, planetary-nebula like object. The line ratios, 
particularly the high values of \nii/\ha\ and \oiii/\ha, toward both 
of these regions are consistent with planetary nebula spectra.

We now proceed to compare the spectral characteristics of these
classical \hii\ regions with those of the much of larger scale  
emission features revealed by the WHAM \ha\ survey, including the diffuse WIM.

\section{THE WARM IONIZED MEDIUM}
\label{sec:wim}

\begin{figure}[p]
\center{
\includegraphics[scale=0.8]{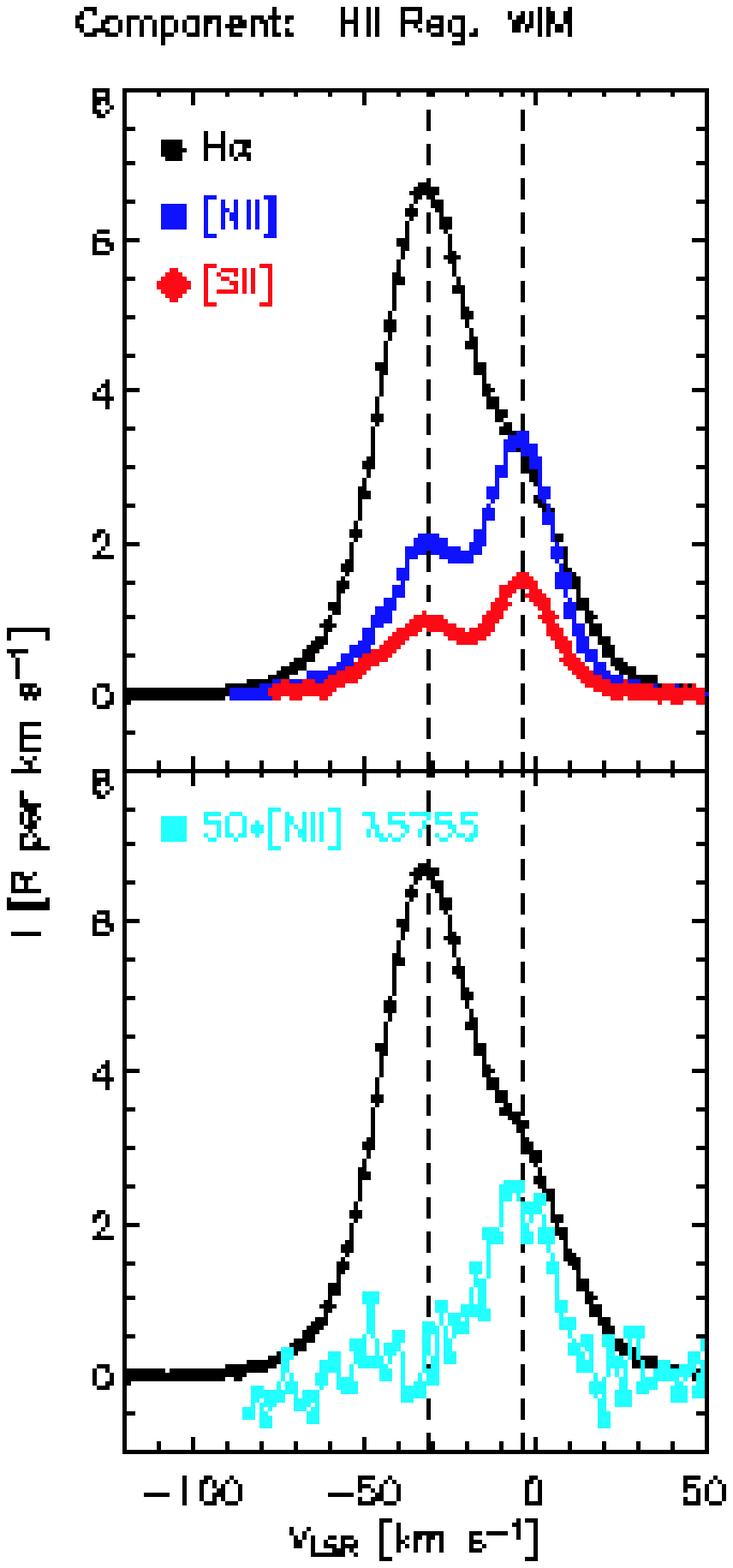}
}
\caption[Emission line spectra toward Sivan 2]{Emission line spectra
  toward the \hii\ region Sivan 2. Two velocity components are
  indicated by the dashed vertical lines, one near -31 \kms\
  associated with the \hii\ region and the other near -3 \kms\
  associated with the fainter, diffuse WIM.  The \niiblue\ spectrum
  has been multiplied by 50 to facilitate the comparison with the \ha\
  profile. Note that the \nii\ and \sii\ emission lines are brighter
  relative to \ha\ for the WIM component, suggesting that the WIM gas
  is warmer. The elevated \niiblue/\ha\ ratio confirms that the WIM
  gas is $\gtrsim$ 3000 K warmer than the \hii\
  region. \label{fig:wimspectra}} 
\end{figure}

As mentioned in \S\S\ref{sec:intro} and \ref{sec:physconds}, the spectral 
characteristics of the WIM differ significantly 
from the classical \hii\ regions.  This has been discussed in detail in 
earlier studies (e.g., Haffner et al 1999).  
As an illustration of the principal difference between WIM and \hii\ 
region spectra, we show \ha, \nii, and \sii\ observations toward the O 
star \hii\ region Sivan 2 in Figure~\ref{fig:wimspectra}.  Two radial 
velocity components are present along this line of sight. Velocity channel 
maps from the WHAM survey show that the emission near \vlsr = 0 \kms\ is 
associated with diffuse foreground emission (i.e., the local WIM).  This 
emission was well separated in velocity from the \hii\ region emission, 
allowing a direct comparison of the relative intensities of the lines from 
the two different sources in the same spectrum.  

The velocity of each component was determined from a least-squares fit of a sum of Gaussian profiles to the \sii\ spectrum.  The \sii\ spectrum was chosen because of its narrow, well-resolved component profiles and its high signal-to-noise.  The resulting component velocities are shown as vertical dashed lines in the Figure. The strength of each component was calculated from a two-component Gaussian fit to each spectrum in which the velocities for each component were fixed as determined from the \sii\ spectrum. 

These spectra clearly reveal that \nii\ $\lambda$6584/\ha, \nii\ 
$\lambda$5775/\ha, and \sii/\ha\ are significantly higher in the diffuse 
gas compared to the \hii\ region, with \nii/\ha\ = 0.83 and \sii/\ha = 
0.38 in the WIM, compared to 0.23 and 0.12, respectively, for the \hii\ 
region. The \nii/\ha\ data suggest that the temperature of the diffuse gas 
is $T \approx$\ 9000 K, compared to $\approx$\ 6100 K for the \hii\ 
region.  The high temperature derived for the WIM relative to the \hii\ 
region, from the \nii$\lambda$6584/\ha\ ratio (\S\ref{sec:physconds}), is 
confirmed by observations of the highly temperature sensitive \niiblue\ 
emission line. These observations show relatively bright \niiblue\ 
emission for the fainter (in \ha) WIM component, while the \niiblue\ line 
is not even detected in the cooler \hii\ region component (see also 
\S\ref{sec:niiblue} below).

Additional differences between the \hii\ regions, the WIM, and other faint, 
large-scale emission features in the \ha\ sky are discussed in the 
following sections. For this paper, the `WIM' refers to the gas that
produces the faint, diffuse emission outside the classical \hii\
regions, extended bubbles and superbubbles, and high latitude
filamentary structures. 

\section{\hi\ CAVITIES AND SUPERBUBBLES}
\label{sec:bubbles}

One of the basic questions concerning the nature of the WIM is how 
ionizing photons from O stars are able to travel hundreds of parsecs from 
the stars. One possibility is the existence of enormous \hi-free bubbles 
surrounding some of the O stars. In \S\ref{subsec:ori} and 
\S\ref{subsec:per}, we examine in detail the faint optical line emission 
associated with two very extended bubble-like regions which have diameters 
of $\sim$ 40\dg\ - 60\dg\ (up to 2 kpc in extent) and which are 
ionized by luminous O associations. The line ratios are then compared with 
ratios in the more diffuse WIM to examine whether the WIM could be a 
superposition of such regions.

\subsection{Orion-Eridanus Bubble}
\label{subsec:ori}

One of the largest networks of interconnected \ha-emitting structures in 
the WHAM sky survey appears in the constellations of Orion and Eridanus, 
shown in Figure~\ref{fig:orimap}. The presence of optical and 
radio-emitting filaments in this general direction has been known for many 
years. \citet{RO79} carried out velocity resolved emission-line 
observations of this region, and found that the filaments, loops, and 
enhanced \ha\ emission are all part of an expanding shell of neutral and 
ionized gas with a diameter of $\approx 280$ pc. They suggested that Lyman 
continuum photons from the Ori OB1 association, located near one side of 
the bubble, travel largely unimpeded through the hot ($T \sim 10^6$~K) 
interior cavity and ionize the inner surface of its surrounding outer 
shell.  They estimated the shell has a density near 1 cm$^{-3}$, which 
is significantly higher than the density in the WIM; nevertheless, the 
large extent of the cavity has produced diffuse and filamentary \ha\ 
covering a $40\dg \times 40\dg$ region of the sky and up to 34\dg\ from 
the OB association.  This picture is supported by the detection of diffuse 
X-ray emission interior to the bubble walls, as well as more recent
studies of \hi\ in the region \citep{Burrows+93, BHB95}.  

The shell is expanding at a velocity of about 20 \kms, likely as a result of supernova activity; however, \citet{RO79} have shown that the contribution from 
shocks to the ionization of the walls of the bubble is likely to be 
negligible.  They found that among the most luminous stars within the 
bubble, $\delta$ Ori, an O9.5~I star, is probably responsible for most of 
the ionization. It is the only hot star in the cavity that has no discrete 
\hii\ region around it, implying that most of its ionizing radiation 
travels unimpeded through the cavity.  The Orion-Eridanus bubble, which is 
significantly brighter in \ha\ compared to the more diffuse WIM, is thus 
an excellent environment in which to study the relationship between 
traditional \hii\ regions and the warm ionized medium.  By comparing the 
physical conditions within, around, and outside this bubble, we can assess 
the similarities and differences between gas that is part of a large 
cavity ionized by a known source and the fainter, diffuse WIM.

\subsubsection{Large-Scale Trends}

\begin{figure}[p]
\center{
\includegraphics[scale=0.8]{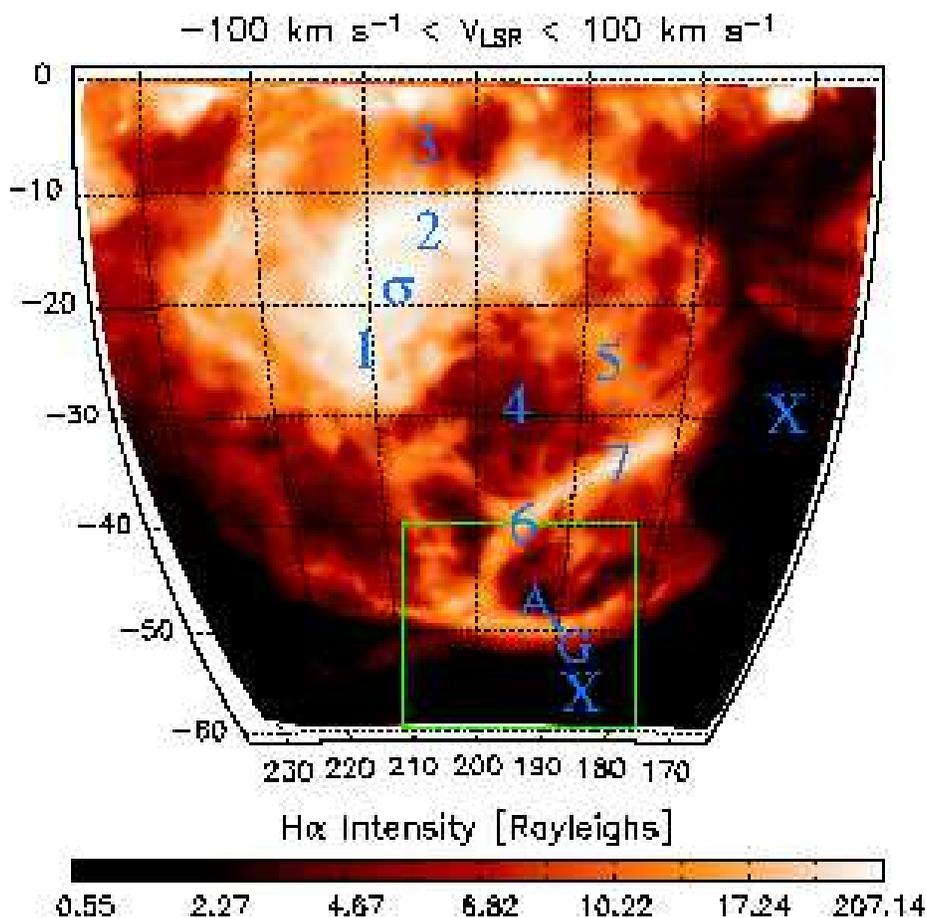}
}
\caption[\ha\ map of the Orion-Eridanus bubble]{ Histogram equalized
  map of \ha\ emission toward the Orion-Eridanus bubble in Galactic
  longitude and latitude, from the WHAM-NSS. The \ha\ emission has
  been integrated over radial velocities within 100\kms\ of the
  LSR. The numbers 1 through 7 on the map refer to the approximate
  locations of the pointed observations summarized in
  Table~\ref{tab:ori}. The $\sigma$ refers to the location of the
  $\sigma$\ Ori \hii\ region, which is near the \ha\ flux-weighted
  center of this asymmetrically-expanding shell. The labels A-G refer
  to the location of series of pointed observations taken along the
  southern edge of the shell. The two Xs show the location of the
  {\sc{OFF}} directions used to remove faint atmospheric and
  background emission from the pointed observations. The green box
  outlines the area  mapped in \nii\ and \sii. \label{fig:orimap}} 
\end{figure}

We have observed this bubble along several lines of sight in the emission 
lines of \ha, \nii, \sii, \oiii, and \hei. Superposed on the map of
this region, presented in Figure~\ref{fig:orimap}, are the approximate
locations of the pointed  
observations.  The bubble is outlined approximately by a circle that goes 
through directions A, 1, 2, 5, 7, and G. Direction 3 is located 
\emph{outside} of the boundary and samples diffuse interstellar gas near 
the Galactic plane. The numbered directions are ordered with increasing 
distance from the Orion OB1 association, located near $\sigma$ Ori and its 
\hii\ region, which is indicated on the diagram with the letter $\sigma$.  
A set of seven closely spaced observations were also made that cut across 
a filament near the lower edge of the bubble, and are labeled A-G.  
Two directions denoted by `X' were used as {\sc{OFF}}s for the reduction of the 
pointed observation spectra. The OFFs have an \ha\ intensity of $\approx$ 0.5 R, and subtracting them from the ONs allows us to isolate emission in the bubble from any background or foreground emission.
The green box in Figure~\ref{fig:orimap} indicates a region of the 
bubble that also was mapped in \nii\ and \sii.

Table~\ref{tab:ori} summarizes our pointed observations of this region, 
with the name of each direction corresponding to the labels in 
Figure~\ref{fig:orimap}.  The columns in the table are similar to those 
in
Tables~\ref{tab:hiibasic} and \ref{tab:hiimulti}. However, the 
fourth column shows the angular distance of each direction from \lb = 
(206.5\dg,~-18.0\dg), the \ha\ flux-weighted center of the bubble, 
as determined by \citet{RO79} and roughly the center of the Orion OB 1 
association.  The spectrum toward direction 1 has two emission 
components; the emission line strengths that appear in the table include 
the total emission from both components.  Also, because of the weak 
dependence of the line ratios on \av, and the low extinction to this 
nearby region \citep[\av $\sim$ 0.15 mag;][]{WL78}, no extinction 
correction was applied to the data.  As discussed in \S\ref{sec:obs}, {\sc{OFF}} 
spectra were subtracted from each of these pointed observations, and all 
of the results in the table represent `background subtracted' values.  
This procedure isolates emission that is only associated with the 
Orion-Eridanus bubble.

A graphical summary of the line ratios is shown in 
Figure~\ref{fig:oridiag}.  The panel on the left includes the 
observations that are within or on the boundary of the bubble (1-7), except for direction 3. The 
panel on the right shows the results for the series of observations that 
cut across the outermost filament at $b \approx -50\dg$ (A-G), which 
appears to be an edge-on projection of the cavity's outer shell.  These 
pointings, spaced about 1\dg\ apart, begin on one side of the \ha\ 
filament (the side toward the O association), cross the filament, and end 
just outside the ionized part of the bubble (see 
\S\ref{subsubsec:orisurv}). The column density of \hi\, from 
\citet{HIAtlas}, is also shown in the panel on the right (\emph{green}), 
and indicates the location of the neutral portion of the shell.  The name 
of each observation direction appears above the panels. The data are shown 
as a function of $D_\theta$, the angular distance from the center of the OB1 
association, which is useful in order to search for potential changes in 
the physical conditions of the gas with increasing distance from the 
ionizing source. The top plot in each panel shows the \ha\ intensity 
toward each direction on a logarithmic scale.  The second plot from the 
top shows the values of \nii/\ha\ and \sii/\ha, while the third and fourth 
plots give \oiii/\ha\ and \hei/\ha, respectively. The horizontal line in 
each plot represents the average value for the ratio in \hii\ regions 
(Table~\ref{tab:hiimulti}; Figure~\ref{fig:hiisummary}); for 
\oiii/\ha, the \hii\ region average, 0.094, is off scale. 

For directions 1-7, we see that \nii/\ha\ varies between $\approx$ 0.2 to 
0.3, and there is no significant correlation with $D_\theta$ out to 
$25\dg$ from the association. The ratio is weakly anti-correlated with 
\iha, with brighter regions of the bubble having lower values of \nii/\ha. 
This is a common behavior seen in these bubble structures and in the WIM, 
both for the Milky Way and other galaxies \citep{HRT99, Rand98, CR01}. 
\nii/\ha\ is near the value of the average \hii\ region, and slightly 
lower than in the $\sigma$ Ori \hii\ region, which appears to reside 
inside (or perhaps on the wall of) the cavity.  The variation in \sii/\ha\ 
($\approx$ 0.15 - 0.25) is larger than for \nii/\ha, and there is a very weak 
trend in which \sii/\ha\ is higher at larger distances from the 
association, where it becomes larger than the average ratio in classical \hii\ 
regions. The ratio \sii/\nii\ also increases with $D_\theta$, from 0.55 for 
direction 1 to 0.95 in direction 6.  As will be seen later, \nii/\ha\ and 
\sii/\ha\ in the bubble are generally lower and exhibit less scatter than the ratios observed in the WIM.

The \oiii/\ha\ ratio is extremely low throughout the bubble, $\lesssim 
0.02$, with the exception of direction 4 (direction 3 is outside 
the bubble), and is an order of magnitude below the average value of 0.18 
for the \hii\ regions (Table~\ref{tab:hiimulti}). Similarly, the 
\hei/\ha\ ratios are low ($\lesssim 0.015$) relative to the \hii\ region 
average.

Based upon the discussion presented in \S\ref{sec:hii}, the \nii/\ha\ and 
\sii/\nii\ data suggest that the temperature of the ionized gas is between 
$6000~K \lesssim T \lesssim 6500~K$ with $0.4 \lesssim \rm{S}^+/\rm{S} 
\lesssim 0.6$. In addition, the \hei\ data suggest that He$^+$/He
$\lesssim 0.3$. We can quantify the  
hardness of the radiation field by assuming from the above He$^+$/He 
ratio that the volume of the He$^+$ zone along these lines of sight is 
smaller than the volume of the H$^+$ zone by a factor of $\lesssim$\ 0.3. 
The ratio of total number of He-ionizing photons $Q$(He$^0$) to H-ionizing 
photons $Q$(H$^0$) along the line of sight is then proportional to this 
volume ratio, specifically $Q$(He$^0$)/$Q$(H$^0$) $\lesssim$ 0.03 
\citep{Osterbrock89}. This corresponds to a star with an effective 
temperature $T_* \lesssim$ 35,000 K, equivalent to O8.5~I or O9.5~V or 
cooler \citep{VGS96}. This is consistent with the likely ionizing source 
of the shell, $\delta$\ Ori, which is an O9.5I star, although further
analysis using recent stellar atmosphere models is needed to confirm
this scenario \citep{MSH05}. The low He$^+$/He is  
consistent with the apparently low ionization state of oxygen, where 
\oiii/\ha\ suggests that O$^{++}$/O $\lesssim$\ 0.04 (exception for 
direction 4, where O$^{++}$/O $\approx$\ 0.1). Outside the bubble and 
closer to the Galactic plane (direction 3), He$^+$/He $\approx 0.7$, 
suggesting that the radiation field outside the shell is significantly 
harder, consistent with a continuum source with $T_*$ $\gtrsim$ 40,000 K, 
an O7 star or earlier.

\subsubsection{The \hii\ to \hi\ Transition through the Shell}
\label{subsubsec:orisurv}

\begin{figure}[p]
\center{
\includegraphics[scale=0.7]{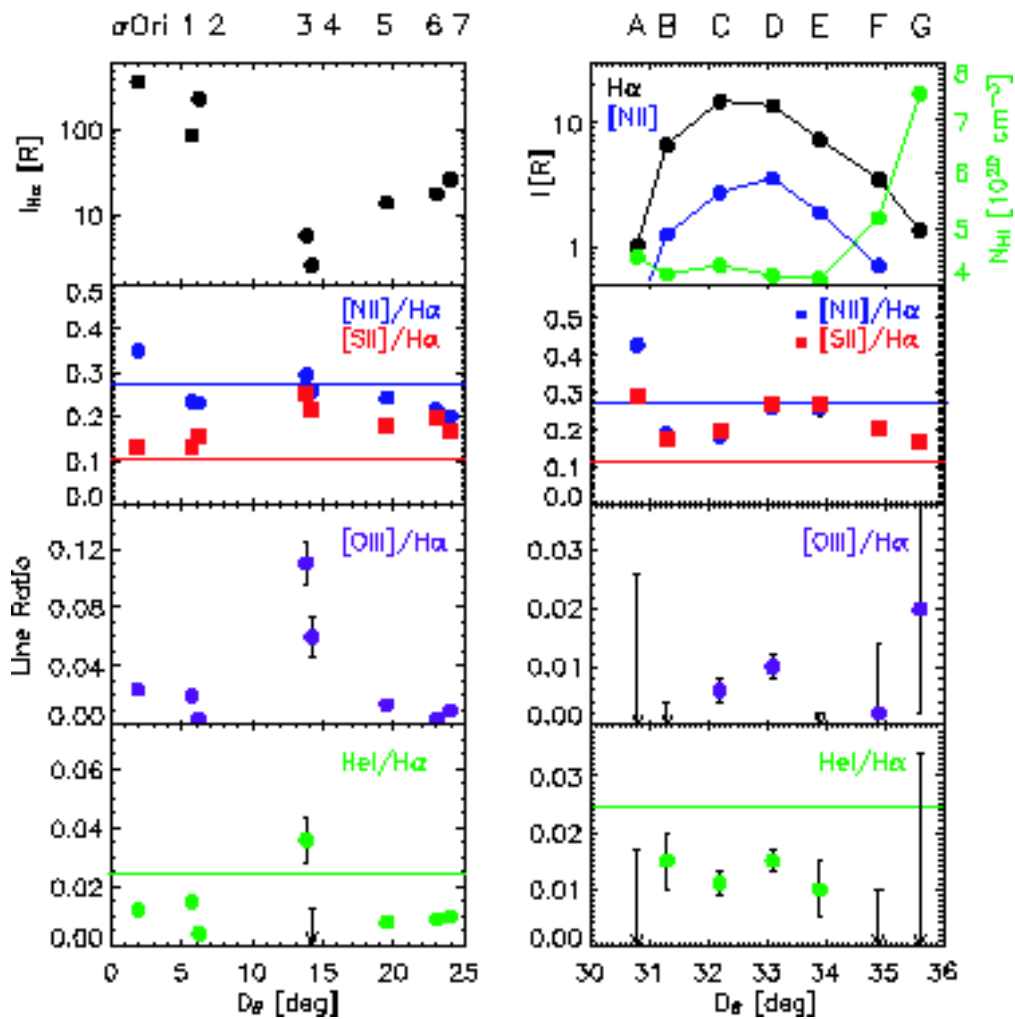}
}
\caption[Emission line ratios toward Orion-Eridanus bubble]{ Summary
  plot of several emission line ratios toward the Orion-Eridanus
  bubble. The labels on the top of both panels refer to the positions
  labeled in the map of Figure~\ref{fig:orimap}.  The left panel shows
  the observations taken within and around the bubble, and the right
  panel shows the observations taken across the outer edge of the
  shell. The top plots show the \ha\ (\emph{black}), \nii\
  (\emph{blue}), and \hi\ 21 cm (\emph{green}) intensity of each
  direction, as a function of distance from the OB1 association.  The
  bottom three plots show the variation in the line ratios of
  \nii/\ha, \sii/\ha, and \oiii/\ha\ with angular distance. The solid
  horizontal lines are the average line ratios for the O-star \hii\
  regions listed in Table~\ref{tab:hiimulti}. The average \oiii/\ha,
  0.18, is off of the range of the plots.  Data points without
  visible errors bars have uncertainties smaller than the symbol size,
  and arrows are used to indicate upper limits. \label{fig:oridiag}} 
\end{figure}

The data on the right panel in Figure~\ref{fig:oridiag} show the 
variation in the line ratios across the outer edge of the bubble, where 
the hydrogen in the shell is making a transition from fully ionized 
(on the shell's inside surface) to completely neutral. Here we see small, 
but significant trends in the ratios of \nii/\ha\ and \sii/\ha\ across
this outer filament, which is a projection enhancement from an edge-on
view of the shell. The ratio  
of \sii/\ha\ follows the variation in \nii/\ha\ very closely, with 
\sii/\nii\ $\approx$ 1.0, except for the direction inside (A), 
where \sii/\nii\ $\approx$ 0.7. Interestingly, the data show that this 
filament is brightest in \nii\ and \sii\ (is highest in temperature?) at a 
different location than the brightest part in \ha, with \nii\ and \sii\ 
both peaking about $\approx 0.5\dg = 3.5 (d/400)$\ pc further away from 
the ionizing source(s). The peak brightness of \hi\ is even farther to the 
outside, near direction G. These trends 
are also shown in the line ratio maps of this region presented below. Both 
\oiii\ and \hei\ are very weak and show no clear trends across the 
filament.

We note that while there are statistically significant trends in \nii/\ha\ 
and \sii/\ha\ across the outer edge of the shell, the magnitude of the 
variations are small.  As a result, 
changes in the physical conditions, as suggested by the line ratios, are 
not very dramatic. Potential complications introduced by the unknown 
geometry and multi-phase nature of the bubble could be important when 
using the observed line ratios to infer changes in the actual physical 
conditions.  For example, small variations in the volume-averaged 
ionization fractions of both N and S along the lines of sight may 
complicate our interpretation that \nii/\ha\ is tracing the temperature of 
the gas. However, the constancy of \sii/\nii\ strongly indicates that the 
variations in \nii/\ha\ and \sii/\ha\ are dominated by variations in 
temperature (see eq. \ref{eq:siieq}). If we assume that N$^+$/N = 0.8 
everywhere across the shell, then the \nii/\ha\ data suggest that the 
temperature of the warm ionized gas associated with the edge of this 
bubble is $T \approx 7300$ K toward direction A, falling to $\approx$\ 
6000 K at the inside edge of the filament, and rising back up to about 
6300~K before falling back down to near 6000~K just outside the shell.  
This apparent rise in temperature toward the backside of the \ha\ filament 
(farther away from the ionizing source) could be due to a slight hardening 
of the radiation as it penetrates into the shell and just before it is 
completely absorbed where the shell becomes neutral \citep{WM04}.

\begin{figure}[p]
\center{
\includegraphics[scale=0.8]{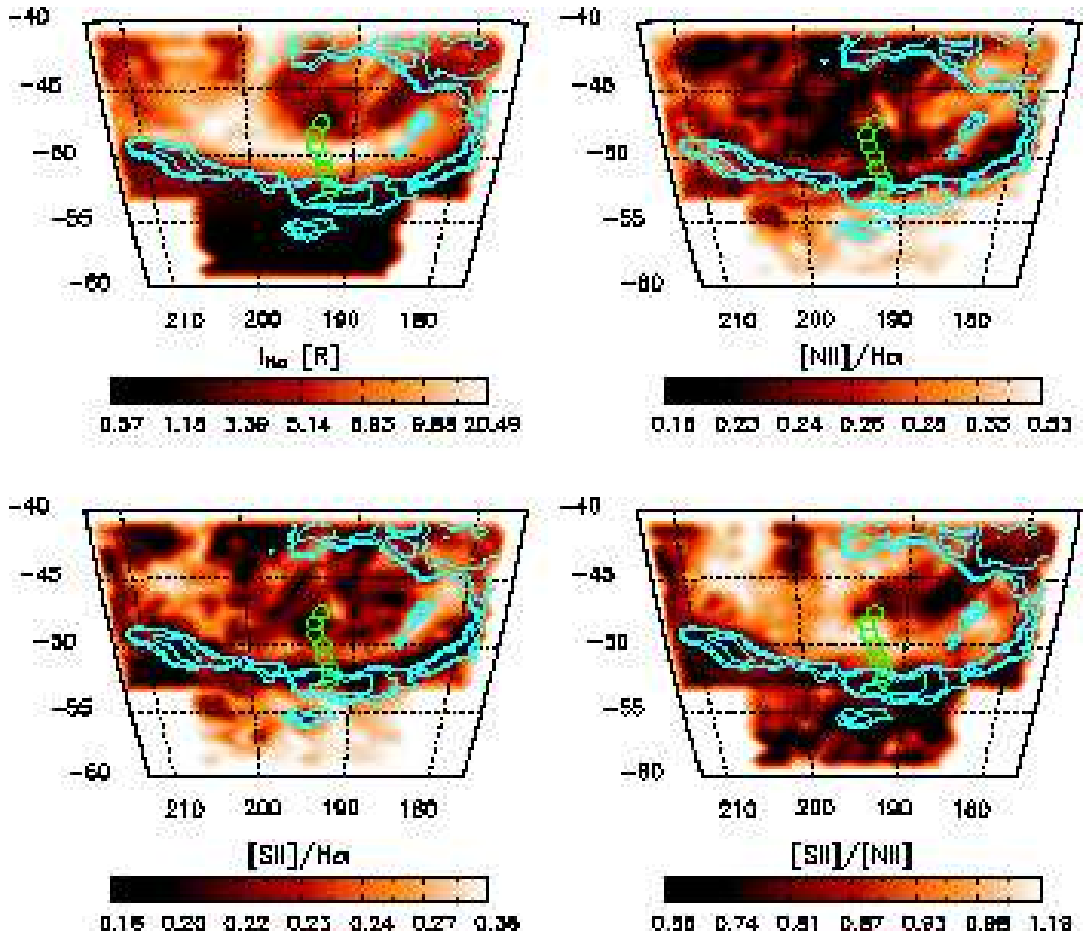}
}
\caption[\ha, \nii/\ha, \sii/\ha, and \sii/\nii\ maps of southern part
of Orion-Eridanus bubble]{ Histogram equalized maps of \ha, \nii/\ha,
  and \sii/\ha, and \sii/\nii\ emission from a portion of the
  Orion-Eridanus bubble. The green circles show the positions of the
  pointed observations A-G summarized in Table~\ref{tab:ori}.  The
  blue contours show the location of bright \hi\ 21 cm emission, from
  \citet{HIAtlas}. The contour levels are at $N_{HI} = 5.6,6,7,9,11\
  \times 10^{20}$~cm$^{-2}$. \label{fig:oriratiomap}} 
\end{figure}

Figure~\ref{fig:oriratiomap} shows a more comprehensive picture of the 
\nii\ and \sii\ emission in this region, obtained in survey mode. The
figure contains maps of the \ha\ and \hi\ intensity, \nii/\ha,
\sii/\ha, and \sii/\nii. The maps were created by integrating  
the emission lines over their entire profiles, between $\pm~50~\kms$ of 
the LSR. The blue contours outline the location of strong 21 cm emission 
from \citet{HIAtlas}, the outermost, neutral portion of the shell, with 
the contour levels corresponding to column densities of 5.6, 6, 7, 9, and 
11$\times 10^{20}$~cm$^{-2}$.  The green circles represent the pointed 
observations A-G discussed above.  This $20\dg \times 20\dg$ view also
includes a large region outside of the shell to  
higher (more negative than -52\dg) Galactic latitude, dominated by the faint, 
diffuse WIM.

The trends that were found for the pointed directions that traverse the 
outer edge of the bubble are also apparent in these survey mode 
observations. However, the maps in Figure~\ref{fig:oriratiomap} reveal 
that these trends hold across the entirety of the shell edge, and not just 
for one slice through it. The \ha\ map in Figure~\ref{fig:oriratiomap} shows that the 
ionized gas associated with the edge of the bubble lies a few degrees 
inside of the neutral \hi\ edge, and decreases in intensity before the 
peak in \hi\ emission, as was also shown in Figure~\ref{fig:oridiag}.
Also, note the ridge of enhanced \nii/\ha\ and \sii/\ha\ at $b\approx
-51\dg$ that runs parallel to, and just between, the \ha\ and 21 cm
bright parts of the outer shell. 
 
The \nii/\ha\ map shows a striking anti-correlation between the \ha\ line 
strength and the \nii/\ha\ line ratio.  For example, the very faint WIM 
emission outside the bubble at $b \lesssim -53\dg$ is the ``brightest" 
region in the \nii/\ha\ map. In addition, several individual features 
\emph{inside} of the bubble show this anti-correlation, such as the region 
of depressed \ha\ emission near \lb = (187\dg, -47\dg) and the region of 
enhanced \ha\ emission near \lb\ = (198\dg,~-41\dg).  Interestingly, a 
spatially coherent depression in \nii/\ha\ follows the shape of the outer 
edge of the bubble, where the \hi\ is getting brighter (i.e., directions F and G). This suggests a decrease in temperature (or a decrease in N$^+$ relative to H$^+$) in the transition 
region from \hii\ to \hi\ in this outer shell. The data are consistent 
with this bubble being a large, hot cavity with the inside walls ionized 
by hot stars within the cavity, as originally suggested by \citet{RO79}.

The \sii/\ha\ map is very similar to the \nii/\ha\ map. We see that the 
diffuse background is significantly brighter in \sii/\ha\ than the rest of 
the bubble.  We also see a dark filament that is nearly co-spatial with 
the \hi\ shell.  This dark filament is similar to the feature in the 
\nii/\ha\ map, but seems to be of a higher contrast relative to its 
surroundings.  The \sii/\nii\ map shows that the interior of the shell has 
an elevated \sii/\nii\ line ratio (i.e., a lower ionization of sulfur) 
than in the WIM at $b \lesssim -53\dg$.  We note that the coherent 
features in both the \nii/\ha\ and \sii/\ha\ maps are well above the 
noise, although the numerical variation in the ratios is small (67\% of 
the data span a range of $\approx$ 0.10 in the ratio).

The interpretation of these maps is complicated by the presence of 
the background (WIM) emission. These maps are as they appear in the sky, 
and no 
background emission has been subtracted from them as was done for the
pointed observations.  Because \nii/\ha\ and \sii/\ha\ are relatively
high in the background, the faint \nii\ and \sii\ emission inside and
near the edge of the bubble is not a representation of emission from
the bubble alone.  
The average strength of the \ha, \nii, and \sii\ emission in the
background is 0.99, 0.35, and 0.26 R, respectively.   
Therefore, because the \nii/\ha\ and \sii/\ha\ ratios are higher in the 
background than within the Orion-Eridanus bubble, the values for the
actual ratios associated with the bubble features in
Figure~\ref{fig:oriratiomap} are slightly lower than that given by the
color bars. That is, there is an even greater difference between the
ionized gas associated with the bubble and the WIM than these maps
indicate. 

\begin{figure}[p]
\center{
\includegraphics[scale=0.8]{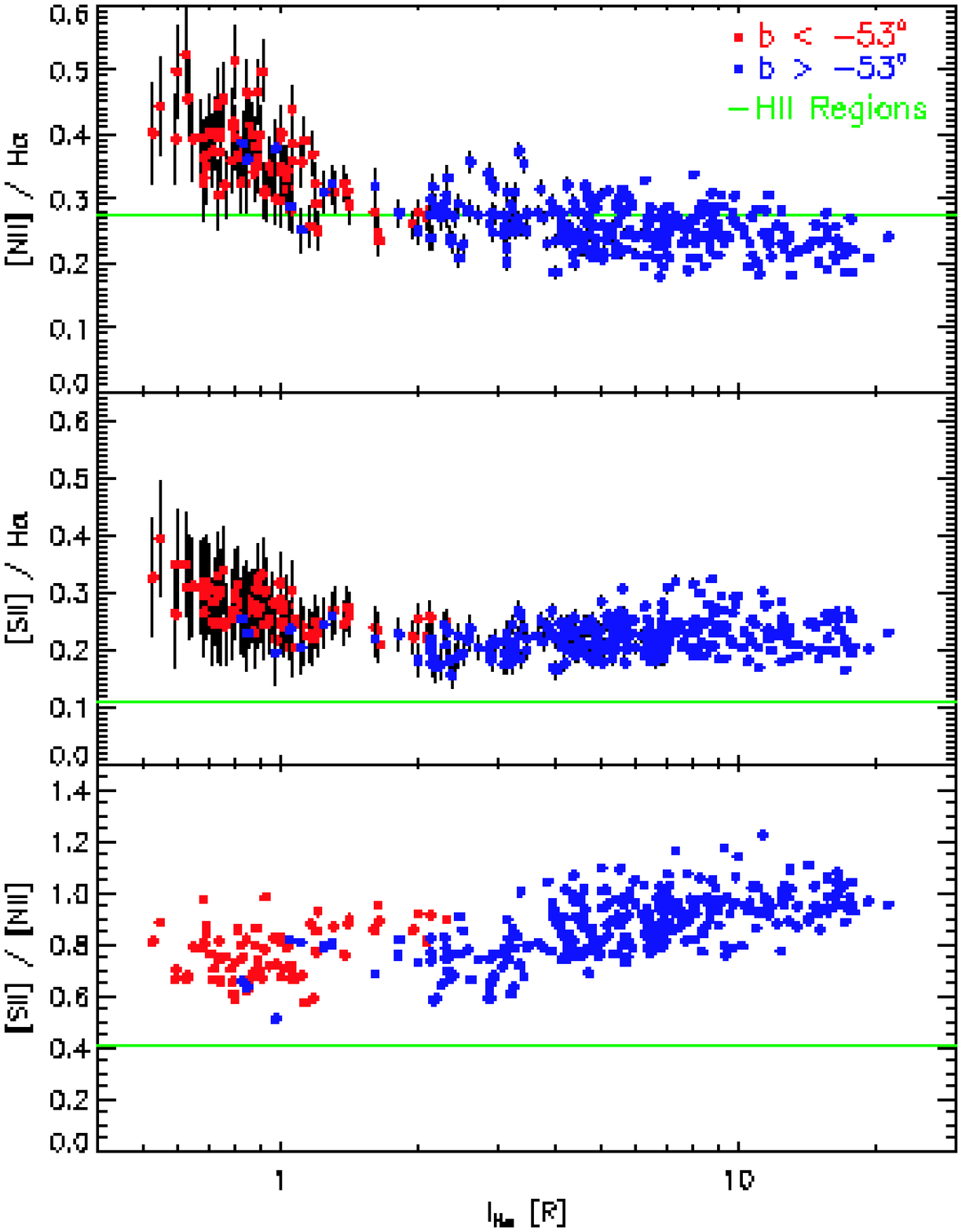}
}
\caption[\nii/\ha\ and \sii/\ha\ versus \iha\ for southern part of
Orion-Eridanus bubble]{ \nii/\ha, \sii/\ha, and \sii/\nii\ as a function of \iha\
  for each direction in the maps toward the Orion-Eridanus bubble,
  with points outside ($b < -53\dg$) and inside ($b > -53\dg$) the
  bubble shown in red and blue, respectively.  The average value of
  for all O-star \hii\ regions is shown as a horizontal green line. 
  Note the increase in \nii/\ha\ and \sii/\ha\ toward
  regions of fainter \ha\ emission (i.e., in the
  WIM).\label{fig:oriratioplot}} 
\end{figure}

Another visualization of the data in these maps is presented in 
Figure~\ref{fig:oriratioplot}. Here, the ratios \nii/\ha, \sii/\ha, and \sii/\nii\ 
are plotted against \ha\ intensity for every direction in 
Figure~\ref{fig:oriratiomap}. The total (random and systematic) uncertainties 
in the data points are shown in the top two panels, with their origin and magnitude discussed in 
\S\ref{sec:errors}. The uncertainty in \sii/\nii\ has been omitted for clarity. 
The data have been separated by latitude, with 
observations outside the bubble ($b < -53\dg$) shown in red (WIM) and 
observations inside the bubble ($b > -53\dg$) shown in blue. For 
comparison, the average values of these line ratios for all of the O-star \hii\ regions is shown as a green solid line. Note that the abscissa 
is on a logarithmic scale.  These plots show an increase in \nii/\ha\ and 
\sii/\ha\ in parts of the map with the faintest \ha\ emission outside of 
the bubble, and that \sii/\nii\ is significantly higher than the average \hii\ region. 
We also see that the scatter in these ratios at low \iha\ is 
similar to the scatter at higher \iha. This scatter is significantly 
larger than the random uncertainties in the data.

\begin{figure}[p]
\center{
\includegraphics[scale=0.8]{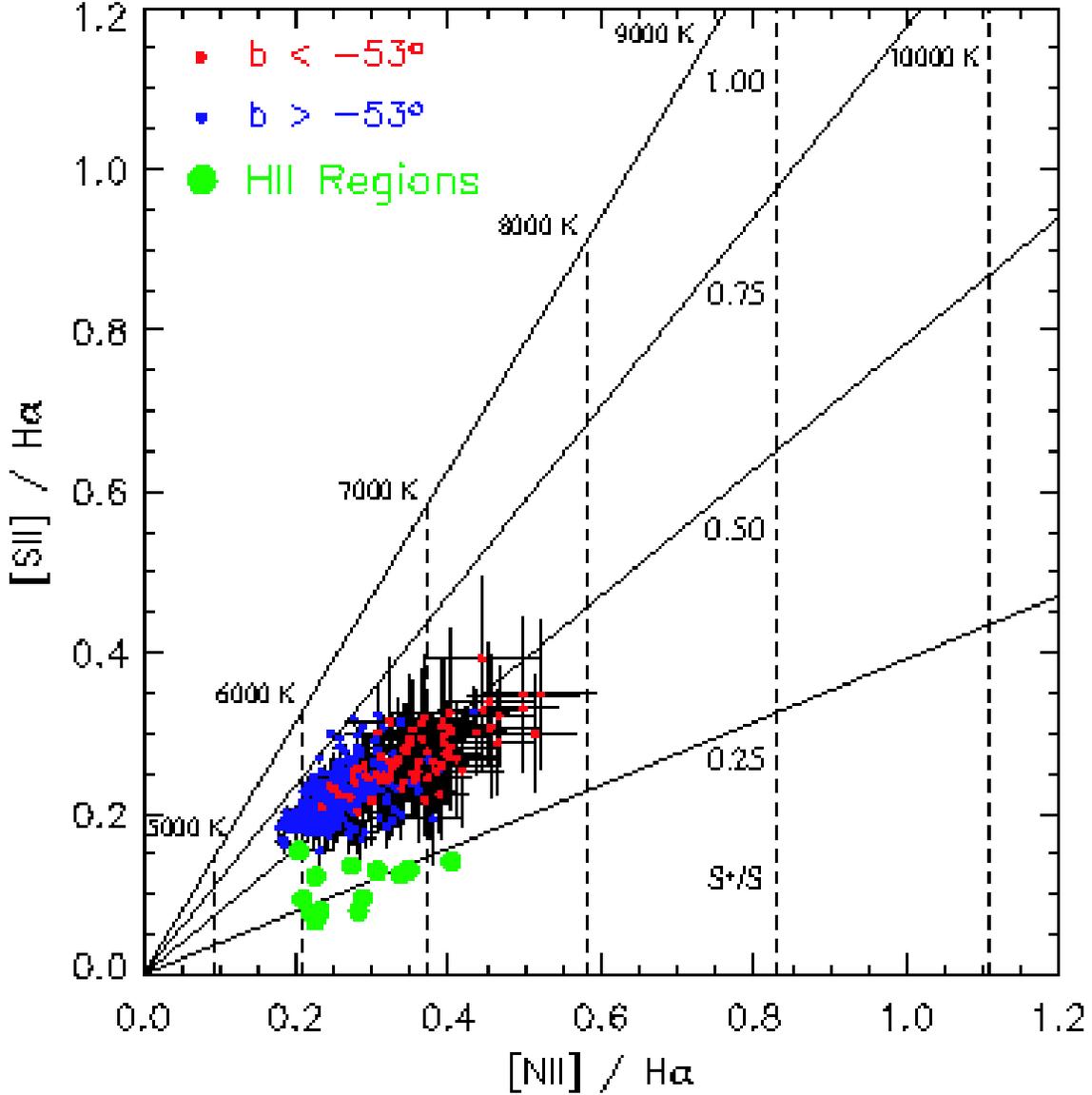}
}
\caption[\nii/\ha\ versus \sii/\ha\ for southern part of
Orion-Eridanus bubble]{ Relationship between \nii/\ha\ and \sii/\ha\
  for each direction in the emission line maps toward the
  Orion-Eridanus bubble.  The symbols are the same as in
  Figure~\ref{fig:oriratioplot}, except that data for all of the
  O-star \hii\ regions are shown. The dashed vertical lines represent
  lines of constant temperature from equation \ref{eq:niieq}, with
  $5000 \rm{K} \le T \le 10000 \rm{K}$. The solid sloped lines
  represent values of constant ionization fraction of S from equation
  \ref{eq:siieq}, with $0.25 \le \rm{S}^+/\rm{S} \le 1.0$. Note that
  the observations within the bubble (\emph{blue}) occupy the same the
  range of values of \nii/\ha\ as the \hii\ regions, but with
  systematically higher values of \sii/\ha. Also note that the
  observations outside the bubble (\emph{red}) have systematically
  higher values of \nii/\ha, suggesting the faint, diffuse WIM is at a
  higher temperature. \label{fig:orirationvs}} 
\end{figure}

A useful diagram that separates the effects of temperature from the 
effects of 
ionization state is shown in Figure~\ref{fig:orirationvs}, where 
\sii/\ha\ is plotted against \nii/\ha, using equations \ref{eq:niieq} 
and \ref{eq:siieq} after the fashion presented in \citet{HRT99}. The 
vertical dashed lines represent the expected ratio of \nii/\ha\ for 
temperatures from 5000~K to 10,000~K. The solid 
lines 
represent the expected ratio of \sii/\nii\ for an increasing fraction of 
S$^+$/S between zero and 1.0.  The symbols are the same as 
Figure~\ref{fig:oriratioplot}, 
except that the ratios for the O star \hii\ regions from
\S\ref{sec:hii} have also been added and are shown individually in
green.  
The data suggest that on average, the ionized gas in the bubble
(\emph{blue})  has a larger fraction of S in the form of S$^+$
compared to the \hii\ regions (50\% vs. 25\%), and is similar to that
in the WIM (\emph{red}).   
The data also suggest that 
most of the gas within the bubble is at \hii\ region-like 
temperatures (6000 K $< T <$ 7000 K), significantly lower than in the 
fainter, more diffuse WIM gas outside the bubble.

\subsubsection{Comparison to the WIM}

In summary, at this point we conclude that although the low 
ionization state of the warm ionized gas in the Orion-Eridanus bubble 
(i.e., enhanced \sii/\ha\ and low \oiii/\ha) is similar to that in the 
fainter, more diffuse WIM, the temperature (\nii/\ha) appears to be 
consistently lower than that in the WIM and close to the value found 
for the classical \hii\ regions.  The lower ionization state of the gas 
compared to the average classical \hii\ region is probably the result of 
the late spectral type (soft ionizing spectrum) of the primary ionizing 
star and the low ionization parameter associated with the dilution of the 
ionizing radiation as it travels to the distant walls of the cavity.  The 
decrease in ionization parameter could explain the weak rise in \sii/\ha\ 
with distance from Ori OB1. On the other hand, we find no evidence for the 
\nii/\ha\ ratio becoming more WIM-like with increasing distance from the O 
association, even at the outer edge, $33\dg$ (at least 250 pc) distant.  
While there is a small increase in both \nii/\ha\ and \sii/\ha\ at the 
transition from ionized to neutral gas within the outer shell, \nii/\ha\ 
never approaches the high values found in the more diffuse ionized gas 
outside the bubble.  Thus it appears that the spectral characteristics of 
the WIM are not explained by the leakage of O star radiation onto cavity 
walls like those of the Orion-Eridanus bubble.

\subsection{Perseus Superbubble and the Local Foreground}
\label{subsec:per}

\begin{figure}[p]
\center{
\includegraphics[scale=0.8]{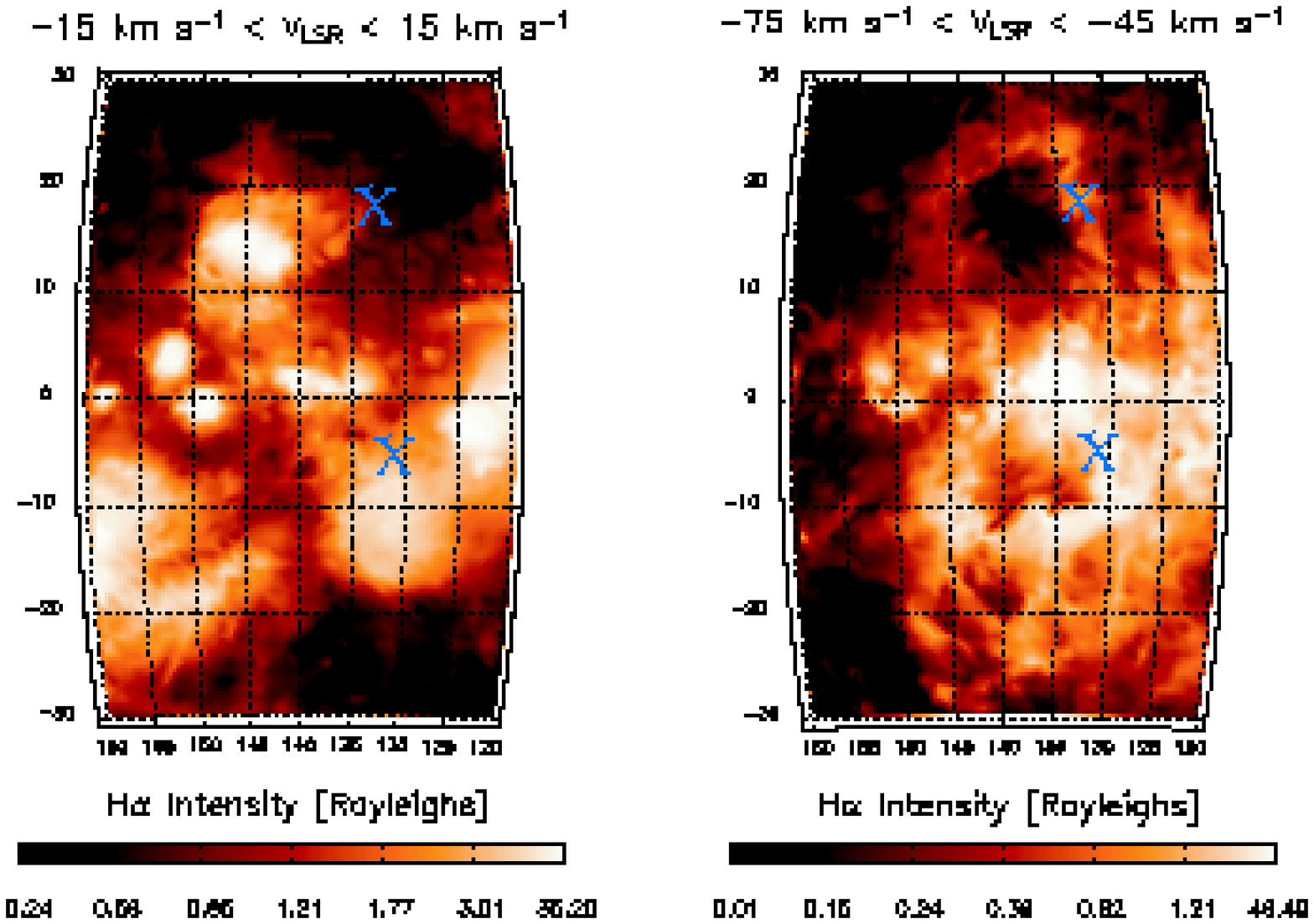}
}
\caption[\ha\ velocity channel maps toward Perseus]{ \ha\ maps of a
  large area of the Galaxy in the direction of the Perseus spiral arm,
  from the WHAM \ha\ sky survey. The left panel is a map of foreground
  emission with $|\vlsr| < 15\ \kms$. Note the presence of several
  large \hii\ regions superposed on a fainter, diffuse
  background. The panel on the right shows emission from the Perseus
  spiral arm with $-75\ \kms < \vlsr < -45\ \kms$.  Note the bipolar
  `superbubble' feature as two loops centered on the W4 \hii\ region
  near (135\dg, 0\dg).  The two `X' labels are the approximate
  location of the two pointed observations that have been observed in
  several emission lines. \label{fig:bowtiesurv}} 
\end{figure}

To investigate whether the size of the bubble influences the emission line 
ratios, we examine next a much larger (2000 pc $\times$ 800 pc) and 
fainter superbubble ionized by the Cas OB6 association and covering much 
of the constellations Perseus, Cassiopeia, and Camelopardalis.  We will 
refer to this enormous structure as the ``Perseus superbubble''. Compared 
to the Orion-Eridanus bubble, this superbubble has nearly nine times the 
linear extent, 1/10 the \ha\ surface brightness, and 1/5 -- 1/10 the gas 
density in its shell (i.e., 0.1 -- 0.2 cm$^{-3}$, comparable to densities 
in the WIM). Figure~\ref{fig:bowtiesurv} shows two velocity interval 
maps of this region covering about 2400 deg$^2$ from the WHAM sky survey.  
Most \ha\ spectra in this region have two or more components, one centered 
near \vlsr = 0~\kms\ and one near $-50$~\kms. 
The map on the left shows foreground \ha\ emission from the solar
neighborhood at $-15~\kms < \vlsr < +15\ \kms$,  
while the map on the right shows emission from the same piece of sky, but over 
the velocity interval $-75\ \kms < v_{LSR} < -45\ \kms$. The emission at 
these more negative velocities is from the Perseus spiral arm, 2-2.5 kpc 
distant, and is dominated by the Perseus superbubble \citep{RSH01}. 

\begin{figure}[p]
\center{
\includegraphics[scale=0.8]{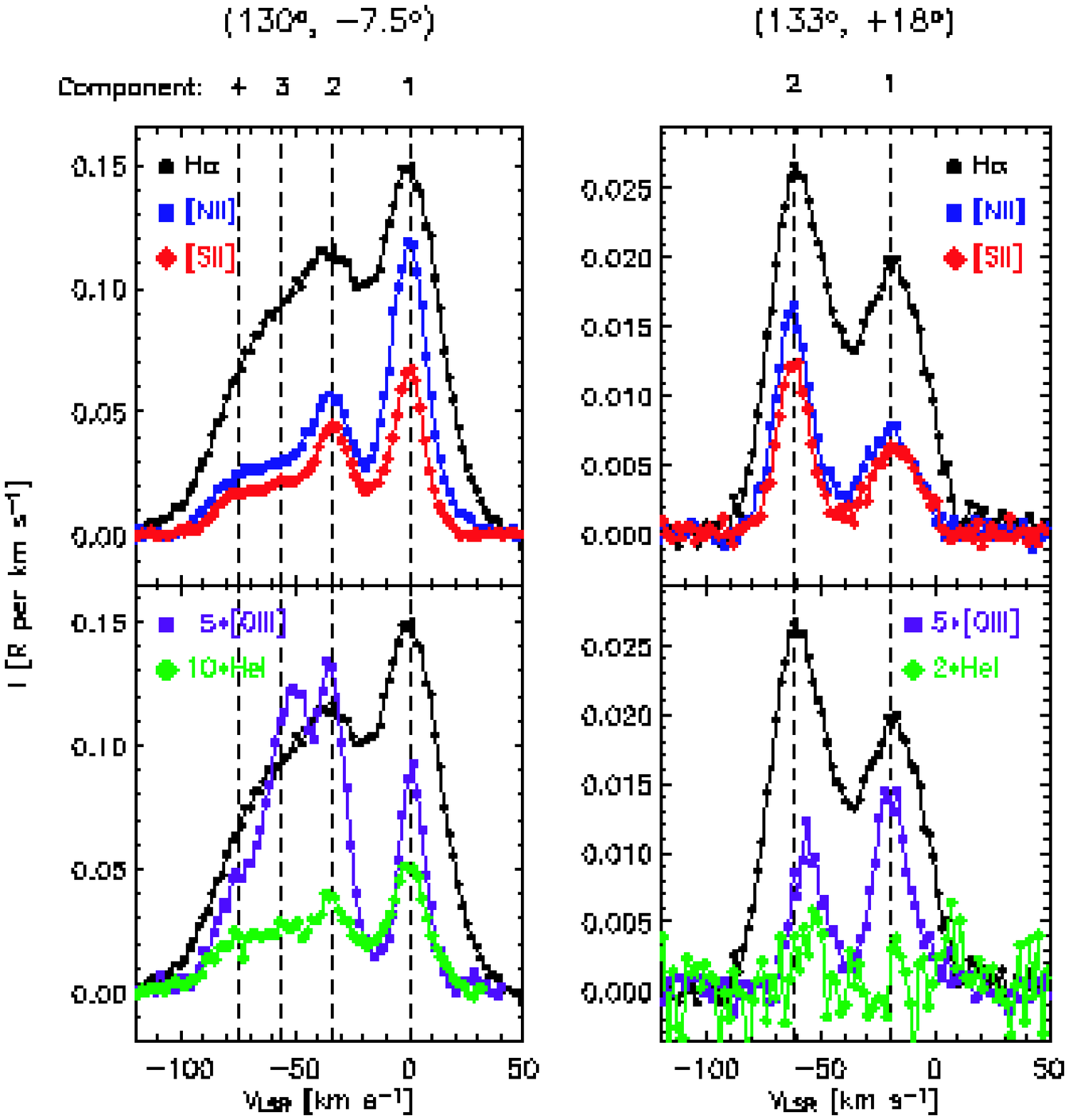}
}
\caption[Emission line spectra toward (130\dg, -7.5\dg) and (133\dg,
+18\dg)]{Emission line spectra along two sightlines toward the Perseus
  spiral arm. The left and right panels shows spectra toward (130\dg,
  -7.5\dg) and (133\dg, +18\dg), respectively (denoted by `X's in
  Figure~\ref{fig:bowtiesurv}). The top panels show the \ha, \nii, and
  \sii\ spectra. The bottom panels show the \ha, \oiii, and \hei\
  spectra. The \oiii\ and \hei\ data have been multiplied by the
  indicated values to facilitate the comparison with the \ha\
  profiles. The vertical dashed lines are at the locations of the
  velocity components that compose the emission profiles. Note the
  strong variation in the relative strengths of the components between
  the different emission lines, especially for \oiii\ in the left
  panel. Maps of the emission associated with components 1 and 3
  toward (130\dg, -7.5\dg) and components 1 and 2 toward (133\dg,
  +18\dg) are shown in the top and bottom panels of
  Figure~\ref{fig:bowtiesurv},
  respectively. \label{fig:bowtiespectra}} 
\end{figure}

Figure~\ref{fig:bowtiespectra} shows the emission line spectra from the 
pointed observations obtained toward the two sightlines at (130\dg, 
-7.5\dg) and (133\dg, +18\dg), denoted by `X's in 
Figure~\ref{fig:bowtiesurv}. These sightlines pass through the 
bipolar loop structure near the outer boundary of the superbubble.  
\ha, \nii, and \sii\ spectra appear in the top plots, with \ha, \oiii\ and 
\hei\ spectra on the bottom. Note that the \oiii\ and \hei\ spectra have 
been multiplied by the indicated values to facilitate the comparison of 
the relative strengths of the lines.  Several fainter {\sc{OFF}} directions, used to 
remove the atmospheric lines and background emission, were located at $|b| 
\gtrsim 20\dg$ near the longitude of the pointed observations.  

Each of the sight lines contains two or more distinct components, whose 
velocities were determined by least-squares fits of a sum of Gaussian 
profiles to the \sii\ spectrum in each direction.  The \sii\ spectrum was 
chosen because of its narrow, well-resolved component profiles and its 
high signal-to-noise.  The resulting component velocities are shown as 
vertical dashed lines in Figure~\ref{fig:bowtiespectra}, with component 
identification numbers shown above the top plot.  The maps on the left and right 
in Figure~\ref{fig:bowtiesurv} correspond to emission from components 1 
and 3, respectively, for (130\dg, -7.5\dg), and components 1 and 2, 
respectively, for (133\dg, +18\dg). Therefore, we identify component 3 
toward (130\dg, -7.5\dg) and component 2 toward (133\dg, +18\dg) with the 
Perseus superbubble.  Component 1 in each direction is associated with 
ionized gas near the outer edge of extended \hii\ regions surrounding 
$\phi$ Per and $\alpha$ Cam, respectively.  The nature of components 2 and 
4 toward (130\dg, -7.5\dg) is not known.

\begin{figure}[p]
\center{
\includegraphics[scale=0.8]{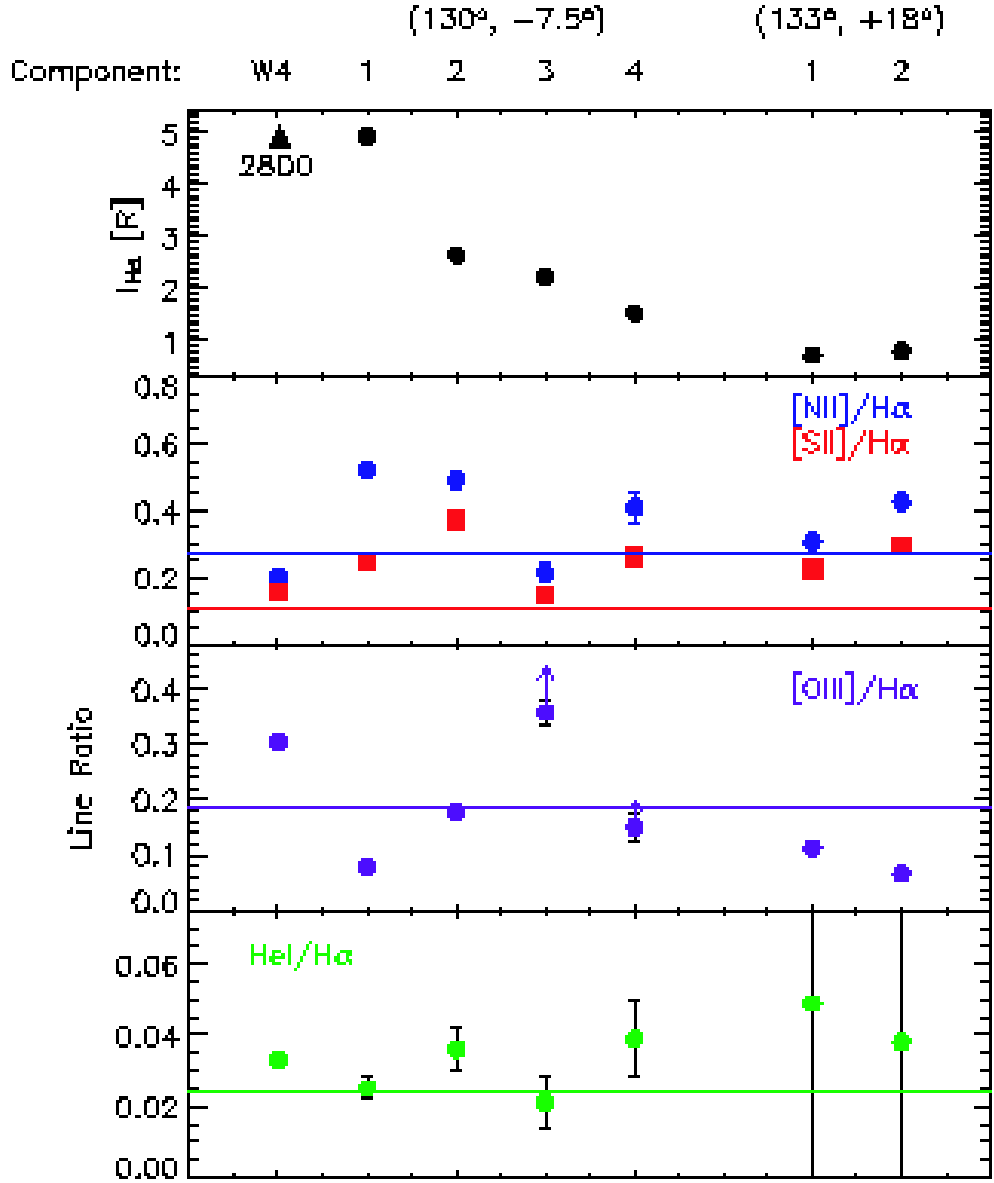}
}
\caption[Emission line ratios toward (130\dg, -7.5\dg) and (133\dg,
+18\dg)]{ Emission line strengths and ratios toward the W4 \hii\
  region and two lines of sight, (130\dg, -7.5\dg) and (133\dg,
  +18\dg), which pass through the loop structures in the Perseus
  spiral arm. The names of the components appear above the top plot,
  and are the same as in Figure~\ref{fig:bowtiespectra}. The layout of
  the plots is the same as Figure~\ref{fig:oridiag}. The upward
  pointed arrows indicate the change in the location of the \oiii/\ha\
  ratio if an extinction correction of \av = 0.7 and 0.9 mag are
  applied to components 3 and 4, respectively (see text). The \ha\
  intensity of the W4 \hii\ region (2800 R; Table~\ref{tab:hiibasic})
  is far off scale in the top plot. \label{fig:bowtiediag}} 
\end{figure}

The strength of each component was calculated from a multi-component 
Gaussian fit to each spectrum in which the velocities for each component 
were fixed as determined from the \sii\ spectra.  A summary of the line 
strengths and their ratios is presented in Table~\ref{tab:per}.  The 
columns of the table are the same as for Table~\ref{tab:ori}, except 
that $D_\theta$ refers to the angular distance of the pointing from the 
W4 \hii\ region (Cas OB6), the presumed source of the ionizing radiation 
for the superbubble.  A graphical representation of the data in 
Table~\ref{tab:per} is shown in Figure~\ref{fig:bowtiediag}. The 
layout of Figure~\ref{fig:bowtiediag} is the same as 
Figure~\ref{fig:oridiag}, with the solid horizontal lines representing 
the average line ratios of the O-star \hii\ regions. The \ha\ intensity 
for the W4 \hii\ regions (2800 R) is far off the scale of the plot.

\hb\ observations toward (130\dg, -7.5\dg) suggest that the extinction is 
generally low, with \av\ $\approx$\ 0.1, 0.4, 0.7, and 0.9 mag for 
components 1, 2, 3, and 4, respectively. This implies a maximal correction 
of $\lesssim$ 30\% for the \oiii\ data (for component 4), and a much 
smaller correction for the other lines. No correction has been applied to 
the data in Figure~\ref{fig:bowtiespectra} or Table~\ref{tab:per}.  
However, the tips of the upward pointed arrows in 
Figure~\ref{fig:bowtiediag}, for the \oiii/\ha\ data, denote the change 
in this ratio if the above extinction corrections are applied.  No 
extinction correction was applied to the data toward the much fainter 
direction of (133\dg, +18\dg), where the uncertainty in the \ha/\hb\ did 
not allow for an accurate determination of \av. However, the low 
extinction for (130\dg, -7.5\dg), suggests that the correction for the 
higher latitude direction is likely to be negligible.

Figures~\ref{fig:bowtiespectra} and \ref{fig:bowtiediag} both show 
significant variations in the relative strengths of the lines, especially 
toward (130\dg, -7.5\dg). The emission component identified with the 
superbubble (component 3) has spectral characteristics similar to the W4 
\hii\ region immediately surrounding Cas OB6, that is, significantly lower 
\nii/\ha\ and \sii/\ha, and higher \oiii/\ha\ than the other components in 
this direction. On the other hand, toward (133\dg, +18\dg) the gas associated with 
the superbubble (component 2), is more WIM-like, with high 
\nii/\ha\ and \sii/\ha\ and low \oiii/\ha\ compared to W4. In the 
following section, we explore the variations in \nii\ and \sii\ using the 
line ratio maps of this region. These maps reveal subtle differences in 
ionization conditions within the low velocity foreground gas, within the loops and filaments of the more distant superbubble, and between the superbubble and the WIM.

\subsubsection{Foreground Emission}

This area of the Galaxy is well suited for emission line studies of the 
diffuse ionized gas because the presence of two well separated velocity 
components allows one to probe two potentially different environments in 
the same spectrum.  At radial velocities near the LSR, there are several 
large classical \hii\ regions, as shown in Figure~\ref{fig:bowtiesurv} 
(left panel). The largest of these are Sivan 3 near (145\dg, +15\dg) 
associated with the O9.5~Ia star $\alpha$\ Cam, Sivan 4 near (155\dg, 
-15\dg) associated with the O7.5~III star $\xi$\ Per, and an \hii\ region 
near (130\dg, -10\dg) associated with the B0.5+sdO star $\phi$\ Per.  

\begin{figure}[p]
\center{
\includegraphics[scale=0.8]{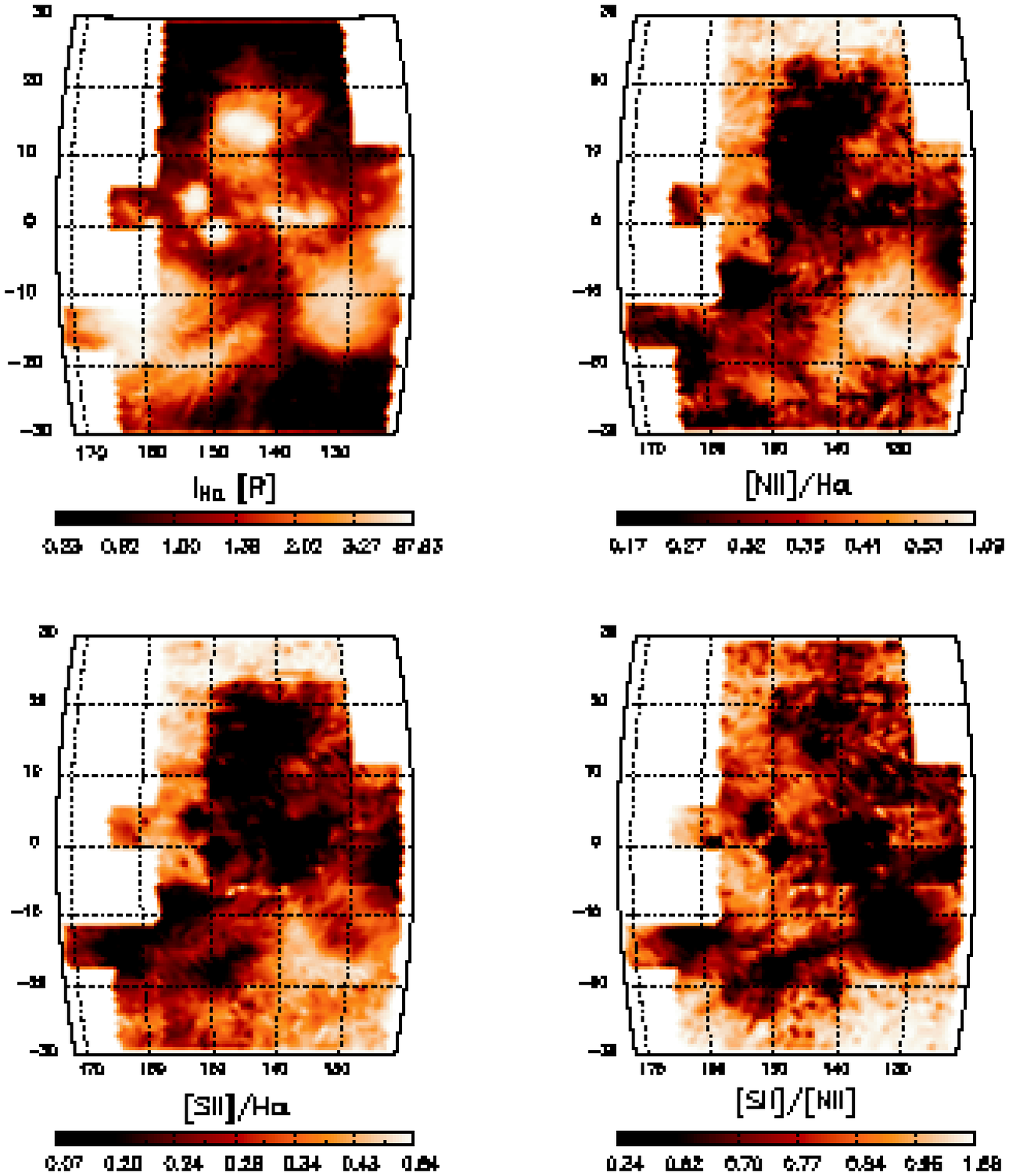}
}
\caption[\ha, \nii/\ha, \sii/\ha, and \sii/\nii\ for local emission
toward Perseus]{ Histogram equalized emission line maps of \ha,
  \nii/\ha, \sii/\ha\ and \sii/\nii\ with $|\vlsr| < 15\ \kms$ toward
  most of the region shown in Figure~\ref{fig:bowtiesurv}.  The bright
  emission from this nearby gas is dominated by several large, bright
  O-star \hii\ regions which appear as dark regions in the line ratio
  maps.  The faint, diffuse background (WIM) appears as ``bright"
  regions in the \nii/\ha\ and \sii/\ha\ maps. Note the exceptional
  appearance of the \hii\ region around $\phi$\ Per, near (130\dg,
  -10\dg), which is ionized by a B0.5+sdO binary
  system. \label{fig:bowtiemaplocal}} 
\end{figure}

Maps of \iha, \nii/\ha, \sii/\ha, and \nii/\sii\ for this relatively nearby gas with $|\vlsr| <$ 
15~\kms\ are shown in Figure~\ref{fig:bowtiemaplocal}.  Many of the 
pointed observations toward the classical, relatively bright \hii\ regions 
reported in \S\ref{sec:hii} are in this map.  The map of \nii/\ha\ reveals 
that the faintest regions of \ha\ (i.e., the WIM) are ``bright" in 
\nii/\ha, with \nii/\ha\ $\gtrsim$ 0.5, while most of the bright \hii\ 
regions appear dark, with \nii/\ha\ $\lesssim$ 0.3. The exception is the 
15\dg\ diameter circular region near (130\dg, -10\dg). This is the $\phi$\ 
Per \hii\ region, which is ionized by a B0.5+sdO binary system 
\citep{HRT99}.  The \sii/\ha\ map is similar to the \nii/\ha\ map, 
suggesting that the \nii/\ha\ and \sii/\ha\ line ratio variations are 
dominated by temperature changes (see equations \ref{eq:niieq} and 
\ref{eq:siieq}).  \sii/\ha\ is high, $\gtrsim$\ 0.4, in regions of 
faint \ha\ emission and low, $\lesssim$\ 0.2, toward \hii\ regions.  The 
$\phi$\ Per \hii\ region is elevated in \sii/\ha, relative to the 
background, but not as much as it is in \nii/\ha.  This is shown clearly 
on the map of \sii/\nii, where the gas associated with classical \hii\ 
regions is depressed in \sii/\nii, especially for $\phi$ Per.

As discussed in \S\ref{sec:physconds}, features on these ratio maps 
can be interpreted as changes in the physical conditions of the gas, where 
\nii/\ha\ follows the temperature of the gas and \sii/\nii\ traces 
S$^+$/S. These maps thus indicate that the faint, diffuse WIM is 
significantly warmer and in a lower ionization state (more S is in the 
form of S$^+$) than in the traditional \hii\ regions. The \hii\ region 
surrounding the B star $\phi$\ Per differs from the other \hii\ regions in 
that it has a significantly higher temperature (i.e., \nii/\ha\ ratio). 
This may be due to the hard ionization ionizing radiation from its hot sdO 
companion.

\begin{figure}[p]
\center{
\includegraphics[scale=0.8]{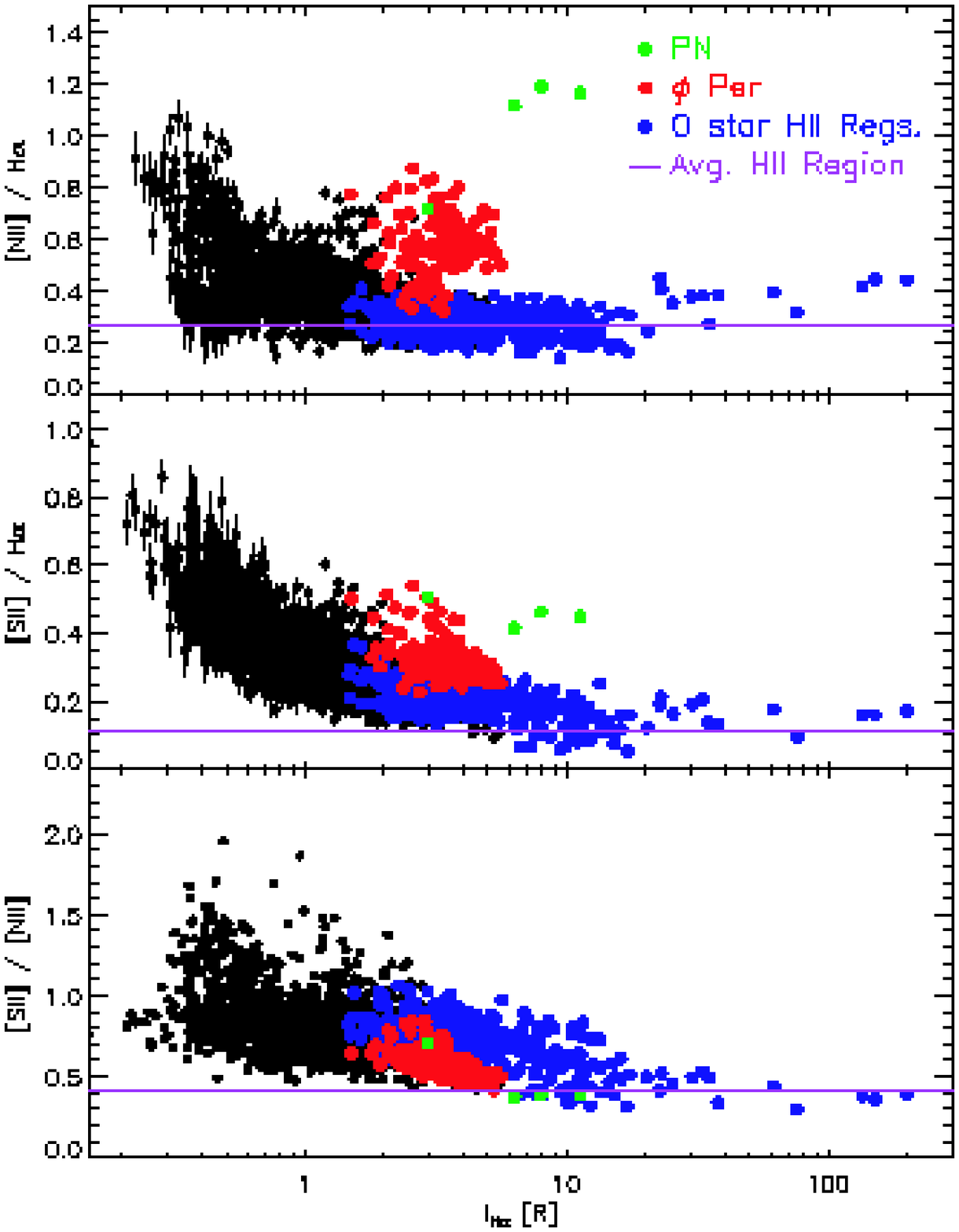}
}
\caption[\nii/\ha, \sii/\ha, and \sii/\nii\ versus \iha\ for local emission
toward Perseus]{ \nii/\ha, \sii/\ha, and \sii/\nii\ as a function of \ha\
  intensity, from the maps of the nearby gas shown in
  Figure~\ref{fig:bowtiemaplocal}.  The horizontal scale is logarthmic
  to show the full range of \iha\ spanned by the data. 
  The WIM observations are denoted by solid black circles. Directions near
  the bright O-star \hii\ regions and the B0.5+sdO $\phi$\ Per \hii\
  region are shown in blue and red, respectively. A few directions
  near the unusual planetary nebula, S216, are shown in green. The
  horizontal lines indicate the average values for the O star \hii\
  regions in Table~\ref{tab:hiimulti}. We see that \nii/\ha, \sii/\ha, and \sii/\nii\
  increase with decreasing \ha\ intensity. \label{fig:bowtieratiolocal}} 
\end{figure}

\begin{figure}[p]
\center{
\includegraphics[scale=0.8]{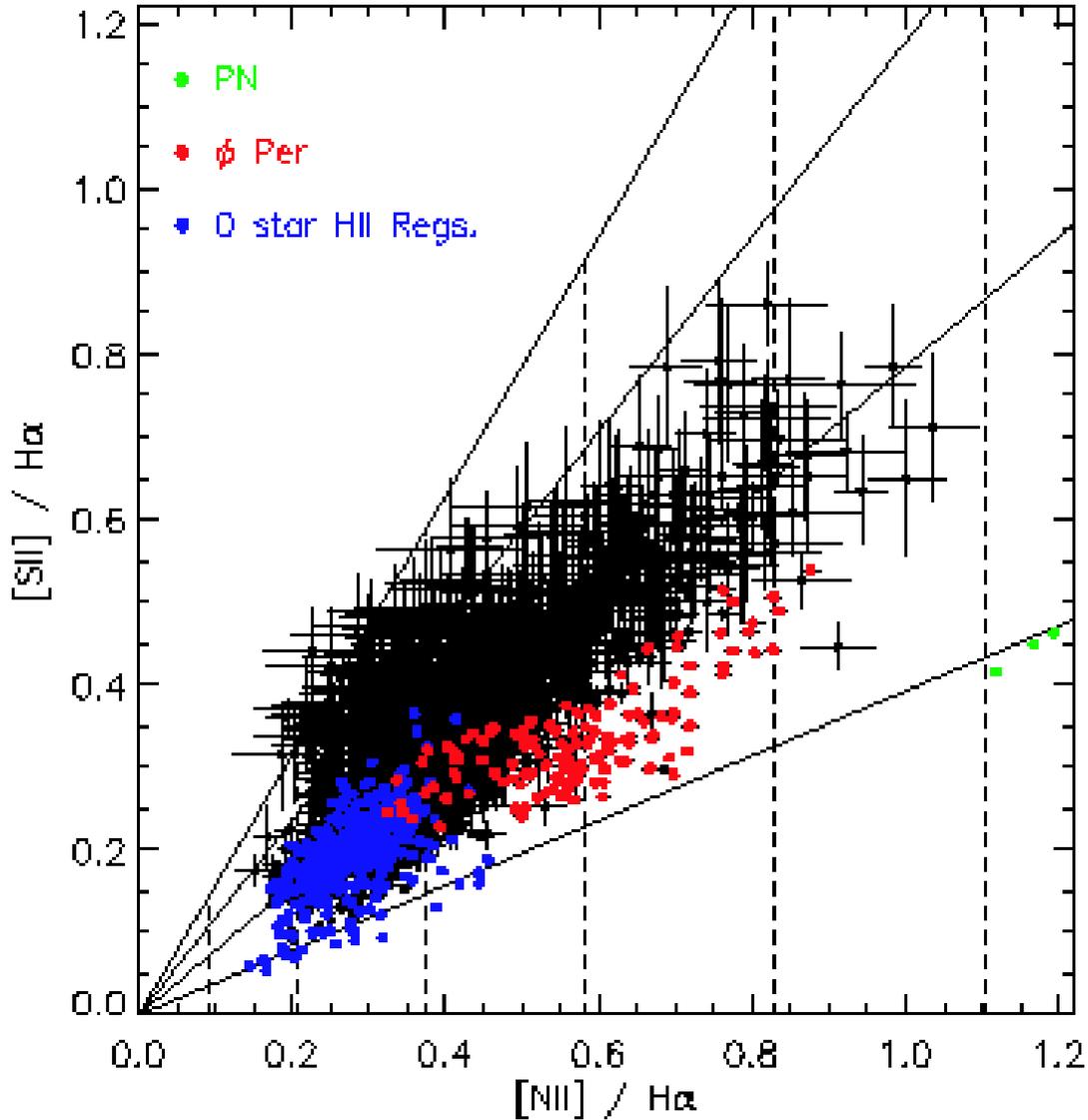}
}
\caption[\nii/\ha\ versus \sii/\ha\ for local emission toward
Perseus]{ \nii/\ha\ versus \sii/\ha\ for nearby ionized gas toward
  Perseus, from the data points in Figure~\ref{fig:bowtieratiolocal}.
  The axes and solid and dashed lines are the same as in
  Figure~\ref{fig:orirationvs}. The symbols have the same meaning as
  in Figure~\ref{fig:bowtieratiolocal}.  The diagram suggests that the
  O-star \hii\ regions are at temperatures $6000 \rm{K} < T < 7000$\
  K, with the gas in the WIM and $\phi$\ Per \hii\ region generally
  warmer with T ranging up to 9000 K. \label{fig:bowtieratiolocalnvs}}

\end{figure}

These variations in line ratios are illustrated in 
Figures~\ref{fig:bowtieratiolocal} and 
\ref{fig:bowtieratiolocalnvs}. Figure~\ref{fig:bowtieratiolocal} 
displays the \nii/\ha\, \sii/\ha, and \sii/\nii\ measurements in 
Figure~\ref{fig:bowtiemaplocal} as function of \ha\ intensity. Data 
points with uncertainties greater than 0.1 in the ratio are omitted. 
Observations toward the large O-star \hii\ regions are shown in blue, 
while observations within 6\dg\ of $\phi$\ Per are shown in red. The data 
points with \iha\ $> 20$ R are all close to the $\xi$\ Per \hii\ region 
near (160\dg, -10\dg). A few additional points, near (158\dg, 0\dg), are 
associated with an unusually large and old planetary nebula S216 
\citep{Reynolds85pn}, and are in green. All other directions (the WIM) are 
indicated by black data points.  The increase in \nii/\ha\ and \sii/\ha\ 
with decreasing \ha\ intensity is apparent, confirming the trend found by 
\citet{HRT99} for a smaller sample of this region of the sky. We also see that \sii/\nii\ in the fainter gas is elevated compared to regions of brighter emission associated with \hii\ regions. 

The scatter in the data points is much larger than the measurement errors, 
implying real variations in the temperature and ionization state of the 
gas. These variations are displayed more quantitatively in 
Figure~\ref{fig:bowtieratiolocalnvs}, with the same data points and 
symbols as Figure~\ref{fig:bowtieratiolocal} and with the same scaling 
and labels as Figure~\ref{fig:orirationvs}. The \nii/\ha\ data suggest 
that most of emission from the \hii\ regions is relatively cool (6000 K $< 
T <$ 7000 K), and that the faintest regions of the WIM extend to 
temperatures $T > 9000 K$.  This diagram also suggests that emission from 
the $\phi$\ Per \hii\ region has a similarly high temperature as the
WIM, but with a lower S$^+$/S.  The large, evolved planetary nebula
S216 (green points) has the highest temperature ($\approx$ 11,000 K)
and a low S$^+$/S ($\approx$ 0.24),  
consistent with the high photospheric temperature of its ionizing star \citep{CudR85, 
TN92}.

\subsubsection{Perseus Arm Emission}

These spectral characteristics of the WIM and the \hii\ regions in the 
local gas near the LSR can now be compared to those of the Perseus 
superbubble in the $-75\ \kms < v_{LSR} < -45\ \kms$ maps. 
The emission from the Perseus arm (right panel in
Figure~\ref{fig:bowtiesurv}) shows a remarkable bipolar,  
closed loop structure centered near (135\dg, 0\dg)  that extends almost 
30\dg\ above and below the Galactic plane (i.e., $\pm 1200$ pc).  The W4 
star-forming region, ionized by the Cas OB6 association, is located at a 
distance of $\approx$\ 2.2 kpc, and is near the center of this structure 
close to the Galactic plane.  Several observational and theoretical 
studies have suggested that a large `chimney' has been carved out near the 
Galactic plane, allowing radiation and hot gas to move out into the 
Galactic halo \citep{NTD96, DTS97, BJM99, Terebey+03}. The \ha\ emission 
from the large arc-shaped feature above the plane at $0\dg < b < +30\dg$ 
was explored by \citet{RSH01}, who found that the size and shape of this 
loop is consistent with a sequence of star-forming events over a period of 
$\sim$\ 10$^7$ yr that have carved out a 1 kpc-scale cavity in the ISM.  
They concluded that the ionized hydrogen in the loop, the upper parts of 
which are more than 1 kpc from the O stars in Cas OB6, appeared to be 
produced by ionizing radiation escaping the association. 
\citet{HRT99} mapped the lower part of this region ($b < -5\dg$), in \nii\ 
and \sii. Here we report on new survey observations of this entire region 
in \nii\ and \sii, as well as pointed observations in two directions along 
the edge of the structure in \hb, \hei, \oiii, and \niiblue.

\begin{figure}[p]
\center{
\includegraphics[scale=0.8]{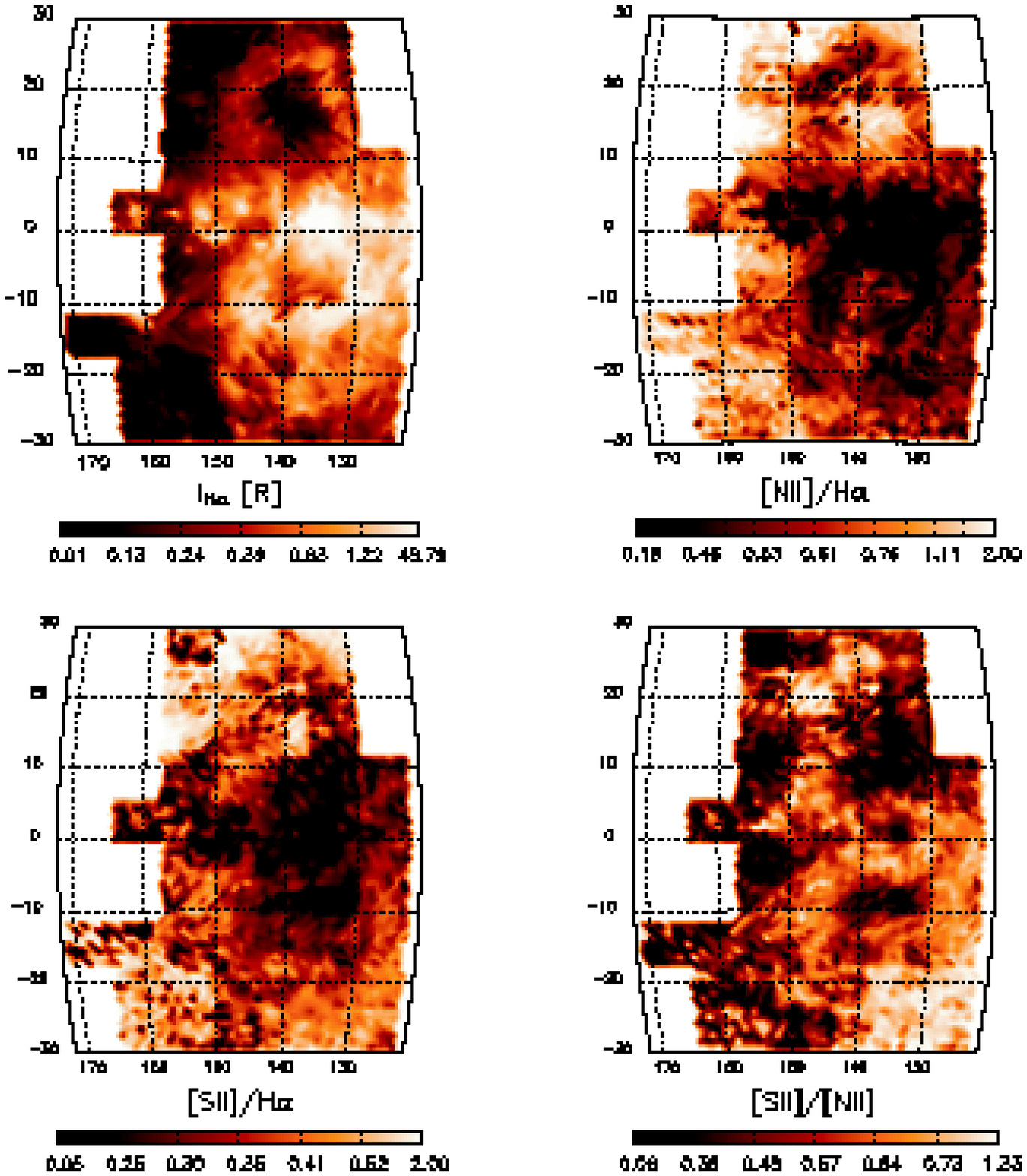}
}
\caption[\ha, \nii/\ha, \sii/\ha, and \sii/\nii\ maps of Perseus
superbubble]{ Histogram equalized emission line maps of \ha, \nii/\ha,
  \sii/\ha\ and \sii/\nii\ with $-75\ \kms < \vlsr < -45\ \kms$ toward
  the same region shown in Figure~\ref{fig:bowtiemaplocal}.  The \ha\
  emission from this distant gas in the Perseus arm ($\sim$ 2 kpc)
  shows a large bipolar loop structure centered on the W4 \hii\ region
  near (135\dg, 0\dg).  These loops appear as depressions in the
  \nii/\ha\ and \sii/\ha\ maps, suggesting they are regions of lower
  temperature compared to the diffuse background. The anti-correlation
  between \iha\ and \nii/\ha\ can be traced along several individual
  filaments with $b < -10\dg$. \label{fig:bowtiemapper}} 
\end{figure}

The \ha\ and 
\nii\ and \sii\ line ratio maps of this radial velocity interval are shown 
in Figure~\ref{fig:bowtiemapper}.  
The \nii/\ha\ map reveals that regions of higher \ha\ intensity have a 
lower \nii/\ha, and that this anti-correlation holds not only for the 
large scale features but also for the smaller filamentary structures of 
the superbubble. The faintest areas of \ha\ (e.g., the WIM at $l > 
150\dg$) have the highest \nii/\ha\ ($\gtrsim$\ 0.7). Along the loops, 
\nii/\ha\ is depressed relative to the regions interior and exterior to 
the loops. Several smaller filaments in the region $130\dg < l < 150\dg, 
-30\dg < b < -10\dg$ show the same detailed anti-correlation of \nii/\ha\ 
and \iha. As expected, the area within about 5\dg\ of the W4 \hii\ region 
has the lowest \nii/\ha\ ratios ($\lesssim$\ 0.4). Because sightlines 
toward the loops and filaments contain weak foreground and background 
emission at the same velocity, \nii/\ha\ from the loops and filaments 
themselves is even more depressed compared to the adjacent regions than 
these maps suggest. The lower \nii/\ha\ ratios in the filamentary 
structures, particularly in the brighter \ha\ regions south of the plane 
where contamination by the fainter WIM along the line of sight is 
negligible, provides strong evidence that these filaments are not 
enhancements resulting from folds and edge projections of the ionized 
outer skin of the superbubble but instead are regions of cooler 
temperature.  Furthermore, the generally lower \nii/\ha\ within the entire 
superbubble compared to that in the more diffuse WIM at $l > 150\dg$, for 
example, implies that the WIM cannot be explained entirely by the 
superposition of regions like the Perseus superbubble.

In the \sii/\ha\ map the contrast between the filaments and the background 
is not as high.  Since the line ratio \sii/\ha\ is a combination of 
temperature and ionization effects, this decrease in contrast may be due 
to changes in the ionization fraction of S$^+$/S that `wash out' the 
temperature effects we see in the \nii/\ha\ map.  This is further evidence 
that the \ha\ enhanced filaments and loops are not simply due to 
geometrical projection effects of an ionized shell and that the 
ionizations conditions (S$^+$/S) within the superbubble differ from that 
in the WIM.

\begin{figure}[p]
\center{
\includegraphics[scale=0.8]{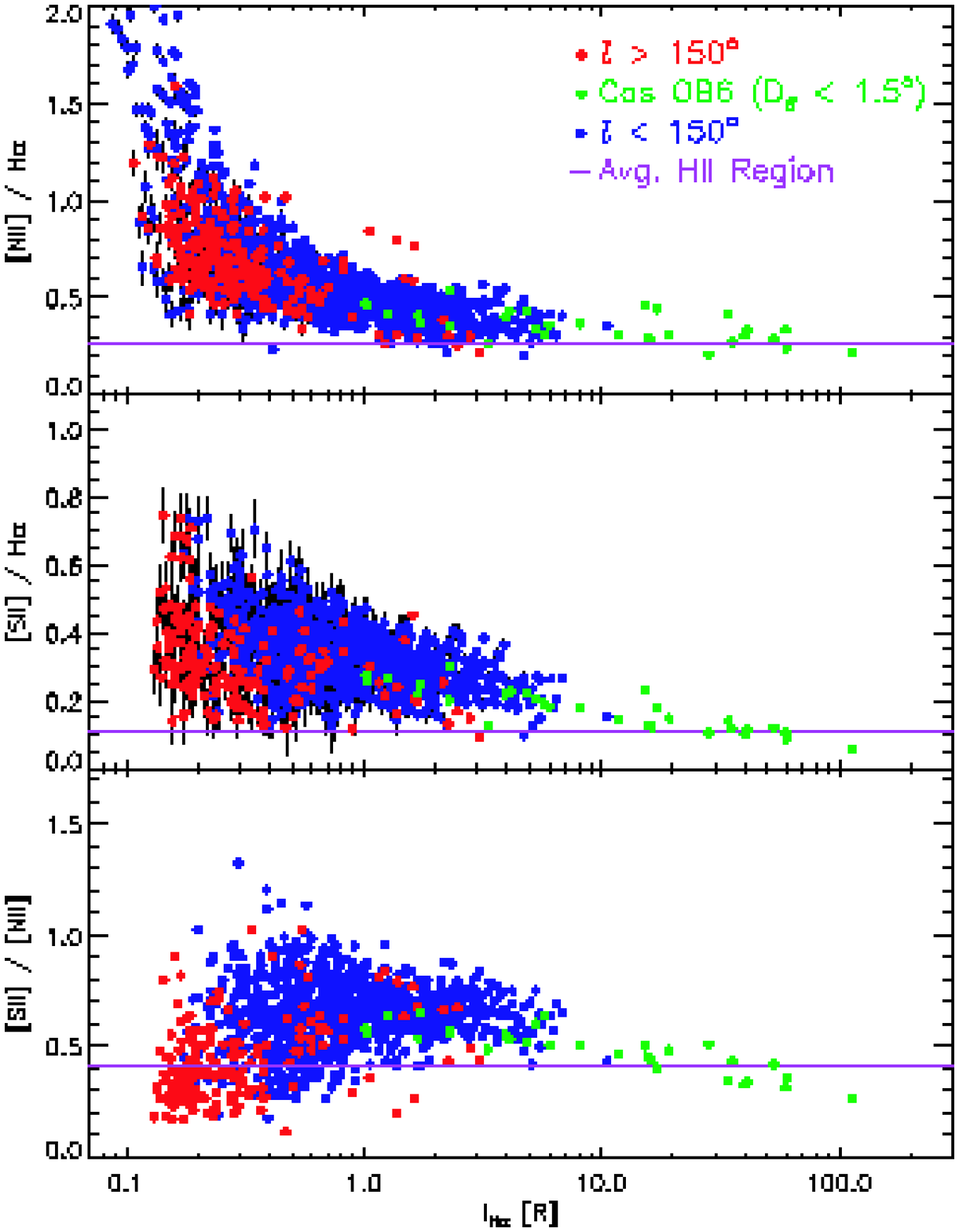}
}
\caption[\nii/\ha\ and \sii/\ha\ versus \iha\ for Perseus
superbubble]{ \nii/\ha, \sii/\ha, and \sii/\nii\ as a function of \ha\
  intensity, from the maps in Figure~\ref{fig:bowtiemapper}.
  Observations within 1.5\dg\ of spectroscopically confirmed members
  of the Cas OB6 association are shown in green.  Observations with
  longitudes $l > 150\dg$ and $l < 150\dg$ shown in red and blue,
  respectively. The horizontal lines indicate the average values for
  the O star \hii\ regions in Table~\ref{tab:hiimulti}. We see that
  both \nii/\ha\ and \sii/\ha\ increase with decreasing emission
  measure, as seen elsewhere in the WIM. Also note that the \nii/\ha\
  and \sii/\ha\ ratios toward the faint diffuse emission regions
  (\emph{red}) are higher, on average, than the emission regions that
  include the bipolar superbubble
  (\emph{blue}). \label{fig:bowtieratioper}} 
\end{figure}

\begin{figure}[p]
\center{
\includegraphics[scale=0.8]{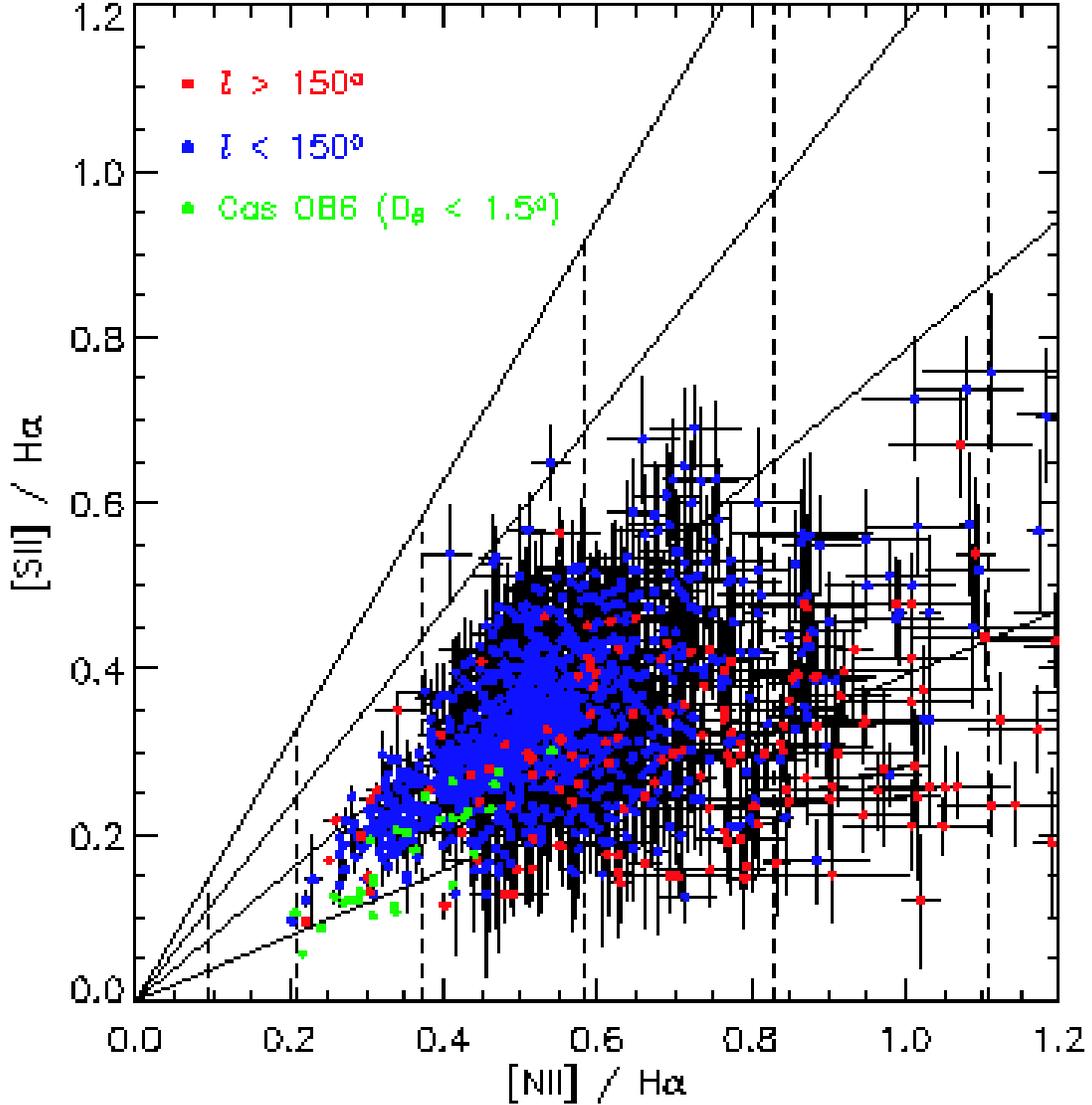}
}
\caption[\nii/\ha\ versus \sii/\ha\ for Perseus superbubble]{
  \nii/\ha\ versus \sii/\ha\ for the more distant ionized gas in the
  Perseus spiral arm, from the maps in Figure~\ref{fig:bowtiemapper}.
  The axes and solid and dashed lines have are the same as in
  Figure~\ref{fig:orirationvs}. 
We see that the dense gas ionized by O stars in the Cas OB6
association (\emph{green}) lie at lower temperatures compared to the
rest of the emission (\emph{blue} and \emph{red}).  
We also see that  the observations of faint diffuse emission at $l >
150\dg$ (\emph{red}) tend to have higher temperatures relative to the
observations at $l < 150\dg$ (\emph{blue}) that include the brighter
loops and filaments.   
 \label{fig:bowtieratiopernvs}}
\end{figure}

Quantitative comparisons between the W4 \hii\ region, the superbubble, and 
the diffuse WIM outside the boundary of the superbubble are shown in 
Figures~\ref{fig:bowtieratioper} and \ref{fig:bowtieratiopernvs}. 
Data points with uncertainties larger than 0.1 in the ratio are not shown.  
Observations within 1.5\dg\ of spectroscopically confirmed members of the 
Cas OB6 association are shown in green, sightlines through the 
superbubble, defined as all points with $l < 150\dg$, are shown in red, 
and the WIM, defined as all points with $l > 150\dg$, are in blue. In 
general we see that \nii/\ha\ is higher in the diffuse 
background where \iha\ is lower.  We also see that the brighter regions near 
W4 have line ratios close to that of the average \hii\ region, while the 
superbubble tends to have \nii/\ha\ ratios closer to that of the
nearby WIM, but weighted to somewhat lower values. 

We note one caveat about our analysis of the complicated component structure toward these lines of sight. 
The method we used to calculate the line strengths, integrating the 
profiles over fixed velocity intervals, is not ideal for exploring 
the full range of physical conditions that may be present in the ionized 
gas in these maps. Most of the spectra in this part of the Galaxy have two 
components centered near 0 \kms\ and $-50$ \kms\ (LSR) with line widths of 
$\approx 20-30$~\kms. However, some spectra have profiles that peak at 
somewhat different velocities.  This is especially true in the region near 
the southern loop with $b \lesssim -10\dg$.  Spectra separated by only 
1\dg\ in this area show variations in the component strengths by a factor 
of 2 and shifts in the velocity centroids of 10 \kms\ or more. Therefore, 
in some cases a map that covers a fixed radial velocity interval may 
sample the entire emission component in one part of the map, but only the 
wing of that component in another part of the map. This complicates the 
interpretation of the line ratios.  A detailed examination of the 
emission, including a study of the dynamics of the gas, would require 
Gaussian fits to all of the components in all of the approximately 2400 
spectra, which is beyond the scope of this work.  Our focus here is 
limited to the general trends that appear among spatially coherent 
features in these maps, which are insensitive to these kinematic 
variations.

\subsubsection{The Perseus Superbubble and the WIM}

Compared to the smaller and brighter Orion-Eridanus bubble, the Perseus 
superbubble is clearly more WIM-like.  Compare, for example, the 
distribution of blue points in Figures~\ref{fig:oriratioplot} and 
\ref{fig:orirationvs} (Orion-Eridanus bubble) with the blue points in 
the corresponding Figures~\ref{fig:bowtieratioper} and 
\ref{fig:bowtieratiopernvs} (Perseus superbubble).  Because the surface 
brightness of the Perseus superbubble is not much brighter than that of 
the WIM, particularly far from the Galactic plane, searching for a trend 
in the line ratios with distance from Cas OB6 is problematic.  However, 
a comparison of the background subtracted pointed observation toward 
(130\dg, $-7.5$\dg), which is 10\dg\ ($\approx$ 350 pc) from Cas OB6, with that 
toward (133\dg, $+18$\dg), which is 17\dg\ ($\approx$ 700 pc) from Cas OB6, indicates 
that the gas in the latter direction is more WIM-like (i.e., higher \nii, 
\sii/\ha\ and lower \oiii/\ha) than in the former (see Table~\ref{tab:per}).  
Nevertheless, even the 
most distant parts of the superbubble are still clearly distinguishable 
from the background WIM, indicating that although this structure does 
have ratios similar to the WIM in some regions of the sky (e.g., outside 
the Orion-Eridanus bubble at $b < -53\dg$), there is still a significant 
difference (which we interpret primarily as a temperature difference) between this superbubble and the adjacent WIM.

\section{HIGH LATITUDE FILAMENTS}
\label{sec:hlfil}

There are other regions of the sky that exhibit \ha\ enhancements with no 
identified O stars or O associations as their source of ionization. Many 
of these are high Galactic latitude filamentary structures that have no 
visible connection to a superbubble.  The emission characteristics of 
four such regions are examined in detail in this section to determine 
where they fit empirically with respect to the classical \hii\ regions, superbubbles, 
and the more diffuse WIM.

\subsection{The Northern Filaments}
\label{subsec:nfil}

\begin{figure}[p]
\center{
\includegraphics[scale=0.8]{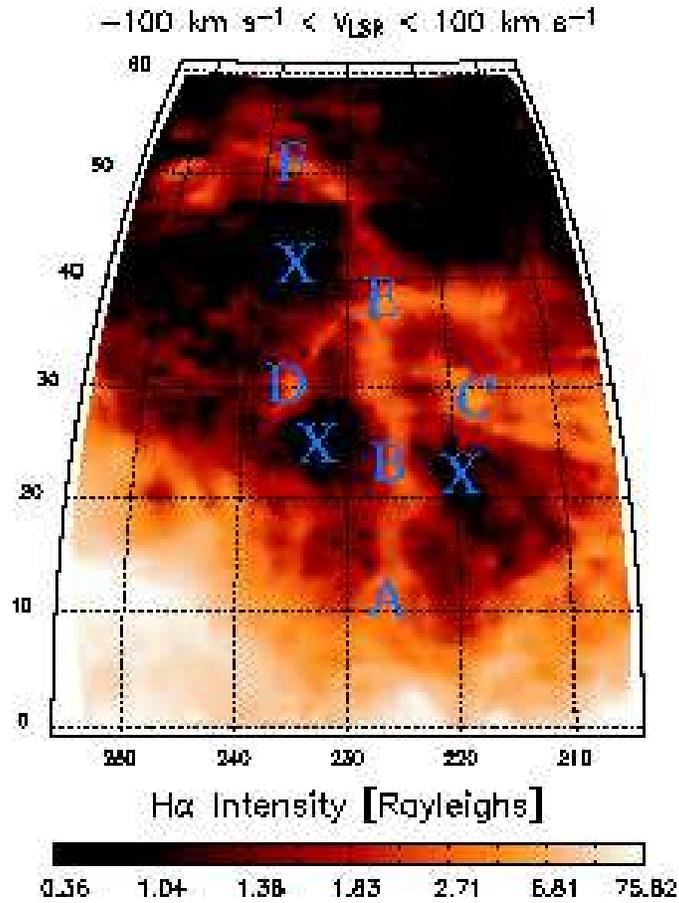}
}
\caption[\ha\ map of Northern Filament]{ \ha\ map of the `northern
  filaments', from the WHAM \ha\ sky survey. The emission has been
  integrated over \vlsr = $\pm$ 100 \kms. The labels A-F indicate the
  approximate location of the pointed observations summarized in
  Table~\ref{tab:nfil}. The three 'X' labels refer to the location of
  the {\sc{OFF}} directions. The narrow vertical filament near $l =
  225\dg$  stretches from the Galactic plane up to $b \approx$
  50\dg. \label{fig:nfilmap}} 
\end{figure}

A remarkable, $\sim 2\dg \times 60\dg$, \ha\ filament rises vertically 
more than 50\dg\ from the Galactic plane near longitude $l=225\dg$. As 
shown in Figure~\ref{fig:nfilmap}, another filament, about 30\dg\ long, 
traverses this longer filament at a right angle near $+37\dg$ latitude. We 
refer to these as the `northern filaments' to distinguish them from a 
comparably sized feature on the WHAM survey map near $l \approx 75\dg$ 
that extends south of the Galactic plane from $b \approx -15\dg$ to $b 
\approx -55\dg$. (This `southern filament' was not included in this study 
because of the complex kinematics of the gas toward and near the 
filament.) \citet{HRT98} examined the region of the northern filaments and 
found no correspondence of the \ha\ emission with observational tracers of 
any other phases of the interstellar medium. A radial velocity gradient 
along the length of the vertical filament, from $+18$ \kms\ near the 
midplane to $-25$ \kms\ at the highest latitude, suggests that it is a 
coherent strand of gas and not simply an enhancement resulting from an 
increase in geometrical path through the edge of a very low surface 
brightness shell.  They also noted that the lower parts of the filament 
are at the same longitude as the \hii\ region S292 surrounding the CMa OB 
1 association in the Galactic plane, and that the emission from this part 
of the filament appears at the same radial velocity as the \hii\ region.  
If the long filament is at the same distance as the OB association 
($\approx$\ 1 kpc), then it reaches a vertical height of $\approx$\ 1.2 
kpc above the plane and has a density near 0.3 cm$^{-3}$. \citet{HRT98} 
suggest that the diffuse ionized gas in these filaments is not likely to 
be material that has been ejected from the star-forming region below it. 
Instead, they argue that the relatively constant \ha\ surface brightness 
along the filament suggests that it is ionized by ambient Lyman continuum 
radiation.

\begin{figure}[p]
\center{
\includegraphics[scale=0.8]{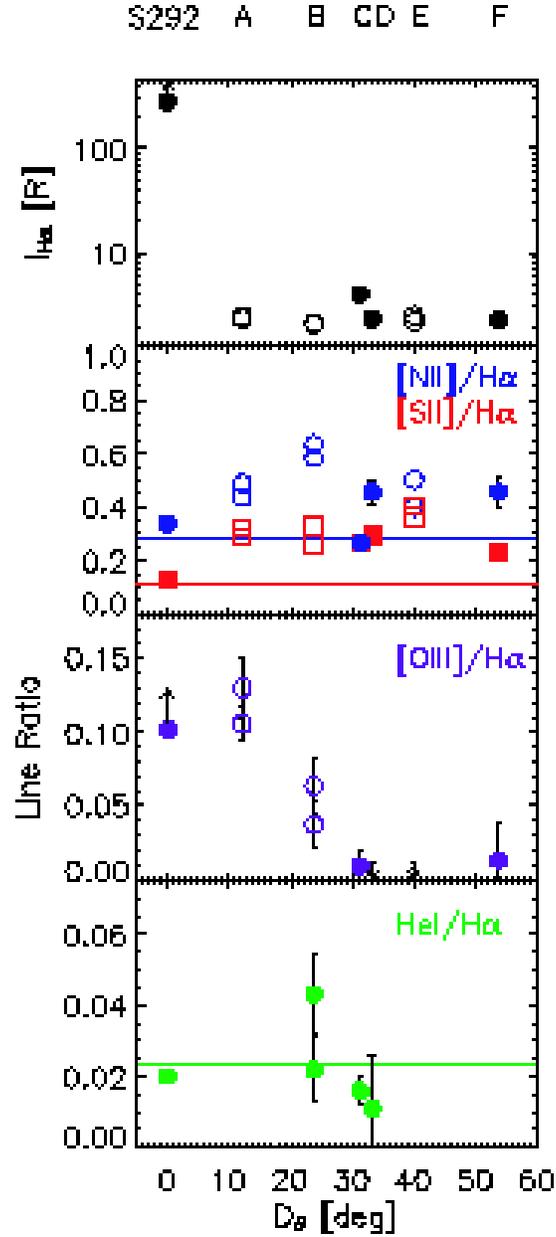}
}
\caption[Emission line ratios toward Northern Filament]{ Emission line
  intensities and ratios toward the `northern filaments', as a
  function of distance from the S292 \hii\ region near the
  midplane. The name of each observation is located above the diagram.
  The layout is the same as for Figure~\ref{fig:oridiag}. Open symbols
  represent observations in which spectra from two different
  {\sc{OFF}} directions were removed. Note the strong decrease in
  \oiii/\ha\ with distance above the Galactic
  plane.\label{fig:nfildiag}} 
\end{figure}

We obtained spectra of \nii, \sii, \oiii, and \hei\ with pointed 
observations along and near these two filaments. A summary of the results 
appears in the top ten rows of Table~\ref{tab:nfil}.  The columns are similar to those in 
Table~\ref{tab:ori}, except that column 4 is the angular distance of 
each observation from the S292 \hii\ region near the Galactic midplane.  
The labels A-F in Figure~\ref{fig:nfilmap} show the 
location of the observations listed in the table, and are sorted by 
increasing distance from S292. The three `X' labels show the location of 
{\sc{OFF}} directions used to remove atmospheric lines and emission from 
the diffuse background. The removal of the background emission was 
particularly important for these observations, because the background 
(\iha $\approx 0.6$ R) is only slightly fainter than the total intensity 
toward the filamentary structures ($\iha \approx 3$ R).  The background 
emission was removed from the pointed observations by choosing the closest 
of the three {\sc{OFF}} directions. However, there is a variation in the strength 
of the \ha\ emission among each of the {\sc{OFF}}s, with \iha\ = $0.4-0.9$ R.  To 
assess the impact of this variation on our results, two different 
background directions were subtracted from the three observations A, B and 
E.  This is reflected in the multiple entries for these observations in Table~\ref{tab:nfil}.
A graphical summary of our results is shown in 
Figure~\ref{fig:nfildiag}, which is similar to 
Figure~\ref{fig:oridiag}.  The open symbols represent the two results 
obtained by subtracting two different {\sc{OFF}} directions from the same pointed 
observation and thus provide a measure of the uncertainty resulting from 
the background subtraction.  None of the data were corrected for 
extinction. However, the upward arrows on the data points for the \hii\ region S292 show 
the shift in these data points if a correction for a visual absorption of 
$\av = 0.81$ is applied to the \hii\ region (see \S\ref{sec:hii}) All 
other data points have shifts smaller than the symbol size.  

Observations at A, B, E, and F along the vertical filament show that the 
\ha\ surface brightness of this filament is nearly constant and 
independent of distance from the midplane, as first noted by 
\citet{HRT98}.  \nii/\ha\ also shows little systematic variation 
($\approx$ 0.4 - 0.5) along the filament's length, except for a somewhat 
elevated value of 0.6 toward direction B. \nii/\ha\ is significantly 
higher than the average (0.27) for \hii\ regions and similar to what is 
observed in the diffuse WIM.  Similarly, we find that \sii/\ha\ ($\approx 
0.2 - 0.4$) is significantly higher than the \hii\ regions ($\approx 
0.1$) and comparable to parts of the WIM.  \sii/\nii\ ($\approx 0.48 - 
0.84$) has significant variations, but is within the scatter of values 
seen in the WIM and has no trend with distance from the plane.  Directions C, 
D, and E, which sample the cross filament were found to have
spectral characteristics similar to those of the vertical filament at the
same latitude (Table~\ref{tab:nfil}).

In contrast to the other emission lines, \oiii/\ha\ shows a very strong 
trend with distance, with \oiii/\ha\ $\approx 0.06$ near the midplane, 
where its value is the same as that toward the \hii\ region S292, falling 
to $\approx$\ 0.025 at B, 24\dg ($\approx$\ 400 pc) above the midplane, 
and dropping to values below or at 0.01 for the two highest latitude 
directions. This implies that O$^{++}$/O $\approx$ 0.1 at $z \approx 
200$\ pc, $\approx$ 0.04 at 400 pc, and $\lesssim$~0.02 above 700 pc. This 
is contrasted with the generally observed behavior of \oiii\ in external 
galaxies, where the \oiii/\ha\ ratios are much higher ($\gtrsim$\ 0.2) and 
in some cases increase with distance above the plane \citep{CR01, MV03}.  
If the filament is photoionized and its temperature is not changing with 
height above the plane, as the constant \nii/\ha\ suggests, then these 
\oiii/\ha\ data suggest that an already low flux of photons having $h\nu 
\gtrsim $\ 35 eV rapidly diminishes with increasing height above the 
plane. The \hei/\ha\ ratio for direction B, while quite uncertain, implies 
that there are a significant number of He-ionizing photons ($h\nu \gtrsim 
24$~eV) at $z \approx 400$\ pc, with He$^+$/He $\approx$\ 0.6 $\pm$ 0.2.  
This is higher than what is seen for many \hii\ regions, and implies that 
the ionizing radiation field at this location has a spectrum that is
similar to an O7 star or earlier.

\subsection{Observations Toward The High Latitude Arc}
\label{subsec:arc}

\begin{figure}[p]
\center{
\includegraphics[scale=0.8]{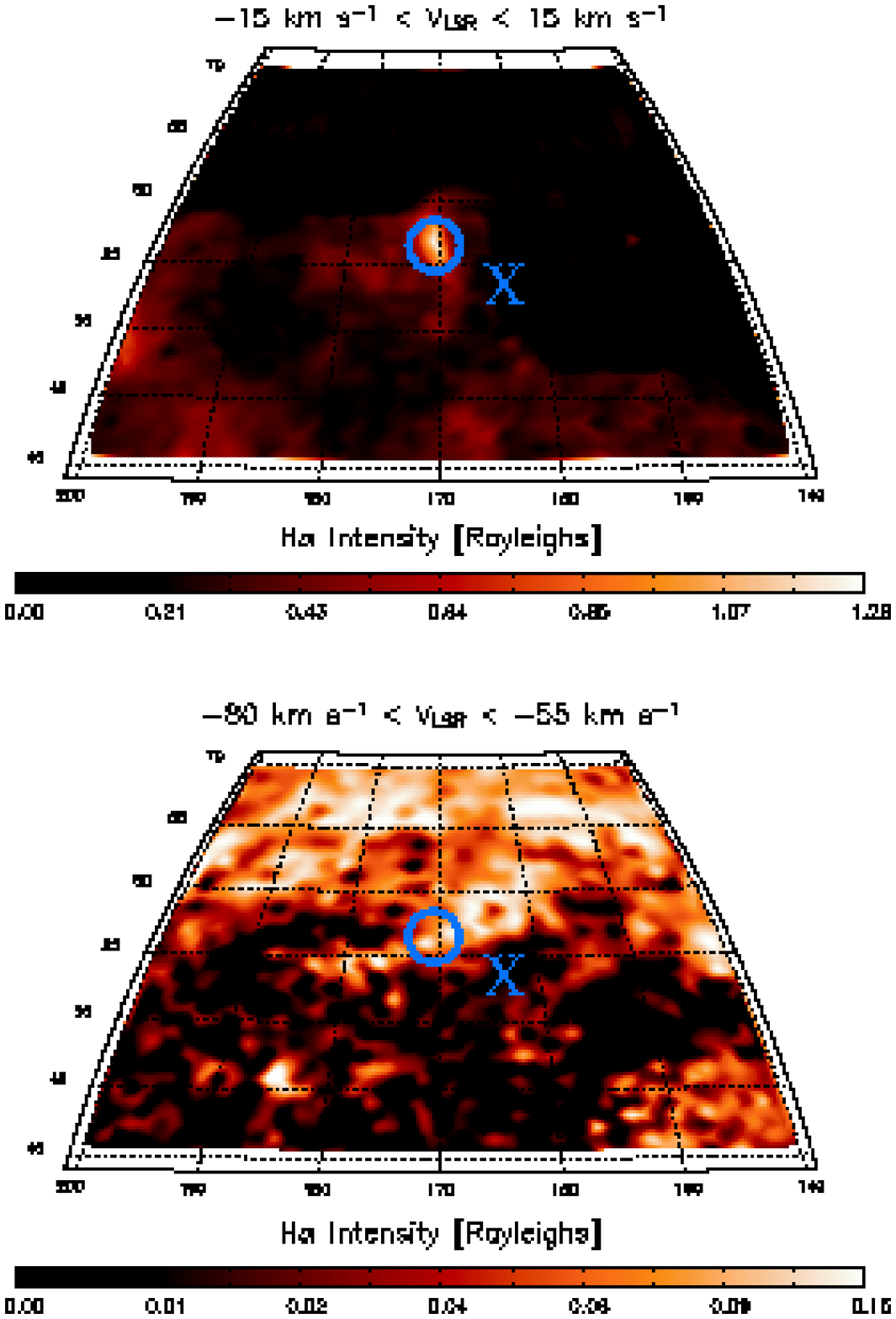}
}
\caption[\ha\ velocity channel maps of High Latitude Arc]{ \ha\ map of
  the Galaxy at high latitude, from the WHAM \ha\ sky survey. The top
  panel shows emission with $|\vlsr| < 15\ \kms$. The lower panel
  shows emission at higher velocity with $-80\ \kms < \vlsr < -55\
  \kms$. The circle is centered on the approximate location of the
  `High Latitude Arc' observations discussed in \S\ref{subsec:arc};
  the `X' shows the location of the {\sc{OFF}} direction. The emission
  in the lower map is spatially coincident with an
  intermediate-velocity \hi\ cloud (IVC) known as the IV Arch
  \citep{Wakker01}. The brightest part of this arch is near $b \approx
  +65\dg$, although there is a fainter spur of \hi\ and \ha\ emission
  that passes through the direction of the High Latitude
  Arc.\label{fig:lockmap}} 
\end{figure}

We also investigated a much shorter filamentary feature, a $\sim$\ 3\dg\ 
long arc of \ha\ emission located near \lb = (171\dg, +57\dg) about 
$10-15$\dg\ away from the faintest directions in the sky in both \ha\ and 
\hi\ \citep{Hausen+02-apj, LJM86}. A WHAM \ha\ sky survey map of the 
region surrounding this arc is shown in Figure~\ref{fig:lockmap}. The 
arc, visible in the top, $|\vlsr| = \pm15\ \kms$ panel, has an \ha\ 
intensity of about 1 R ($\approx$ 3 cm$^{-6}$\ pc), approximately 3-4 
times brighter than the surrounding \ha\ background.  The location of the 
pointed observations is indicated by a circle (the 1\dg\ WHAM beam is much 
smaller than this circle). The location of the {\sc{OFF}} direction is denoted by 
an `X'.

The \ha\ spectrum in the direction of this arc actually consists of two 
emission components, one associated with the arc at \vlsr\ $\approx$\ 0\ 
\kms, and another at \vlsr\ $\approx$\ $-70$ \kms. The lower panel of 
Figure~\ref{fig:lockmap} shows the same piece of sky, but for the 
velocity interval $-80\ \kms < \vlsr < -55\ \kms$.  Fortuitously, the line 
of sight through the arc also passes through a faint portion of an 
intermediate-velocity \hi\ cloud (IVC) known as the IV Arch located 
approximately $500 - 700$ pc above the Galactic midplane \citep{Wakker01, KD96}.  
The brightest \hi\ emission from the IV Arch is across the top of the map 
near $b=+65\dg$, where its associated \ha\ emission is also relatively 
bright. In \ha, the IV Arch is typically $3\dg - 5\dg$ wide and extends 
more than 60\dg\ across the northern Galactic hemisphere. A 
small section of the IV Arch extends down away from the main section and 
passes through the direction of the lower velocity high-latitude \ha\ arc.  
The \ha\ intensity of the IV Arch in this direction is approximately 
0.1~R, making it the faintest \ha\ emission structure in this study, 
with a density of $0.1 - 0.2$ cm$^{-3}$.

We have observed emission lines of \ha, \nii, \sii, \hei, and \oiii\ and 
measured the line strengths for these two velocity components. A summary 
of the observations appears in the bottom two rows of Table~\ref{tab:nfil}.

\subsubsection{The Arc}

The low velocity arc has \nii/\ha~=~0.72 and \sii/\nii~=~0.65, which are 
comparable to some of the highest values observed in the WIM.  The 
detection of \niiblue\ confirms that these high ratios are primarily a 
result of an elevated temperature in the gas (see \S\ref{sec:niiblue}).  
The relatively low \oiii/\ha\ ($\approx$ 0.06) is also consistent with 
the WIM.  However, \hei/\ha\ ($\approx 0.05 \pm 0.01$) is the highest ratio among all the observations in this study including the classical \hii\ regions.  Together, these line ratios suggest that this arc is ionized by a hard radiation source with a low flux that is local to the arc.  The very high \hei/\ha\ ratio is not seen elsewhere in the WIM, and indicate the arc is not a density enhancement ionized only by the diffuse interstellar radiation field. The \hei\ data imply the presence of a hard ionizing spectrum.

A search for a potential ionizing source yielded one candidate, 
the DA white dwarf WD1026+453. This star has a visual magnitude $m_V 
\approx 16.1$ and is near the arc at (170.92\dg, +56.60\dg). It is the 
only spectroscopically confirmed hot star that lies within the arc (in 
projection) and within the field of view of the pointed observations 
discussed above. Ultraviolet spectroscopy suggests that the temperature of 
the star is $T_* \approx$ 35,000 K and that it is at a distance of 
$\approx$ 200 pc \citep{Vennes+97}. It has also been detected in extreme 
ultraviolet by ROSAT and EUVE \citep{Pye+95, Bowyer+96}. \hi\ emission 
line maps in this region show a slight enhancement in 21 cm line emission 
at the velocities of the ionized gas on the lower right side of the arc.
If we adopt the above distance, the emission measure suggests the density of the gas in the arc is approximately 0.9 cm$^{-3}$. 

The identification of WD1026+453 as the potential ionizing source for the 
arc raises the question about the contribution from hot evolved stellar 
cores to the diffuse ionizing radiation field above the Galactic disk.  
The high \hei/\ha\ for this region suggests that a future, more 
comprehensive study of \hei\ emission at high latitudes could provide 
insights about the role of such stars.  On the other hand, the rapid 
decrease in \oiii/\ha\ with latitude along the vertical northern filament 
strongly suggests that the radiation field, at least in that part of the 
Galaxy, actually softens with distance from the midplane.

\subsubsection{The IV Arch}

The detections of \nii\ and \sii\ in the $-67$ \kms\ component are the 
first detections of these diagnostic emission lines from the IV Arch. No 
corresponding \oiii\ and \hei\ emission were detected. The very high value 
of \sii/\nii\ ($\approx 1.8$) is unique. We caution that because the \ha\ 
emission in this component is only $\approx$\ 0.1 R, the uncertainty in 
this result is substantial. However, at the 1$\sigma$ limits, for both 
\nii/\ha\ and \sii/\ha, \sii/\nii\ $\gtrsim$\ 1.0.  These ratios suggest 
that the temperature of the emitting gas is about 7000~K and that almost 
all of the S is singly ionized, with S$^+$/S $\gtrsim 0.7$.  The 
abundances that we use to infer these physical conditions 
(\S\ref{sec:physconds}) are supported by UV absorption lines studies 
\citep{Wakker01}, which show that the IV Arch has a gas phase abundances 
similar to that in the local ISM. The ionization state could be 
unusually low due to the weak flux of ionizing radiation (and thus low 
ionization parameter) inferred from its very low \ha\ surface brightness 
(0.1 R).

In summary, the northern filaments, the high latitude arc, and the IV Arch 
are all characterized by high \nii/\ha\ and \sii/\ha\ ratios that are not 
unlike the ratios observed in the more diffuse WIM.

\section{OBSERVATIONS OF \niiblue}
\label{sec:niiblue}

Many observations show that in faint, diffuse emission regions, 
the ratio \nii$~\lambda6584$/\ha, tends to be significantly 
larger than in classical \hii\ regions and increases with decreasing 
\ha\ 
surface brightness. 
This is a general result that is commonly observed 
in the Galaxy \citep[e.g.,][]{HRT99} as well as in external galaxies 
\citep[e.g.,][]{CR01}. The interpretation is that ionized gas at lower
densities (specifically at n$_e \lesssim 0.1$ cm$^{-3}$) is at 
higher temperatures.  Furthermore, this increase in temperature may be 
beyond what can be attributed to photoionization alone \citep{RHT99, WM04, ED05}, 
which has important implications for the role of other heating processes 
that may be operating in the WIM.  However, the \nii$~\lambda6584$/\ha\ line ratio is a function of several physical parameters, not just temperature.
Therefore it is important to test 
this interpretation of elevated temperatures using other observational 
information.

As discussed in \S\ref{sec:physconds}, the ratio 
\nii$~\lambda6584$/\niiblue\ provides a direct measure of the electron 
temperature of the emitting gas. Here we describe our observations of 
these lines toward a collection of classical O star \hii\ regions as well 
as sightlines that sample the diffuse WIM. Table~\ref{tab:niiblue} 
summarizes the observations.  The first 11 rows show the data for the 
\hii\ regions, including the five \hii\ regions reported earlier by 
\citet{Reynolds+01}.  The last 6 rows show the data obtained for 
sightlines that sample the much fainter diffuse ionized medium. One WIM 
sightline, (130\dg, $-7.5$\dg), was reported previously by 
\citet{Reynolds+01}.  For some directions, \niiblue\ was not detected, and 
only upper limits to the line ratio of \niiblue/\nii$~\lambda6584$ are 
given, as described in \S\ref{sec:errors}. The temperatures inferred by these line ratios appear in the last 
two two columns on the right. $T_{6584}$ is the temperature suggested by 
the \nii$~\lambda6584$/\ha\ ratio; $T_{5755}$ is the temperature inferred 
from the \niiblue/\nii$~\lambda6584$ observations.

\begin{figure}[p]
\center{
\includegraphics[scale=0.8]{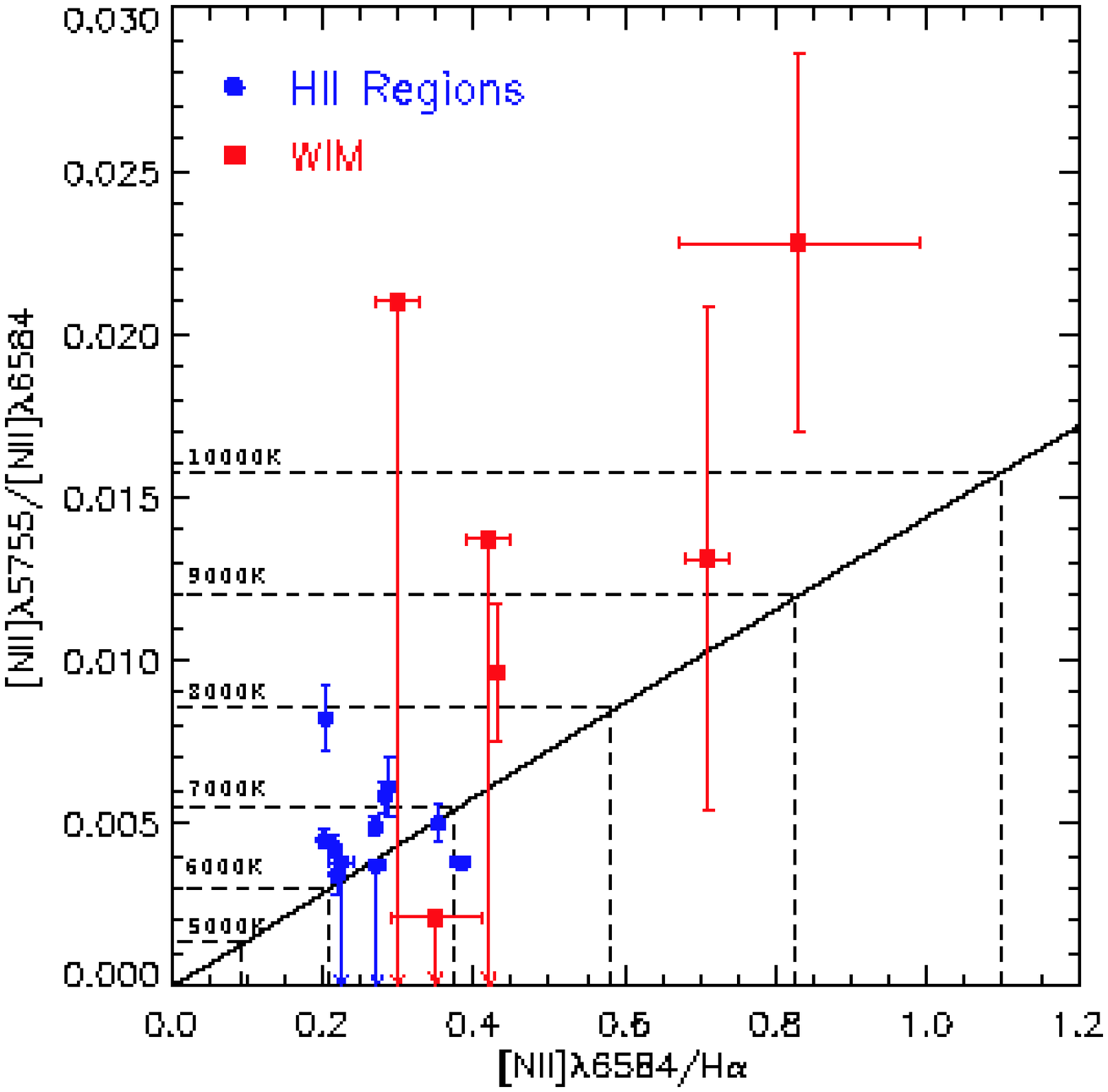}
}
\caption[\niiblue/\nii$~\lambda6584$ versus \nii$~\lambda6584$/\ha\
for WIM and \hii\ regions]{ Ratio of \niiblue/\nii$~\lambda6584$
  versus \nii$~\lambda6584$/\ha\ for several sightlines that sample
  \hii\ regions (blue) and the diffuse WIM (red). Upper limits are
  denoted by arrows. The solid black line shows the expected
  relationship for these ratios, from equations \ref{eq:niieq} and
  \ref{eq:niiblueeq}. The dashed lines indicate the expected ratios
  temperatures between 5000 K and 10,000 K. This diagram shows that a
  significant fraction of gas in the WIM is 2000-3000 K warmer than
  \hii\ regions. \label{fig:bluenii}} 
\end{figure}

These results are shown in Figure~\ref{fig:bluenii}, where 
\niiblue/\nii$~\lambda6584$ is plotted against \nii$~\lambda6584$/\ha.  
Data for the \hii\ regions have blue symbols and the WIM red symbols. 
Upper limits to the line ratios are indicated by arrows. The solid black 
line is the locus of expected ratios of these lines from equations 
\ref{eq:niieq} and \ref{eq:niiblueeq}. The dashed lines show the 
predicted ratios for temperatures between 5000 and 10000 K. For the three 
WIM sightlines in which \niiblue\ is detected, both of the line ratios are 
significantly higher than those for classical O star \hii\ regions. This 
provides convincing confirmation that 1) the WIM is approximately $2000 - 
3000$~K warmer than classical O star \hii\ regions, and 2) higher 
\nii/\ha\ intensity ratios are due at least in large part to higher 
temperatures.

The data points tend to lie preferentially above the expected relationship 
(solid line).  This could be explained if N$^+$/N in the WIM were 
significantly lower than the assumed value of 0.8. A ratio of N$^+$/N = 0.6 would bring the two temperature diagnostics into average agreement with one another. However, both 
photoionization modeling \citep{Sembach+00} and observations of elevated 
\sii/\ha\ ratios suggest that the WIM, and in particular the WIM nitrogen, 
is not highly ionized.  Another reason could be that the gas phase 
abundance of nitrogen in the WIM is lower than in the \hii\ regions by a 
factor of $\gtrsim$~2, which also seems improbable. The most likely 
explanation is that there is a range of temperatures in the WIM 
\citep{Reynolds+01}, as is indicated by the large scatter in \nii/\ha\ (e.g., Figure~\ref{fig:alldiag}b). 
The metastable level of the \niiblue\ line lies higher above ground (about 4 
eV) than that of the \nii$~\lambda6584$\ line (about 2 eV), which means 
that \niiblue\ is preferentially produced in regions of higher temperature 
along the line of sight.  The deviation of the points from the solid line 
can be explained with an appropriate (non-unique) range of temperatures 
\citep{Reynolds+01}.  The conclusion that variations in \nii/\ha\ trace 
variations in temperature has also been confirmed by the recent detection 
and study of [\ion{O}{2}] $\lambda$3727 emission from the WIM \citep{Mierkiewicz+04}.

\section{SUMMARY AND CONCLUSIONS}
\label{sec:summary}

We have presented a large number of new observations of several optical 
emission lines toward \ha-emitting features in the Galaxy that span a wide 
range in surface brightness, angular scale, environment, and morphology.  
We have explored the relative intensities of these emission lines to infer 
the physical conditions of the emitting gas, and we have compared these 
conditions with those of traditional, O star \hii\ regions in an attempt 
to gain insight into the nature of the WIM and its relationship to hot 
stars and large-scale bubbles and filaments within the interstellar 
medium. We found significant variations in the temperature and ionization 
state among these emission features, revealing that warm ionized gas is 
heterogeneous in nature. We have strengthened the general assertion that 
the WIM is warmer and less ionized compared to classical \hii\ regions and 
found significant variations in temperature and ionization state 
\emph{within} the diffuse WIM.

\begin{figure}[p]
\center{
\includegraphics[scale=0.7]{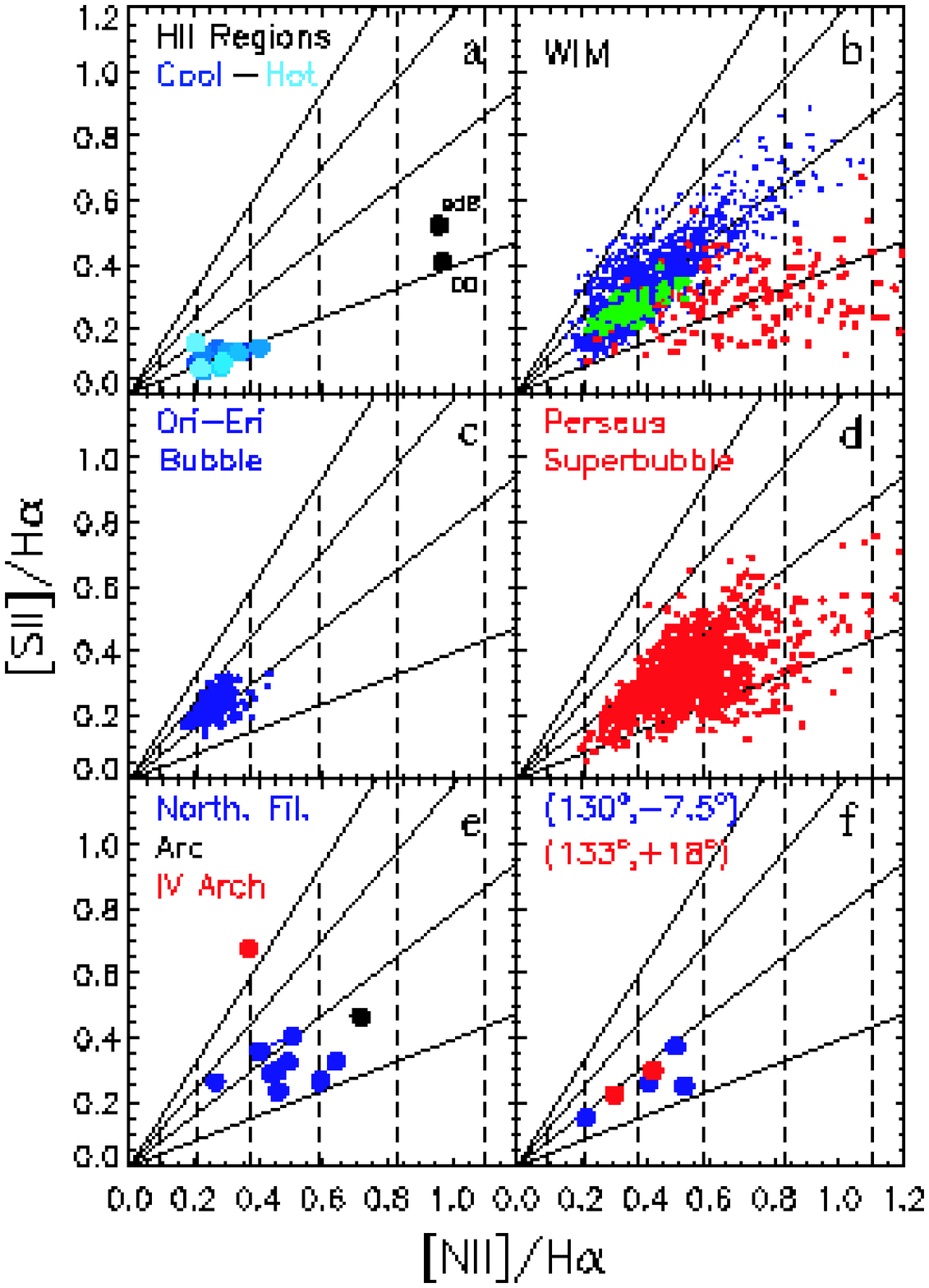}
}
\caption[\nii/\ha\ versus \sii/\ha\ for every observation]{ \nii/\ha\
  versus \sii/\ha\ for every observation in this study, separated into
  several categories.  The dashed vertical lines represent lines of
  constant temperature from equation \ref{eq:niieq}, with $5000 \rm{K}
  \le T \le 10000  
\rm{K}$. The solid sloped lines represent values of constant
ionization fraction of S from equation \ref{eq:siieq}, with $0.25 \le
\rm{S}^+/\rm{S} \le 1.0$.  
Panel (a) shows the data for all \hii\ regions, with the O-star \hii\
regions in increasingly lighter shades of blue for an increasing
photospheric temperature of the ionizing star(s). Panel (b) shows
emission from the WIM in the local gas toward Perseus ({\it{blue}}),
Orion-Eridanus ({\it{green}}), and in the more distant Perseus spiral
arm ({\it{red}}).  
Panels (c) and (d) show emission associated with the Orion-Eridanus
bubble and Perseus superbubble, respectively. Panel (e) includes data
toward the northern filaments ({\it{blue}}), the high latitude arc
({\it{black}}) and the IV Arch ({\it{red}}). Panel (f) shows the line
ratios for different velocity components toward (130\dg, -7.5\dg)
({\it{blue}}) and (133\dg, +18\dg) ({\it{red}}). 
\label{fig:alldiag}}
\end{figure}

An overview of some of these observations is presented in 
Figure~\ref{fig:alldiag}, which shows the diagnostic plots of \nii/\ha\ 
vs. \sii/\ha\ for various emission regions observed in pointed and 
survey mode.  The dashed vertical and solid sloped lines 
represent lines of constant temperature and constant S$^+$/S, respectively, and 
have the same values as in Figures~\ref{fig:orirationvs}, 
\ref{fig:bowtieratiolocalnvs}, and \ref{fig:bowtieratiopernvs}. The 
range of physical conditions is indicated by the different
distributions and by the scatter in the data points. Different types
of emission regions occupy different areas of the  
diagram, and even within a given catagory of emission region, variations 
can be significant. The classical O star \hii\ regions (except for the two 
regions ionized by hot stellar cores) show the least variation 
(Figure~\ref{fig:alldiag}a), with temperatures between 6000~K and 
7000~K and S$^+$/S $\approx 0.25$.  In contrast, the faint diffuse WIM 
(Fig.~\ref{fig:alldiag}b) has the most variation; that is, it occupies 
the largest area, with temperatures ranging between 7000~K and 10,000~K 
and S$^+$/S ranging from about 0.1 to 1. Figure~\ref{fig:alldiag}b also 
shows that the mean properties of the WIM can be slightly different in 
different regions of the Galaxy.  For example, in the Perseus arm S$^+$/S 
in the WIM has an average near 0.25, while in the more nearby gas
toward that direction, its average is near 0.6. 

From this study, we make the following closing statements:

1. The temperature of diffuse ionized gas is higher in regions of 
   lower 
   emission measure. This is implied by the elevated \nii/\ha\ and 
   \sii/\ha\ ratios toward the relatively faint Perseus 
   superbubble, high latitude filaments, and diffuse background WIM, 
   compared to the relatively bright Orion-Eridanus bubble and the even 
   brighter O star \hii\ regions. The relation between 
   temperature (traced by \nii/\ha) and \ha\ intensity also holds within 
   the diffuse WIM itself (e.g., Figures~\ref{fig:oriratioplot}, 
   \ref{fig:bowtieratiolocal}, and \ref{fig:bowtieratioper}).
   The elevated temperatures in the WIM are confirmed by the 
   \niiblue/\nii$~\lambda6583$ intensity ratios
   (Figure~\ref{fig:bluenii}) and recent observations of [O~II]
   \citep{Mierkiewicz+04}. \\ 

2. The ionization state in diffuse ionized gas is generally lower than that in
   classical \hii\ regions.
   Several new observations of \oiii\ and \hei\ indicate that, in general,
   the fraction of O$^{++}$/O and He$^+$/He in the WIM and in the
   large bubble structures is low compared to 
   \hii\ regions, implying a lower ionization state due to a softer
   ionizing
   radiation field and probably a lower ionization parameter (e.g.,
   Figures~\ref{fig:bowtiediag} and \ref{fig:nfildiag}). 
   The high ionization of sulfur (i.e., the low values of S$^{+}$/S) in the WIM of the Perseus arm (Figs. 15, 16, 21b) appears to be an exception to this trend.
   The data also suggest that the diffuse low density gas close
   to the Galactic plane may be more highly ionized than gas at larger
   distances from the plane. In the inner Galaxy, this trend appears to
   reverse \citep{MR05}. \\

3. Conditions within the WIM vary significantly. The mean 
   temperature and ionization state can change considerably from one 
   sight line to the next and even along a single sight line.  Moreover, 
   the mean properties of the WIM change from one 
   region of the Galaxy to another (e.g., see Fig.~\ref{fig:alldiag}b).
   Within the WIM, values of T and S$^+$/S extend to significantly
   higher values than are found within classical \hii\ regions or even
   the extended bubbles. 
   \\

4. High latitude filaments superposed on the faint WIM
   have spectral characteristics similar to the WIM. This suggest a
   close relationship between these high latitude structures and the more
   diffuse background. \\

5. The Perseus superbubble provides strong evidence that a luminous
   O star cluster near the midplane can produce wide-spread, nearly 
   WIM-like ionization conditions to distances of 1000 pc or more from the ionizing stars.
   This superbubble has spectral
   characteristics similar to those
   observed in portions of the diffuse WIM (Figs.~\ref{fig:alldiag}b 
   and d) and quite
   different from the spectral characteristics of the bright \hii\ region 
   that
   immediately surrounds its source of ionization, the Cas OB6 star 
   cluster and other classical \hii\ regions (Fig.~\ref{fig:alldiag}a 
   and d;
   Table~\ref{tab:hiimulti}). The fact that the smaller, brighter, 
   denser
   Orion-Eridanus bubble has low \nii/\ha\ ratios, similar to those in
   classical \hii\ regions, implies that bubble size, gas density within
   the ionized shell, and/or the flux and spectrum of the radiation 
   escaping
   O star clusters may be important in setting the conditions
   within the ionized gas. For example, the presence of an extra source of
   non-ionizing heat can raise the gas temperature in low density 
   ($\lesssim 0.1$ cm$^{-3}$)
   ionized gas \citep{RHT99}, which could account for the elevated, more
   WIM-like temperatures in the low density Perseus superbubble, as
   well as the more general relationship between temperature and
   emission measure (point 1 above). \\

6. Filamentary structures within the Perseus superbubble have 
   physical conditions that 
   differ from conditions
   in the fainter, more diffuse parts of the bubble. This implies 
   that these 
   filaments are discrete entities, likely regions of higher density, 
   and not just directions of increased 
   pathlength through folds or edge 
   projections of a shell or sheet. The slightly depressed \nii/\ha\ and 
   \sii/\ha\ ratios in these filaments, compared to the fainter 
   superbubble emission adjacent to them, 
   suggest that they are in fact cooler than the gas along adjacent 
   sightlines through the superbubble. \\

In conclusion, we believe that high spectral resolution emission line 
observations at visible wavelengths open a new window on the study of 
interstellar matter and processes not available through other techniques 
at other wavelengths. In particular, the application of nebular line 
diagnostics to the study of the warm ionized component of the interstellar 
medium provides an opportunity to understand better, through both 
observations and modelling, the large-scale effects of hot stars on the 
ionization and morphology of the interstellar medium within the Galaxy's 
disk and halo.

\section{ACKNOWLEDGEMENTS}

We gratefully acknowledge the anonymous referee for a thorough review which improved the paper. 
We thank Kurt Jaehnig for his outstanding technical support in the
continuing operation of the WHAM instrument. This work was funded by
the National Science Foundation through grants AST-0204973 and
AST-0401416. GJM acknowledges additional support from the Wisconsin
Space Grant Consortium. This research has made use of the SIMBAD
database, operated at CDS, Strasbourg, France. 

{\it Facilities:} \facility{WHAM ()}

\newpage


\clearpage

\clearpage

\clearpage


\begin{deluxetable}{lllrrrrrcl}
\rotate
\tablecaption{Observations of HII Regions\label{tab:hiibasic}}
\tablewidth{0pt}
\tablehead{ \colhead{Name} & \multicolumn{2}{c}{Ionizing Star(s)} & \colhead{$l$} &
  \colhead{$b$} & \colhead{$V_{LSR}$} & \colhead{$A_V$} & \multicolumn{3}{c}{$I_{H\alpha}$} \\
  \colhead{ } & \colhead{Name} & \colhead{Sp. Type} &
  \colhead{($^\circ$)} & \colhead{($^\circ$)} & \colhead{(km s$^{-1}$)} & 
  \colhead{(mag)} & \multicolumn{3}{c}{(R)}  }
\startdata
S132\tablenotemark{a}    & GP Cep     & WN6+O6I    & 102.8 &  -0.9 & -51 & 2.32 $\pm$ 0.03 &    820 &$\pm$& 20  \\
W4        & Cas OB6      & O4I+...    & 134.7 &  +0.9 & -46 & 3.13 $\pm$ 0.02              &   2800 &$\pm$& 50 \\
S142-3\tablenotemark{a}  & HD 215835  & O6V        & 107.0 &  -1.0 & -43 & 1.84 $\pm$ 0.03 &    425 &$\pm$& 10 \\
S184\tablenotemark{a}    & HD 5005    & O6.5V+O8V  & 123.2 &  -6.3 & -34 & 1.00 $\pm$ 0.01 &    329 &$\pm$& 4 \\
S292      & CMa OB1      & O6.5V+...  & 224.0 &  -1.7 & +14 &                              &    279 &$\pm$& 1 \\
S220      & $\xi$ Per    & O7.5III    & 159.7 & -12.4 &  -4 & 0.52 $\pm$ 0.01              &    496 &$\pm$& 3 \\
Sivan 4A  & $\xi$ Per    & O7.5III    & 160.4 & -13.1 &  -1 & 0.87 $\pm$ 0.01              &    135 &$\pm$& 2 \\
Sivan 4B  & $\xi$ Per    & O7.5III    & 161.1 & -13.9 &  +5 & 0.60 $\pm$ 0.02              &     25 &$\pm$& 1 \\
S264      & $\lambda$ Ori& O8III      & 195.1 & -12.0 &  +6 &                              &    176 &$\pm$& 1 \\
Sivan 2   & AO Cas       & O9III      & 117.6 & -11.1 & -31 & 0.37 $\pm$ 0.06              &     13 &$\pm$& 1 \\
S126      & 10 Lac       & O9V        &  96.7 & -17.0 & -10 & 0.27 $\pm$ 0.03              &     26 &$\pm$& 1 \\
Sivan 3   & $\alpha$ Cam & O9.5Ia     & 144.1 & +14.1 &  -9 & 0.97 $\pm$ 0.08              &     41 &$\pm$& 2 \\
S276      & $\sigma$ Ori & O9.5V      & 207.0 & -17.0 & +12 &                              &    364 &$\pm$& 1 \\
& & & & & & & & & \\                                                                                    
sdB       & PG 1047+003  & sdB        & 250.9 & +50.2 & -13 &                              &    2.4 &$\pm$&  0.1   \\
DO        & PG 1034+001  & D0         & 247.6 & +47.7 & -11 &                              &    2.0 &$\pm$&  0.1   \\

\enddata
\tablenotetext{a}{These \hii\ regions do not fill the WHAM 1\dg\ beam, and
  therefore actual \ha\ intensity for these objects may be
  significantly higher.}
\end{deluxetable}

\begin{deluxetable}{lrrrr}
\tablecaption{HII Region Line Ratios \label{tab:hiimulti}}
\tablewidth{0pt}
\tablehead{ \colhead{Name} &
  \colhead{[N II]/H$\alpha$} &   \colhead{[S II]/H$\alpha$} & 
  \colhead{[O III]/H$\alpha$} &   \colhead{He I/H$\alpha$} \\
  \colhead{ } &  \colhead{(energy)} & \colhead{(energy)}& \colhead{(energy)}& \colhead{(energy)} }
\startdata
S132\tablenotemark{a}  &  0.22 $\pm$  0.01 &  0.08 $\pm$  0.01 &  0.49 $\pm$  0.02\phn &  0.037 $\pm$  0.001  \\ 
W4                     &  0.20 $\pm$  0.01 &  0.15 $\pm$  0.01 &  0.30 $\pm$  0.01\phn &  0.033 $\pm$  0.001  \\ 
S142-3\tablenotemark{a}&  0.29 $\pm$  0.01 &  0.09 $\pm$  0.01 &  0.24 $\pm$  0.01\phn &  0.031 $\pm$  0.001  \\ 
S184\tablenotemark{a}  &  0.28 $\pm$  0.01 &  0.08 $\pm$  0.01 &  0.349 $\pm$  0.006 &  0.036 $\pm$  0.001  \\ 
S292       &  0.34 $\pm$  0.01 &  0.12 $\pm$  0.01 &  0.099 $\pm$  0.002 &  0.020 $\pm$  0.001  \\ 
S220       &  0.40 $\pm$  0.01 &  0.14 $\pm$  0.01 &  0.035 $\pm$  0.002 &  0.019 $\pm$  0.001  \\ 
Sivan 4A   &  0.23 $\pm$  0.01 &  0.07 $\pm$  0.01 &  0.271 $\pm$  0.006 &  0.030 $\pm$  0.001  \\ 
Sivan 4B   &  0.31 $\pm$  0.01 &  0.13 $\pm$  0.01 &  0.105 $\pm$  0.004 &  0.018 $\pm$  0.002  \\ 
S264       &  0.23 $\pm$  0.01 &  0.08 $\pm$  0.01 &  0.084 $\pm$  0.002 &  0.013 $\pm$  0.001  \\ 
Sivan 2    &  0.23 $\pm$  0.02 &  0.12 $\pm$  0.01 &                     &                      \\ 
S126       &  0.27 $\pm$  0.01 &  0.14 $\pm$  0.01 &  0.125 $\pm$  0.006 &  0.018 $\pm$  0.002  \\ 
Sivan 3    &  0.21 $\pm$  0.02 &  0.09 $\pm$  0.01 &  0.076 $\pm$  0.008 &                      \\ 
S276       &  0.35 $\pm$  0.01 &  0.13 $\pm$  0.01 &  0.025 $\pm$  0.002 &  0.011 $\pm$  0.001  \\ 
& & & &  \\ 
sdB        &  0.96 $\pm$  0.05 &  0.52 $\pm$  0.03 &  0.12 $\pm$  0.02 &                      \\ 
DO         &  0.97 $\pm$  0.05 &  0.41 $\pm$  0.02 &  2.57 $\pm$  0.14 &                      \\ 

\enddata
\tablenotetext{a}{These \hii\ regions do not fill the WHAM 1\dg\ beam.}
\end{deluxetable}

\begin{deluxetable}{lrrrrrrrrclrcl}
\rotate
\tablecaption{Observations Toward the Orion - Eridanus Bubble \label{tab:ori}}
\tablewidth{0pt}
\tablehead{ \colhead{Name}  & \colhead{$l$} &
  \colhead{$b$} & \colhead{$D_\theta$} & \colhead{$V_{LSR}$} & \colhead{$I_{H\alpha}$} & 
  \colhead{[N II]/H$\alpha$} &   \colhead{[S II]/H$\alpha$} & 
  \multicolumn{3}{c}{[O III]/H$\alpha$} &  \multicolumn{3}{c}{He I/H$\alpha$} \\
  \colhead{ } & 
  \colhead{($^\circ$)} & \colhead{($^\circ$)} & \colhead{($^\circ$)} & \colhead{(km s$^{-1}$)} &
  \colhead{(R)} & \colhead{(energy)} & \colhead{(energy)}& \multicolumn{3}{c}{(energy)}& \multicolumn{3}{c}{(energy)} }
\startdata
S276 & 207.0 & -17.0 &  1.9 & +12   &   364 $\pm$ 1\phs &  0.35 $\pm$ 0.01 &  0.13 $\pm$  0.01 &  0.023 &$\pm$&  0.002  &  0.012 &$\pm$&  0.001  \\ 
   1 & 211.5 & -22.1 &  5.7 & -16,+8&  86.3 $\pm$  0.1 &  0.23 $\pm$  0.01 &  0.13 $\pm$  0.01 &  0.020 &$\pm$&  0.002  &  0.015 &$\pm$&  0.001  \\ 
   2 & 205.5 & -12.7 &  6.2 & +0    & 228.1 $\pm$  0.2 &  0.23 $\pm$  0.01 &  0.16 $\pm$  0.01 &  0.004 &$\pm$&  0.01   &  0.004 &$\pm$&  0.001  \\ 
   3 & 204.6 &  -5.1 & 13.8 & +1    &   5.7 $\pm$  0.1 &  0.29 $\pm$  0.01 &  0.25 $\pm$  0.01 &  0.11  &$\pm$&  0.01   &  0.036 &$\pm$&  0.008  \\ 
   4 & 194.7 & -28.0 & 14.2 & +10   &   2.6 $\pm$  0.1 &  0.26 $\pm$  0.02 &  0.21 $\pm$  0.01 &  0.06  &$\pm$&  0.02    &        &  $<$&  0.01  \\
   5 & 186.7 & -25.5 & 19.5 & +1    &  13.9 $\pm$  0.1 &  0.24 $\pm$  0.01 &  0.18 $\pm$  0.01 &  0.014 &$\pm$&  0.002  &  0.008 &$\pm$&  0.002  \\
   6 & 195.9 & -39.9 & 23.0 & +7    &  17.8 $\pm$  0.1 &  0.21 $\pm$  0.01 &  0.20 $\pm$  0.01 &  0.004 &$\pm$&  0.002  &  0.009 &$\pm$&  0.001  \\
   7 & 185.5 & -33.9 & 24.0 & +5    &  26.1 $\pm$  0.1 &  0.20 $\pm$  0.01 &  0.17 $\pm$  0.01 &  0.010 &$\pm$&  0.002  &  0.010 &$\pm$&  0.001  \\ \\
   A & 193.0 & -47.5 & 30.8 & +1    &   1.0 $\pm$  0.1 &  0.43 $\pm$  0.03 &  0.29 $\pm$  0.05 &        &$<$  &  0.02  &         &$<$&    0.02  \\ 
   B & 193.8 & -48.4 & 31.3 & -11   &   6.6 $\pm$  0.1 &  0.19 $\pm$  0.01 &  0.18 $\pm$  0.01 &        &$<$  &  0.004  &   0.02  &$\pm$&  0.01  \\ 
   C & 193.6 & -49.2 & 32.2 & -7    &  14.9 $\pm$  0.1 &  0.18 $\pm$  0.01 &  0.20 $\pm$  0.01 &  0.006 &$\pm$&  0.002  &   0.011 &$\pm$&  0.002  \\ 
   D & 192.9 & -50.1 & 33.1 & -2    &  13.7 $\pm$  0.1 &  0.26 $\pm$  0.01 &  0.27 $\pm$  0.01 &  0.010 &$\pm$&  0.002  &   0.015 &$\pm$&  0.002  \\ 
   E & 192.7 & -50.9 & 33.9 & +8    &   7.3 $\pm$  0.1 &  0.26 $\pm$  0.01 &  0.27 $\pm$  0.01 &        &$<$  &  0.002  &   0.010 &$\pm$&  0.005  \\ 
   F & 192.1 & -51.8 & 34.9 & +5    &   3.5 $\pm$  0.1 &  0.20 $\pm$  0.01 &  0.20 $\pm$  0.01 &        &$<$  &  0.01   &         &$<$&    0.01   \\ 
   G & 192.1 & -52.6 & 35.6 & +2    &   1.3 $\pm$  0.1 &  0.16 $\pm$  0.02 &  0.17 $\pm$  0.04 &   0.02 &$\pm$&  0.02   &         &$<$&    0.03  \\ 
 \enddata
\end{deluxetable}

\begin{deluxetable}{lrrrrrrrrr}
\rotate
\tablecaption{Observations Toward the Perseus Superbubble \label{tab:per}}
\tablewidth{0pt}
\tablehead{ \colhead{Name}  & \colhead{$l$} &
  \colhead{$b$} & \colhead{$D_\theta$} & \colhead{$V_{LSR}$} & \colhead{$I_{H\alpha}$} & 
  \colhead{[N II]/H$\alpha$} &   \colhead{[S II]/H$\alpha$} & 
  \colhead{[O III]/H$\alpha$} &   \colhead{He I/H$\alpha$} \\
  \colhead{ } & 
  \colhead{($^\circ$)} & \colhead{($^\circ$)} & \colhead{($^\circ$)} & \colhead{(km s$^{-1}$)} &
  \colhead{(R)} & \colhead{(energy)} & \colhead{(energy)}& \colhead{(energy)}& \colhead{(energy)} }
\startdata
1  & 130.0 &  -7.5 &      &  0  & 4.9 $\pm$ 0.1 & 0.52 $\pm$ 0.02 & 0.25 $\pm$ 0.01 & 0.08 $\pm$  0.01 & 0.025 $\pm$  0.003  \\ 
2  & 130.0 &  -7.5 &      & -33 & 2.7 $\pm$ 0.1 & 0.49 $\pm$ 0.03 & 0.37 $\pm$ 0.02 & 0.17 $\pm$  0.02 &  0.036 $\pm$  0.006  \\ 
3  & 130.0 &  -7.5 &  9.6 & -56 & 2.2 $\pm$ 0.1 & 0.21 $\pm$ 0.03 & 0.15 $\pm$ 0.02 & 0.35 $\pm$  0.02 &  0.021 $\pm$  0.007  \\ 
4  & 130.0 &  -7.5 &      & -75 & 1.5 $\pm$ 0.1 & 0.41 $\pm$ 0.05 & 0.26 $\pm$ 0.03 & 0.16 $\pm$  0.02 &  0.04 $\pm$  0.01\phn  \\ \\
1  & 133.0 &  18.0 &      & -20 & 0.7 $\pm$ 0.1 & 0.30 $\pm$ 0.03 & 0.22 $\pm$ 0.03 & 0.11 $\pm$  0.01 &  0.05 $\pm$  0.06\phn  \\ 
2  & 133.0 &  18.0 & 17.2 & -62 & 0.8 $\pm$ 0.1 & 0.42 $\pm$ 0.03 & 0.30 $\pm$ 0.02 & 0.06 $\pm$  0.01 &  0.04 $\pm$  0.06\phn  \\ 
 \enddata
\end{deluxetable}

\begin{deluxetable}{lrrrrrrrrr}
\rotate
\tablecaption{Observations of High Latitude Filaments \label{tab:nfil}}
\tablewidth{0pt}
\tablehead{ \colhead{Name}  & \colhead{$l$} &
  \colhead{$b$} & \colhead{$D_\theta$} & \colhead{$V_{LSR}$} & \colhead{$I_{H\alpha}$} & 
  \colhead{[N II]/H$\alpha$} &   \colhead{[S II]/H$\alpha$} & 
  \colhead{[O III]/H$\alpha$} &   \colhead{He I/H$\alpha$} \\
  \colhead{ } & 
  \colhead{($^\circ$)} & \colhead{($^\circ$)} & \colhead{($^\circ$)} & \colhead{(km s$^{-1}$)} &
  \colhead{(R)} & \colhead{(energy)} & \colhead{(energy)}& \colhead{(energy)}& \colhead{(energy)} }
\startdata
S292    & 224.0 &  -1.7 &  0.0 & +14 &  279 $\pm$ 1\phs& 0.34 $\pm$ 0.01 & 0.12 $\pm$ 0.01 & 0.099 $\pm$ 0.002 &  0.020 $\pm$  0.001  \\
A \#1   & 226.0 &  10.2 & 12.1 & +17 &  2.5 $\pm$ 0.1 & 0.49 $\pm$ 0.03 & 0.32 $\pm$ 0.02 &  0.10  $\pm$ 0.02\phn  &    \\ 
A \#2   & 226.0 &  10.2 & 12.1 & +17 &  2.3 $\pm$ 0.1 & 0.44 $\pm$ 0.02 & 0.29 $\pm$ 0.02 &  0.14  $\pm$ 0.02\phn  &    \\ 
B \#1   & 225.2 &  22.1 & 23.8 & +10 &  2.1 $\pm$ 0.1 & 0.63 $\pm$ 0.02 & 0.33 $\pm$ 0.01 &  0.06  $\pm$ 0.02\phn  &  0.04 $\pm$  0.01\phn  \\ 
B \#2   & 225.2 &  22.1 & 23.8 & +10 &  2.1 $\pm$ 0.1 & 0.58 $\pm$ 0.02 & 0.26 $\pm$ 0.02 &  0.04  $\pm$ 0.02\phn  &  0.02 $\pm$  0.01\phn  \\ 
C       & 217.1 &  28.9 & 31.3 & -4  &  4.0 $\pm$ 0.1 & 0.27 $\pm$ 0.03 & 0.26 $\pm$ 0.01 &  0.01 $\pm$ 0.01\phn &  0.016 $\pm$  0.004  \\ 
D       & 235.2 &  29.7 & 33.2 & +5  &  2.3 $\pm$ 0.1 & 0.46 $\pm$ 0.05 & 0.29 $\pm$ 0.01 &        $<$   0.004 &  0.01 $\pm$  0.02\phn  \\ 
E \#1   & 226.2 &  38.2 & 40.0 & -1  &  2.5 $\pm$ 0.1 & 0.50 $\pm$ 0.03 & 0.40 $\pm$ 0.02 &        $<$   0.008 &    \\ 
E \#2   & 226.2 &  38.2 & 40.0 & -1  &  2.2 $\pm$ 0.1 & 0.40 $\pm$ 0.04 & 0.36 $\pm$ 0.02 &        $<$   0.01\phn &    \\ 
F             & 236.3 &  50.9 & 53.7 & -18 &  2.3 $\pm$ 0.1   &  0.46 $\pm$ 0.05  &  0.23 $\pm$ 0.01  & 0.02 $\pm$   0.02\phn &    \\ 
\\
Arc  & 170.8 &  56.9 & \phn &  +3 & 1.31 $\pm$  0.02 &  0.72 $\pm$  0.03 &  0.46 $\pm$  0.02 &  0.06  $\pm$   0.02\phn &  0.05  $\pm$  0.01\phn   \\ 
IV Arch        & 170.8 &  56.9 & \phn & -67& 0.11 $\pm$  0.02 &  0.4  $\pm$  0.2\phn  &  0.7  $\pm$  0.2\phn  &                     &                      \\  
 \enddata
\end{deluxetable}

\begin{deluxetable}{lclrl}
\tablecaption{Observations of [N II]$~\lambda5755$ \label{tab:niiblue}}
\tablewidth{0pt}
\tablehead{ \colhead{Name}  & 
  \colhead{[N II]$~\lambda6584$/H$\alpha$} &   \colhead{[N II]$~\lambda5755$/[N II]$~\lambda6584$ } & 
  \colhead{$T_{6583}$} & \colhead{$T_{5755}$} \\
  \colhead{ } & \colhead{(energy)} & \colhead{(energy)} &
  \colhead{(K)} & \colhead{(K)} }
\startdata

S132   &  0.219 $\pm$ 0.006 & \phm{$<$ }0.0034 $\pm$ 0.0006 & 6050 $\pm$  \phm{1}50 & \phm{$<$ }6200 $\pm$ 300  \\ 
    W4 &  0.204 $\pm$ 0.006 & \phm{$<$ }0.0082 $\pm$ 0.0010 & 5950 $\pm$  \phm{1}50 & \phm{$<$ }7900 $\pm$ 300  \\ 
S142-3 &  0.288 $\pm$ 0.010 & \phm{$<$ }0.0061 $\pm$ 0.0009 & 6500 $\pm$  \phm{1}50 & \phm{$<$ }7300 $\pm$ 300  \\ 
S184   &  0.282 $\pm$ 0.004 & \phm{$<$ }0.0058 $\pm$ 0.0005 & 6500 $\pm$  \phm{1}50 & \phm{$<$ }7200 $\pm$ 200  \\ 
S126   &  0.272 $\pm$ 0.010 &           $<0.0037$ & 6400 $\pm$  \phm{1}50 & $<6300$                \\ 
Sivan 2 & 0.226 $\pm$ 0.016 &           $<0.0038$ & 6100 $\pm$ 100 & $<6400$                \\ 
S117\tablenotemark{a}   &  0.202 $\pm$ 0.005 & \phm{$<$ }0.0045 $\pm$ 0.0003 & 5950 $\pm$ \phm{1}50 & \phm{$<$ }6700 $\pm$ 100 \\
S220\tablenotemark{a}   &  0.383 $\pm$ 0.012 & \phm{$<$ }0.0038 $\pm$ 0.0002 & 7050 $\pm$ \phm{1}50 & \phm{$<$ }6400 $\pm$ 100  \\
S252\tablenotemark{a}   &  0.270 $\pm$ 0.009 & \phm{$<$ }0.0049 $\pm$ 0.0003 & 6400 $\pm$ \phm{1}50 & \phm{$<$ }6800 $\pm$ 100 \\
S261\tablenotemark{a}   &  0.353 $\pm$ 0.009 & \phm{$<$ }0.0050 $\pm$ 0.0006 & 6900 $\pm$ \phm{1}50 & \phm{$<$ }6900 $\pm$ 200 \\
S264\tablenotemark{a}   &  0.216 $\pm$ 0.006 & \phm{$<$ }0.0042 $\pm$ 0.0004 & 6050 $\pm$ \phm{1}50 & \phm{$<$ }6500 $\pm$ 200 \\
\\
(130,-7.5)\tablenotemark{a} &  0.43 $\pm$  0.01 &  \phm{$<$ }0.010 $\pm$ 0.002   &  7300 $\pm$ 100  & \phm{$<$ }8300 $\pm$ 600 \\
High Lat. Arc &  0.71 $\pm$  0.03 &  \phm{$<$ }0.013 $\pm$  0.008  &  8500 $\pm$  100 &   \phm{$<$ }9300 $\pm$  2000  \\ 
  WIM(Sivan 2) &  0.8 $\pm$  0.2 &  \phm{$<$ }0.023 $\pm$  0.006  &  9000 $\pm$  600 &  \phs11700 $\pm$  1400  \\
  WIM(S132) &  0.35 $\pm$  0.06 &             $<0.002$  &  6900 $\pm$  400 & $<5500$                \\
(133,+18)\tablenotemark{b}  &  0.30 $\pm$  0.03 &             $<0.02$  &  6600 $\pm$ 200 & $<11300$ \\
(133,+18)\tablenotemark{c}  &  0.42 $\pm$  0.03 &             $<0.01$  &  7300 $\pm$ 100 & $<9500$    \\
 \enddata
\tablenotetext{a}{Data from \cite{Reynolds+01}}
\tablenotetext{b}{Component with \vlsr $\approx$ -20 \kms}
\tablenotetext{c}{Component with \vlsr $\approx$ -60 \kms}

\end{deluxetable}

\end{document}